\documentclass[aps, prl, twocolumn, letterpaper, nofootinbib, superscriptaddress, balancelastpage, longbibliography]{revtex4-1}
\usepackage{graphicx}
\usepackage{url}
\usepackage{xcolor}
\usepackage[colorlinks,citecolor=blue,urlcolor=blue,linkcolor=blue]{hyperref}
\usepackage{bm, bbold, xcolor, slashed}
\usepackage{amsmath}
\usepackage{amssymb}
\usepackage{easy-todo}
\usepackage{stmaryrd}
\usepackage{braket}

\def\Caravel{{\textsc{Caravel}}}
\def\fire{{\textsc{Fire~6}}}
\DeclareMathOperator{\arccosh}{arccosh}
\newcommand{\mA}{m_{A}}
\newcommand{\mS}{m_{\phi}}
\newcommand{\ep}{\epsilon}
\newcommand{\bk}{\bm{k}}
\newcommand{\bp}{\bm{p}}
\newcommand{\br}{\bm{r}}
\newcommand{\bq}{\bm{q}}
\newcommand{\bS}{\bm{S}}
\newcommand{\bL}{\bm{L}}

\newcommand{\bpb}{\bar{\bm{p}}}
\newcommand{\bep}{\bm{\ep}}

\begin{document}
\title{Conservative Binary Dynamics with a Spinning Black Hole at $\mathcal{O}(G^3)$ from Scattering Amplitudes}

\author{Fernando Febres Cordero}
\email{ffebres@hep.fsu.edu}
\affiliation{Physics Department, Florida State University, Tallahassee, FL
32306-4350, USA}

\author{Manfred Kraus}
\email{mkraus@hep.fsu.edu}
\affiliation{Physics Department, Florida State University, Tallahassee, FL
  32306-4350, USA}

\author{Guanda Lin} 
\email{linguandak@pku.edu.cn}
\affiliation{CAS Key Laboratory of Theoretical Physics, Institute of
Theoretical Physics, Chinese Academy of Sciences, Beijing, 100190, China}
\affiliation{Higgs Centre for Theoretical Physics, University of Edinburgh,
James Clerk Maxwell Building, Peter Guthrie Tait Road, Edinburgh, EH9 3FD,
United Kingdom}

\author{Michael S. Ruf}
\email{mruf@physics.ucla.edu} 
\affiliation{Mani L. Bhaumik Institute for Theoretical Physics, University of
California at Los Angeles, Los Angeles, CA 90095, USA} 

\author{Mao Zeng} 
\email{mao.zeng@ed.ac.uk}
\affiliation{Higgs Centre for Theoretical Physics, University of Edinburgh,
James Clerk Maxwell Building, Peter Guthrie Tait Road, Edinburgh, EH9 3FD,
United Kingdom}

\begin{abstract}
We compute the conservative two-body Hamiltonian of a compact binary system
with a spinning black hole through $\mathcal{O}(G^3)$ to all orders in velocity,
including linear and quadratic spin terms.
To obtain our results we calculate the classical limit of the two-loop amplitude
for the scattering of a massive scalar particle with a massive
spin-1 particle minimally coupled to gravity. We employ modern scattering amplitude
and loop integration techniques, in particular numerical unitarity, integration-by-parts
identities, and the method of regions. The conservative potential in terms of rest-frame
spin vectors is extracted by matching to a non-relativistic effective field theory.
We also apply the Kosower-Maybee-O'Connell (KMOC) formalism to calculate the impulse in
the covariant spin formalism directly from the amplitude.
We work systematically in conventional dimensional regularization and explicitly evaluate
all divergent integrals that appear in full- and effective-theory amplitudes,
as well as in the phase-space integrals that arise in the KMOC formalism.
\end{abstract}

\maketitle

\section{Introduction}
After the momentous observation of gravitational waves produced by the merging
of a black hole pair~\cite{LIGOScientific:2016aoc}, the LIGO-Virgo-KAGRA
collaboration continues to explore gravitational waves generated by compact
binary systems. The systems so far detected~\cite{LIGOScientific:2021djp} are
mainly formed by a pair of black holes with masses in the range of 5 to 100
solar masses, although a handful of events have been associated to systems
involving one or two neutron stars.  Extracting such details about the binary
systems that produced the detected gravitational waves hinges on our ability to
precisely predict the structure of such signals based on the parameters of the
systems, like for example the masses of the coalescent compact objects,
especially in the light of the next-generation gravitational wave detectors~\cite{Punturo_2010,Reitze:2019iox}.
A critical theoretical input to produce these predictions is the conservative
Hamiltonian of the two-body system.

Great efforts have been devoted to improve the predictions of
binary dynamics through perturbative means.
We work in the post-Minkowskian (PM) framework \cite{Bertotti:1956pxu,
Kerr:1959zlt, Bertotti:1960wuq, Portilla:1979xx, Westpfahl:1979gu,
Portilla:1980uz, Bel:1981be, Westpfahl:1985tsl, Ledvinka:2008tk,
Damour:2016gwp} which expands in Newton's constant while keeping the full
dependence on velocity.
Long considered difficult, high-order PM corrections to the
conservative dynamics of spinless binary system enjoyed recent progress using
scattering amplitudes and their classical limits \cite{Bjerrum-Bohr:2018xdl, Cheung:2018wkq, Kosower:2018adc}, in combination with effective field theory and advanced integration techniques. The computation of the conservative two-body dynamics at
$\mathcal{O}(G^3)$ \cite{Bern:2019nnu, Bern:2019crd, DiVecchia:2021bdo, Herrmann:2021tct, Bjerrum-Bohr:2021din, Brandhuber:2021eyq} and $\mathcal{O}(G^4)$ 
\cite{Bern:2021dqo, Bern:2021yeh} for spinless binary systems has been completed.\footnote{The $\mathcal{O}(G^3)$ and $\mathcal{O}(G^4)$ results have also been obtained \cite{Kalin:2020fhe, Dlapa:2021npj, Dlapa:2021vgp} in the worldline approach. See also
progress in post-Newtonian  corrections in the last few years in
Refs.~\cite{Foffa:2019hrb,Blumlein:2019zku,
Blumlein:2020pyo,Blumlein:2020pog,Blumlein:2021txe, Blumlein:2020znm,
Blumlein:2021txj, Bini:2020wpo, Bini:2020nsb, Bini:2020uiq, Bini:2020rzn}.}

Including spin effects is expected to be of high relevance in order to describe
some of the detected binary systems (see e.g.
Ref.~\cite{LIGOScientific:2021djp}). This is particularly the case due to known
degeneracies between the mass-ratio of the binary system and their spin, which
leads to strongly biased parameter
inference~\cite{Buonanno:2009zt,Brown:2012qf,Baird:2012cu, Krishnendu:2021cyi}.
Therefore obtaining predictions for spin-dependent terms
in the conservative potential is a necessity.
Related results for up to fifth post-Newtonian (PN) order have appeared employing 
classical~\cite{Barker:1970zr, Barker:1975ae, Kidder:1992fr, Kidder:1995zr, Tagoshi:2000zg, Faye:2006gx, Blanchet:2006gy, Damour:2007nc, Steinhoff:2007mb, Steinhoff:2008zr, Steinhoff:2008ji, Marsat:2012fn, Hergt:2010pa, Hergt:2012zx, Bohe:2012mr, Hartung:2013dza, Marsat:2013wwa, Bohe:2015ana, Bini:2017pee, Siemonsen:2017yux} and effective field theory methods ~\cite{Porto:2005ac, Porto:2006bt, Porto:2007tt, Porto:2008tb, Levi:2008nh, Porto:2008jj, Porto:2010tr, Levi:2010zu, Porto:2010zg, Hartung:2011ea, Levi:2011eq, Porto:2012as, Levi:2014sba, Levi:2014gsa, Levi:2015msa, Levi:2015uxa, Levi:2015ixa, Levi:2016ofk, Maia:2017gxn, Maia:2017yok, Levi:2019kgk, Levi:2020kvb, Levi:2020uwu, Antonelli:2020aeb, Levi:2020lfn, Antonelli:2020ybz, Goldberger:2020fot, Liu:2021zxr, Cho:2021mqw, Kim:2021rfj, Cho:2022syn}.
In the PM approach, progress has been made at the first two PM orders~\cite{Bini:2017xzy, Bini:2018ywr, Vines:2017hyw, Vines:2018gqi, Guevara:2017csg, Guevara:2018wpp, Chung:2018kqs, Arkani-Hamed:2019ymq, Guevara:2019fsj, Chung:2019duq, Damgaard:2019lfh, Aoude:2020onz, Chung:2020rrz, Guevara:2020xjx}.
At $\mathcal O(G^2)$, the scattering amplitudes approach have produced results up to quadratic-in-spin terms~\cite{Bern:2020buy, Kosmopoulos:2021zoq},
up to quartic-in-spin terms~\cite{Chen:2021qkk}, fifth-power in spin~\cite{Bern:2022kto},
and even including all-orders-in-spin contributions~\cite{Aoude:2022trd, Aoude:2022thd}.
Furthermore, recently a study based on the worldline quantum field
theory (QFT) formalism~\cite{Mogull:2020sak,Jakobsen:2021smu,Jakobsen:2021lvp, Jakobsen:2021zvh}
has been presented on the impulse 
and spin kick in the
scattering of two spinning compact objects including quadratic-in-spin contributions
and up to order $\mathcal{O}(G^3)$~\cite{Jakobsen:2022fcj}.
Proposals to express observables for  spinning binaries through a generating function
known as the eikonal exponent have appeared in Refs.~\cite{Bern:2020buy,Cristofoli:2021jas}.

There is not a single definition of conservative dynamics in the literature, and here we adopt the definition used by Damour \textit{et.\ al.}~\cite{Damour:1995kt, Damour:2016gwp} as well as Ref.~\cite{Bern:2021yeh}, based on a time-symmetric prescription for the graviton propagators, thereby excluding contributions from radiation-reaction and in particular contributions from zero-frequency graviton exchanges.
These effects conserve the total energy of the system but in general not the orbital angular momentum. To determine the latter contributions it is efficient to employ complementary approaches based on linear response relations or soft theorems (see e.g.\ Refs.\cite{Jakobsen:2022fcj} and ~\cite{Alessio:2022kwv} for recent applications involving spin; and the latter reference for hidden relations with the spinless case).

In this letter we compute for the first time the conservative binary Hamiltonian with a spinning black hole up to $\mathcal{O}(G^3)$ including
terms linear and quadratic in spin. 
\section{Scattering Amplitudes} 
We make use of the fact that $2\to 2$
scattering amplitudes for processes involving spin-$s$ massive particles can be
used to fully characterize the conservative spinning dynamics through $2s$
powers of spin~\cite{Vaidya:2014kza}. Therefore,
we consider a theory of a massive scalar $\phi$ and a massive spin-1 (vector)
field $A_\mu$ minimally coupled to gravity, which is described by the following
Lagrangian
\begin{align}
	\mathcal{L} ={}& \sqrt{-g}\left[-\frac{2 R}{\kappa^2}+\frac{1}{2}g^{\mu\nu}
	\partial_\mu\phi \partial_\nu\phi - \frac{1}{2}\mS^2\phi^2\right.\nonumber\\
	&\phantom{\sqrt{-g}}\quad\left.-\frac{1}{4}g^{\mu\alpha}g^{\nu\beta}F_{\alpha\beta}F_{\mu\nu} 
	+\frac{1}{2}m_A^2 g^{\mu\nu} A_\mu A_\nu\right]\,,
\label{eq:lagrangian}
\end{align}
where $\kappa=\sqrt{32\pi G}$, $g_{\mu\nu}$ is the metric tensor,
$g=\det(g_{\mu\nu})$,  $R$ is the Ricci scalar associated to $g_{\mu\nu}$, and
${F_{\mu\nu} = \partial_\mu A_\nu - \partial_\nu A_\mu}$.
Furthermore we work in the weak field approximation, i.e. $g_{\mu\nu} =
\eta_{\mu\nu} + \kappa h_{\mu\nu}$, where $h_{\mu\nu}$ is the graviton field.
We consider the elastic scattering process 
  $A(p_1,\ep_1) + \phi(p_2) \to \phi(p_3) + A(p_4,\ep_4),$
where $p_1^2 = p_4^2 = \mA^2$, $p_2^2 = p_3^2 = \mS^2$, and $\ep_{1,4}$ are the
polarization vectors of the incoming and outgoing vector particles.  We work in
the center-of-mass (COM) frame in which the momentum transfer takes the form $q^\mu =
p_2^\mu - p_3^\mu = (0,\bq)$. Momenta are
decomposed~\cite{Landshoff:1969yyn,Parra-Martinez:2020dzs} as $p_1 = \bar{p}_1 -
q/2$, $p_2 = \bar{p}_2 + q/2$, $p_3 = \bar{p}_2 -q/2$, $p_4 = \bar{p}_1 + q/2$,
with $\bar{p}_i\cdot q =0$. In the COM frame the spatial momenta read $\bp_1 =
-\bp_2 = \bpb - \bq/2 \equiv \bp$ and $E_{A,\phi}=\sqrt{m_{A,\phi}^2+\bp^2}$. We express the scattering amplitude in terms
of the variables $\{\mA,\mS,\sigma,q^2\}$ with $\sigma = \frac{p_1\cdot
p_2}{\mA\mS}$.

We numerically compute the tree-level, one- and two-loop scattering amplitudes
employing the multi-loop numerical unitarity method~\cite{Ita:2015tya,
Abreu:2017idw, Abreu:2017xsl, Abreu:2017hqn, Abreu:2018jgq}, as implemented 
in the \Caravel{} framework~\cite{Abreu:2020xvt}, using finite field
arithmetic to allow for the reconstruction of corresponding analytic
expressions~\cite{vonManteuffel:2014ixa, Peraro:2016wsq}. This framework has
already been used in the calculation of two-loop amplitudes in
gravity~\cite{Abreu:2020lyk} and we have extended it here to allow the handling of massive
particles. We have also included the interaction vertices needed in our calculation
from the Lagrangian in Eq.~\eqref{eq:lagrangian} with the
help of the \textsc{xAct}~\cite{Brizuela:2008ra,Nutma:2013zea} package.

We decompose the scattering amplitudes in terms of $5$ form factors
$\mathcal{M} = \sum_{n=1}^5M_n T_n$,
$T_n=\ep_{1,\mu}T_n^{\mu\nu}\ep_{4,\nu}^\star$, with
\begin{equation}
  \{T^{\mu\nu}_1,{\dots}, T^{\mu\nu}_5\}{=}\{\eta^{\mu\nu}, q^\mu q^\nu,q^2 \bar{p}_2^\mu\bar{p}_2^\nu,\,\bar{p}_2^{[\mu} q^{\nu]},\bar{p}_2^{(\mu} q^{\nu)}\}\,,
\label{eqn:tensors}
\end{equation}
where  $a^{(\mu}b^{\nu)}=a^{\mu}b^{\nu}+a^{\nu}b^{\mu}$ and  $a^{[\mu}b^{\nu]}=a^{\mu}b^{\nu}-a^{\nu}b^{\mu}$.
The form factors $M_n$ are obtained by computing $5$ different helicity amplitudes
and solving numerically for them.
Parity invariance and crossing symmetry imply that only $M_{1,2,3,4}$ are present in the amplitude, though $M_5$ is present in the integrand before loop integration.

To perform the loop integration we employ integration-by-parts (IBP)
identities~\cite{Chetyrkin:1981qh}, obtained with the
\fire{}~\cite{Smirnov:2019qkx} program.  This step is highly simplified by
employing finite field values for the kinematic invariants and truncating the
IBP relations by taking into account the power counting of the contributions
relevant to the classical result.  After doing an univariate rational
reconstruction of the master-integral coefficients in $q^2$, we expand in a
Laurent series all master integrals and their coefficients for small momentum
transfer $q^2$ and in $\ep$, using the results of
Ref.~\cite{Parra-Martinez:2020dzs} and keeping terms up to
$\mathcal{O}\big((q^2)^0\big)$. The coefficients in this double expansion are
yet-unknown rational functions in the variables $\{\mA,\mS,\sigma\}$. We then
employ multiple numerical evaluations to perform a multivariate rational
reconstruction~\cite{Peraro:2016wsq}.  We are able to reconstruct
the analytic expressions for the relevant classical terms of all required
amplitudes by employing modest computational resources. 
The resulting amplitudes are given in the supplemental material, which includes
computer-readable files~\cite{SpinBBHResults}.

\section{Effective Field Theory} 
We employ effective field theory
(EFT) techniques~\cite{Neill:2013wsa,Cheung:2018wkq,Bern:2020buy} to extract
the two-body Hamiltonian for our system.  We construct a
non-relativistic EFT with non-local contact interactions described by the
Lagrangian
\begin{align}
 L_{\textrm{EFT}} &= 
 \int_{\bk} \hat{\phi}^\dagger(-\bk)\left( \mathrm{i}\partial_t - \sqrt{\bk^2 + \mS^2}\right)\hat{\phi}(\bk) \nonumber\\
 &+ \int_{\bk} \hat{A}^{\dagger,i}(-\bk)\left( \mathrm{i}\partial_t - \sqrt{\bk^2 + \mA^2}\right)\hat{A}^i(\bk) \\
 &- \int_{\bk,\bk^\prime} \tilde{V}_{ij}(\bk,\bk^\prime)\hat{A}^{\dagger,i}(\bk^\prime)\hat{A}^j(\bk)\hat{\phi}^\dagger(-\bk^\prime)\hat{\phi}(-\bk)\;,\label{eq:EFTLagrangian}\nonumber
\end{align}
with $\int_{\bk} \equiv \int \frac{d^3\bk}{(2\pi)^3}$. The fields $\hat{\phi}$
and $\hat{A}$ are defined in their respective rest frames. The potential
$\tilde{V}_{ij}$ can be decomposed
into different spin operators according to 
\begin{equation}
 \tilde{V}_{ij}(\bk,\bk^\prime) = \sum_{n=1}^{4} \tilde{V}^{(n)}(\bk,\bk^\prime) O^{ij}_n\left(\frac{\bk+\bk^\prime}{2},\bk^\prime-\bk\right)\;,
 \label{eqn:potential}
\end{equation}
where the non-relativistic tensor structures $O_n^{ij}(\bpb,\bq)$ are 
\begin{equation}
\{O^{ij}_1,\dots, O^{ij}_4\}=\{\delta^{ij},\, \bpb^{[i} \bq^{j]},\bq^i\bq^j,\bq^2 \bpb^i\bpb^j\}\,.
\label{eq:operators}
\end{equation}
To connect to the classical spin we relate the non-relativistic tensor
structures to spin operators using the representation  $(\bS_i)_{jk} = -\mathrm{i}\ep_{ijk}$. We find
\begin{equation}
\begin{split}
O_1&=\mathbb{1}\;,\qquad 
O_2 = -\mathrm{i}(\bq\times\bpb)\cdot\bS\;,\\
O_3&=\frac{1}{2}\bq^2\bS^2-(\bq\cdot\bS)^2\;, \\
O_4 &= \bq^2\left( \frac{1}{2}\bpb^2\bS^2 - (\bpb\cdot\bS)^2\right)\;,
\end{split}
\label{eq:Spinoperators}
\end{equation}
where we defined $O_n \equiv O^{ij}_n(\bpb,\bq)\hat{A}^{\dagger,i}(\bpb-\bq/2)\hat{A}^j(\bpb+\bq/2)$.
We ignore the Casimir operator ${O_5=\bq^2 \bm S^2}$ in the amplitude calculation for simplicity in this initial study,
though we determine its coefficient in the potential by requiring consistency
with the 
results of Ref.~\cite{Jakobsen:2022fcj} in the aligned-spin limit\footnote{Up to $O(G^2)$, this could be fixed by conjectured relations between different spin structures in Refs.~\cite{Bern:2022kto} and \cite{Aoude:2022trd}.}. 
Note that the polarization tensor representation of $O_n^{ij}$
in Eq.~\eqref{eq:operators} can be used within conventional
dimensional regularization for computing divergent EFT amplitudes,
while the spin operators in Eq.~\eqref{eq:Spinoperators}
are intrinsically $3$-dimensional and are used to represent the finite
potential obtained by EFT matching.
We normalize the coefficients of the operators in
Eq.~\eqref{eqn:potential} including dimensional regularization factors as
\begin{align}
\tilde{V}^{(n)}(\bk,\bk^\prime) ={}& \frac{(4\pi)^{3/2-\epsilon}}{\bq^2}\sum_{\substack{L,Q=0}}^\infty \left(\frac{\mu^2}{\bq^2}\right)^{L\ep} \frac{G^{L+1}|\bq|^{L+Q}}{2^{(L+1)(1-2\ep)}}  \nonumber\\ 
&\ \times\frac{\Gamma \left[1-\frac{1}{2}\left(1-2\epsilon \right) L\right]}{\Gamma \left[\frac{1}{2} (1-2
	\epsilon ) (L+1)\right]}\tilde{c}^{(n)}_{L+1,Q}(\bk^2)\;,\label{eqn:pot_ansatz}
\end{align}
where $\bq = \bk^\prime - \bk$ and $\mu$ is the dimensional regularization
scale, while  $L$ denotes the loop order and $Q$ the additional powers of
$\hbar$ beyond the classical limit.
The unknown coefficients
$\tilde{c}^{(n)}_{L,Q}(\bk^2)$ are fixed by matching the EFT and full-theory
amplitudes.

\begin{figure}[ht!]
\begin{equation*}
 \includegraphics[height=1.6cm]{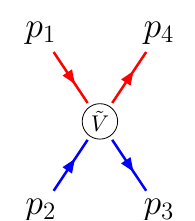} \raisebox{0.7cm}{\Large +}
 \includegraphics[height=1.6cm]{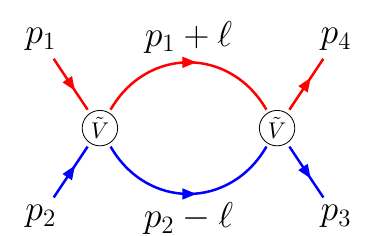}  \raisebox{0.7cm}{\Large +}
 \includegraphics[height=1.6cm]{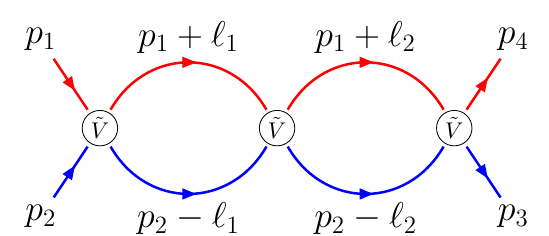}
\end{equation*}
 \caption{The EFT amplitude is given by iterated bubble diagrams. Blue (red)
lines refer to scalar (vector) particles. Each potential $\tilde{V}$ has a perturbative
expansion in terms of $\kappa$.}
 \label{fig:eft_diags}
\end{figure}
The EFT amplitude is given by iterated bubble diagrams~\cite{Cheung:2018wkq}, see
Fig.~\ref{fig:eft_diags}. We start by decomposing the numerator of each diagram in terms of the
operators $O_n$ by explicitly constructing projectors. Then, we follow 
the procedure outlined in Ref.~\cite{Cheung:2018wkq}, i.e. we integrate
out energy components of the loop momenta using the residue theorem. Then the propagators are expanded for in the soft region $\ell\sim |\bq|$, yielding the amplitude in terms of the
\textit{linearized} $(3-2\epsilon)$-dimensional master integrals of Ref.~\cite{Parra-Martinez:2020dzs}.
We then obtain the $L$-loop EFT amplitude as an expansion in  $|\bm q|$,
\begin{equation}
 \mathcal{M}_{\textrm{EFT}}^{(L)} = \left(\frac{\kappa}{2}\right)^{2+2L} \left(\frac{\mu^2}{\bq^2}\right)^{L\ep}
 \sum_{n=1}^{4} \sum_{k=0}^\infty \frac{ d^{(n)}_{L,k}}{|\bq|^{2-k}}~ O_n \;.
 \label{eqn:eft_expansion}
\end{equation}
As explained in Refs.~\cite{Chung:2018kqs,Bern:2020buy} to perform the matching to the relativistic loop amplitude 
the relativistic tensor structures $T_n$ have to be converted into non-relativistic counterparts $O_n$. The necessary relations are given in the supplemental material~\cite{SpinBBHResults}.

Let us discuss the differences between the EFT matching
procedure that we have employed and the one introduced in Ref.~\cite{Cheung:2018wkq} and later used in
Refs.~\cite{Bern:2019nnu,Bern:2019crd} at the two-loop order and
in Ref.~\cite{Bern:2020buy} for the case of spin.  In these references, the
integration of the relativistic amplitude is performed by an expansion and
resummation in the velocity. In this process IR-divergent integrals
are kept unevaluated and explicitly matched to the corresponding EFT integrals,
therefore strictly speaking no dimensional regularization is required.
However, this strategy relies on subtle arguments for the cancellation of evanescent
effects, i.e.\ terms subleading in $\epsilon$ or $|\bq|$ at lower loops being promoted to
physically-significant terms after multiplying divergent iteration integrals.
Here we explicitly evaluate all such subleading terms
in dimensional regularization in a transparent manner.
This makes it natural to use dimensional regularization in an relativistic
approach to conservative classical dynamics as introduced in Ref.~\cite{Parra-Martinez:2020dzs}
in conjunction with treating the EFT in dimensional regularization as described in this letter.

\section{Classical Hamiltonian} 
The result of the EFT calculation described before are the matching
coefficients $\tilde{c}^{(n)}_{L+1}=\tilde{c}^{(n)}_{L+1,0}$. For the spin-orbit coupling we
find
		\newcommand{\Scalar}{\tilde{c}^{(1)}}
	\newcommand{\SpinOrbit}{\tilde{c}^{(2)}}
\begin{align}
	 \SpinOrbit_{L+1} &=\SpinOrbit_{L+1,\mathrm{red}} +\SpinOrbit_{L+1,\mathrm{iter}} +\frac{\Scalar_{L+1,\mathrm{red}}}{m_A^2(\gamma_1+1)}\ ,\label{eq:cecompositionCF}
\end{align}
with $\gamma_1 = E_A/\mA$. 
The coefficients $\Scalar_{L+1,\mathrm{red}}$ are related to the scalar
coupling~\cite{Bern:2019nnu,Bern:2019crd},
while the $\SpinOrbit_{L+1,\mathrm{iter}}$ are given in terms of lower-order coefficients.
We find
\begin{widetext}
\begin{align}
\Scalar_{1,\mathrm{red}}(\bk^2)={}&\frac{m_A^2m_\phi^2}{E^2\xi}\left(1-2\sigma^2\right)\,,\qquad
\Scalar_{2,\mathrm{red}}(\bk^2)={} \frac{3(m_\phi+m_A)m_\phi^2m_A^2}{4E^2\xi}(1-5\sigma^2)\,,\\
\Scalar_{3,\mathrm{red}}(\bk^2)={}&\frac{m_A^2m_\phi^2}{E^2\xi}\left[-\frac{2}{3} m_A m_\phi \left(\frac{\operatorname{arccosh}(\sigma)}{\sqrt{\sigma^2-1}} \left(-12 \sigma ^4+36 \sigma ^2+9\right)+22 \sigma ^3-19 \sigma \right)-2 (m_\phi^2+m_A^2) \left(6 \sigma ^2+1\right)\right]\nonumber\\
&+\frac{3Em_A^2m_\phi^2}{4E^2\xi}(m_A+m_\phi)\frac{(1-2\sigma^2)(1-5\sigma^2)}{(\sigma^2-1)}-\frac{3m_A^4m_\phi^4}{E^2\xi\bk^2}\ ,\\
\SpinOrbit_{1,\mathrm{red}}(\bk^2)={}&-\frac{2\sigma m_\phi}{E\xi}\ ,\qquad
\SpinOrbit_{2,\mathrm{red}}(\bk^2)={} \frac{\mS(4\mA+3\mS)\sigma(5\sigma^2-3)}{4E\xi(\sigma^2-1)}\ ,\\
\SpinOrbit_{3,\mathrm{red}}(\bk^2) ={}& \frac{m_\phi }{E\xi(\sigma^2{-}1)^2}\left[\left.{-}2m_A^2\sigma(3{-}12\sigma^2{+}10\sigma^4){-}\left(\tfrac{83}{6}{+ }27\sigma^2{-}52\sigma^4{+}\tfrac{44}{3}\sigma^6\right)m_Am_\phi- m_\phi^2\sigma\left(\tfrac{7}{2}{-}14\sigma^2{+}12\sigma^4\right)\right.\right.\nonumber\\
  &\qquad\qquad\qquad\left. {+}\frac{(4m_A{+}3m_\phi)E}{4}\sigma(2\sigma^2{-}1)(5\sigma^2{-}3){+}4m_A m_\phi\sigma(\sigma^2{-}6)(2\sigma^2{+}1)\sqrt{\sigma^2{-}1}\operatorname{arccosh}(\sigma)\right]\,,\\
\SpinOrbit_{1,\mathrm{iter}}(\bk^2)={}&0\,,\qquad
\SpinOrbit_{2,\mathrm{iter}}(\bk^2)={} E \xi  {\tilde c}_1^{\text{(2)}} \frac{\partial {\tilde c}_1^{\text{(1)}}}{\partial \bk^2}+{\tilde c}_1^{\text{(1)}} \left(E \xi  \frac{\partial {\tilde c}_1^{\text{(2)}}}{\partial \bk^2}+\frac{{\tilde c}_1^{\text{(2)}} \left(\frac{2 E^2 \xi }{\bk^2}+\frac{1}{\xi }-3\right)}{2 E}\right)\,,\\
\SpinOrbit_{3,\mathrm{iter}}(\bk^2)={}&\left({\tilde c}_1^{\text{(1)}}\right){}^2 \left({-}\frac{2}{3} E^2 \xi ^2 \frac{\partial ^2{\tilde c}_1^{\text{(2)}}}{\partial (\bk^2)^2}{+}\left(\xi  \left(3{-}\frac{E^2 \xi }{\bk^2}\right){-}1\right) \frac{\partial {\tilde c}_1^{\text{(2)}}}{\partial \bk^2}{+}{\tilde c}_1^{\text{(2)}} \left(\frac{\frac{1}{2 \xi }{-}2}{E^2}{+}\frac{3 \xi {-}1}{\bk^2}\right)\right)\\
	&{+}{\tilde c}_1^{\text{(1)}} \left({\tilde c}_1^{\text{(2)}} \left(\left({-}\frac{3 E^2 \xi ^2}{\bk^2}{+}6 \xi {-}2\right) \frac{\partial {\tilde c}_1^{\text{(1)}}}{\partial \bk^2}{-}\frac{4}{3} E^2 \xi ^2 \frac{\partial ^2{\tilde c}_1^{\text{(1)}}}{\partial (\bk^2)^2}\right)\right.\nonumber \\
	&\qquad\qquad\left. {+}\frac{4}{3} E \xi  \left(\frac{\partial {\tilde c}_2^{\text{(2)}}}{\partial \bk^2}{-}2 E \xi  \frac{\partial {\tilde c}_1^{\text{(1)}}}{\partial \bk^2} \frac{\partial {\tilde c}_1^{\text{(2)}}}{\partial \bk^2}\right){+}\frac{E^2 \xi ^2 \left({\tilde c}_1^{\text{(2)}}\right){}^2}{2 \bk^2}{+}{\tilde c}_2^{\text{(2)}} \left(\frac{\frac{2}{3 \xi }{-}2}{E}{+}\frac{E \xi }{\bk^2}\right)\right){-}\frac{1}{6} E^2 \xi ^2 \left({\tilde c}_1^{\text{(2)}}\right){}^3\nonumber\\
	&{+}{\tilde c}_1^{\text{(2)}} \left(\frac{2}{3} E \xi  \left(\frac{\partial {\tilde c}_2^{\text{(1)}}}{\partial \bk^2}{-}2 E \xi  \left(\frac{\partial {\tilde c}_1^{\text{(1)}}}{\partial \bk^2}\right){}^2\right){+}\frac{{\tilde c}_2^{\text{(1)}} \left(\frac{3 E^2 \xi }{\bk^2}{+}\frac{1}{\xi }{-}3\right)}{3 E}\right){+}\frac{2}{3} E \xi  {\tilde c}_2^{\text{(1)}} \frac{\partial {\tilde c}_1^{\text{(2)}}}{\partial \bk^2}{+}\frac{4}{3} E \xi  {\tilde c}_2^{\text{(2)}} \frac{\partial {\tilde c}_1^{\text{(1)}}}{\partial \bk^2}\nonumber\,,
\end{align}
\end{widetext}
where $E = E_A + E_\phi$ and $\xi = E_AE_\phi/E^2$.
These expressions represent our results for the spin-orbit term in the momentum-space potential
in Eq.~\eqref{eqn:pot_ansatz}. The corresponding results for the quadratic-in-spin
terms are given in the supplemental material~\cite{SpinBBHResults}. 
Finally we convert to position space and obtain the Hamiltonian by computing the Fourier transform of  the potential in Eq.~\eqref{eqn:pot_ansatz}.
The resulting Hamiltonian reads
\begin{equation}
\begin{split}
H &= \sqrt{\bp^2+\mA^2} + \sqrt{\bp^2 + \mS^2} + V^{(1)}(\br^2,\bp^2) \\
&+ V^{(2)}(\br^2,\bp^2)\frac{\bL\cdot\bS}{\br^2}
+ V^{(3)}(\br^2,\bp^2)\frac{(\br\cdot\bS)^2}{\br^4} \\
&+ V^{(4)}(\br^2,\bp^2)\frac{(\bp\cdot\bS)^2}{\br^2}
+ V^{(5)}(\br^2,\bp^2)\frac{\bS^2}{\br^2} \;,
\end{split}
\label{eq:hamiltonian}
\end{equation}
where each coefficient is expanded as
\begin{equation}
V^{(n)}(\br^2,\bp^2) = \sum_{L=1}^3 \left(\frac{G}{r}\right)^L~c_L^{(n)}(\bp^2) + \mathcal{O}(G^4)\;.
\end{equation}
The scalar potential $V^{(1)}(\br^2,\bp^2)$ coincides with the
results of Ref.~\cite{Bern:2019nnu,Bern:2019crd}. 
The $c_L^{(n)}(\bp^2)$ coefficients
are provided in the supplemental material~\cite{SpinBBHResults} as linear sums
of the $\tilde c_{L,0}^{(n)} (\bp^2)$ coefficients in
Eq.~\eqref{eqn:pot_ansatz}.

\noindent
\textit{Validation.} 
We performed several checks of our computational framework in dimensional
regularization. We checked that the EFT amplitudes reduce to
their scalar counterparts given in Eq.~(10.6) of Ref.~\cite{Bern:2019crd} when
setting $\tilde{c}^{(n)}_{L,Q}=0$ for $n> 1$. 
Even more, we reproduced the two-loop matching coefficients of
Refs.~\cite{Bern:2019nnu,Bern:2019crd}.
For the EFT
calculation including spin, we cross checked our tree-level and one-loop
matching coefficients against Refs.~\cite{Holstein:2008sx,Kosmopoulos:2021zoq}.
The scalar coefficients up to one-loop are also in agreement with
Ref.~\cite{Cristofoli:2020uzm} to all orders in $\ep$.  Beyond one-loop we
checked explicitly the cancellation of the \emph{super-classical} terms, i.e.\ unphysical terms behaving as
powers of $1/\hbar \sim 1/|\bm q|$, where we
observe a non-trivial interplay between the various matching coefficients.
Furthermore the coefficients extracted in this way are finite in the
non-relativistic limit $|\bp|\ll 1$ (this can be seen explicitly from the
post-Newtonian expanded coefficients provided in the supplemental
material~\cite{SpinBBHResults}).

We solved the equation of motion based on the computed conservative potential
to obtain the impulse in either the limit of an aligned spin or the limit of
a spinless probe moving in a Kerr background. After translating to the
covariant spin formalism using the procedure of Refs.~\cite{Vines:2017hyw, Liu:2021zxr},
the impulse agrees with the recent worldline QFT result of Ref.~\cite{Jakobsen:2022fcj}
in either of the two above limits, when specializing to the Kerr black hole case.
The result is also consistent with the scattering angle in the probe limit obtained
in Ref.~\cite{Vines:2018gqi}.
Furthermore, the velocity expansion of our Hamiltonian is consistent with the next-to-next-to-leading order post-Newtonian results for spin-orbit and spin-squared coupling ~\cite{Levi:2015ixa, Levi:2015uxa} calculated by the EFTofPNG package \cite{Levi:2017kzq}, after finding a canonical transformation.\footnote{We thank Justin Vines for carrying out the check, and thank Jan Steinhoff for sharing a private code used for the canonical transformation and comparison.}

\section{Observables from the KMOC formalism} 
We employ the full-theory amplitudes we obtained to compute classical
gravitational observables using the KMOC formalism \cite{Kosower:2018adc}. In
this formalism, a change of classical variable $O$, associated with any operator
$\mathbb{O}$,  between \textit{in} and \textit{out} states is measured as
$
\Delta O=\Bra{\text{out}}\mathbb{O} \Ket{\text{out}}-\Bra{\text{in}}\mathbb{O} \Ket{\text{in}}.
$
Relating \textit{in} and \textit{out} states by the S-matrix allows to reformulate the problem in terms of scattering amplitudes.
In the classical limit
$\Delta O$ is given by~\cite{Kosower:2018adc}
\begin{equation}
\Delta O=\int \frac{{\mathrm{d}}^{D}q}{(2\pi)^{D-2}} \delta\left(-2 p_{1} \cdot q\right) \delta\left(2 p_{2} \cdot q\right) e^{\mathrm{i} b \cdot q}\left(\mathcal{I}_{O, \mathrm{v}}+\mathcal{I}_{O, \mathrm{r}}\right)
\label{eqn:KMOCkernel}
\end{equation}
where $\mathcal{I}_{O,\rm v}$ is the \textit{virtual} kernel obtained from virtual amplitudes, while $\mathcal{I}_{O,\rm r}$ is the \textit{real} kernel depending on cut amplitudes with phase-space integration. 

We begin with obtaining kernels $\mathcal{I}_{\rm{v}}$ and
$\mathcal{I}_{\rm{r}}$ in order to compute the momentum impulse $\Delta p_i$
(corresponding to the momentum operator $\mathbb{P}_{i}$). The spinless case has been discussed
in~Ref.~\cite{Herrmann:2021lqe} and we highlight new features introduced by
polarizations and spins here. The \textit{virtual} kernel $\mathcal{I}_{\Delta p,\rm
v}^{\mu}$ is simply $q^{\mu}\mathcal{M}$, which can be expanded with the tensors
in Eq.~\eqref{eqn:tensors}.
For the conservative dynamics we only need to consider two-particle cuts and the real kernel takes the form
\begin{equation}
\mathcal{I}_{\Delta p,\rm r}^{\mu}=\int \hskip -2pt  \mathrm{d}\Phi_{2}\sum_{\epsilon_{\rm c}} \ell^\mu \mathcal{M}_{\scriptscriptstyle \mathrm{L}}\mathcal{M}_{\scriptscriptstyle \mathrm{R}}^\star\ ,
\end{equation}
where $\mathrm{d}\Phi_{2}$ is the two-body phase-space measure and $\mathcal{M}_{\scriptscriptstyle \mathrm{R}}$ and $\mathcal{M}_{\scriptscriptstyle \mathrm{L}}^\star$  are amplitudes for the processes
$A(p_1,\ep_1) + \phi(p_2) \to \phi(p_1+\ell) + A(p_2-\ell,\ep_{c}),$
and $\phi(p_1+\ell) + A(p_2-\ell,\ep_{c}))\to \phi(p_3) + A(p_4,\ep_4),$ respectively.
The sum runs over the polarizations of the vector particles in the
cut. These one-loop integrals can be evaluated directly through sub-loop
integration, for details see Ref.~\cite{Herrmann:2021tct}. 
To expand the double-cut in terms of form factors, we need the additional
tensor structures 
\begin{equation}
	\{T^{\mu\nu}_6,\dots, T^{\mu\nu}_9\}=\{ \bar{p}_2^{[\mu} n^{\nu]},\, \bar{p}_2^{(\mu} n^{\nu)},\,q^{[\mu} n^{\nu]},\, q^{(\mu} n^{\nu)}\}\,,
\label{eqn:tensorsforcut}
\end{equation}
where $n^{\mu}=\epsilon^{\mu\nu\sigma\rho}q_{\nu}p_{1,\sigma}p_{2,\rho}$. After combining $\mathcal{I}_{\rm r}$ and $\mathcal{I}_{\rm v}$, we find that all super-classical terms cancel. 

The next step is to translate the polarization vectors into spin vectors.
Following Ref.~\cite{Maybee:2019jus}, we first boost the polarization vector
$\ep_4^{*}$ into the frame of $p_1$.  With the Pauli-Lubanski operator
$\mathbb{W}^{\mu}=\epsilon^{\mu\nu\sigma\rho}\mathbb{P}_{\nu}\mathbb{J}_{\rho\sigma}/2$,
we define the spin vector $s_{1}^{\mu}$ as the expectation value of
$\mathbb{W}_{1}^{\mu}/m_1$. It is straightforward to convert from $\epsilon_1$ to $s_1$ given $s^{\mu}_1=\mathrm{i} \epsilon^{\mu \nu \rho \sigma} p_{1,\nu} \ep_{ \rho}(p_1) \ep^{*}_{{\sigma}}(p_1)/{m_1}$.
Finally, we perform the Fourier transform from the momentum space to the impact
parameter space as in Eq.~\eqref{eqn:KMOCkernel}. We compare our momentum
impulse results up to $O(G^{3})$ with Ref.~\cite{Liu:2021zxr,Jakobsen:2022fcj} and
find agreement in the Kerr black hole case.
This serves as yet another check on the correctness of our amplitude and confirms the
practical value of the KMOC formalism at nontrivial PM orders
beyond the spinless case \cite{Herrmann:2021lqe, Herrmann:2021tct}.
There is no obstruction to computing other observables such as the spin kick using the KMOC formalism, which we leave to future work.
%

\section{Conclusions} 
In this letter we have presented the conservative two-body
Hamiltonian for a compact binary system with a spinning black hole at
$\mathcal{O}(G^3)$ with exact velocity dependence and including 
terms linear and quadratic in spin. Though obtained from scattering, the
Hamiltonian is directly applicable to bound orbits. This is especially
valuable since analytic continuation for observables in these two kinds
of orbits, through the boundary-to-bound map \cite{Kalin:2019rwq, Kalin:2019inp, Cho:2021arx}, has not been worked out for generic spin configurations.
The computation employs powerful modern scattering amplitude techniques,
including numerical unitary, integration-by-parts identities, expansion-by-regions 
and functional reconstruction algorithms.
 
We have shown that the effective field theory approach to study the binary
dynamics with
spin~\cite{Vaidya:2014kza,Bern:2020buy} can be applied in complex high-order
calculations. Furthermore, we have performed matching calculations for the effective
field theory of Ref.~\cite{Cheung:2018wkq} at the two-loop level, for both the
spinless case and spinning case, fully within dimensional regularization for the first time.
We have also demonstrated the power of the KMOC formalism beyond the lowest order in $G$
for spinning binary dynamics.

There are many directions to extend the applicability of our framework.
For instance, one can study Hamiltonian terms with higher powers in spin by
considering minimally-coupled massive higher-spin particles~\cite{Vaidya:2014kza, Arkani-Hamed:2019ymq, Chiodaroli:2021eug}.
Also including
 finite-size ~\cite{Cheung:2020sdj,Kalin:2020lmz, Bern:2020uwk,Cheung:2020gbf,
Bini:2020flp,Haddad:2020que,Aoude:2020ygw,Mougiakakos:2022sic} and
radiation~\cite{Luna:2016due, Goldberger:2016iau, Goldberger:2017frp,
Goldberger:2017vcg, Chester:2017vcz, Goldberger:2017ogt, Shen:2018ebu,
Bautista:2019evw, Goldberger:2019xef, Almeida:2020mrg, Laddha:2018rle,
Laddha:2018myi, Laddha:2018vbn, Sahoo:2018lxl, Ciafaloni:2018uwe,
Laddha:2019yaj, PV:2019uuv, Saha:2019tub, Manu:2020zxl, Sahoo:2020ryf,
Alessio:2022kwv} effects is an interesting possibility of relevance for
phenomenological studies in gravitational wave detectors. Finally, our
framework can be extended beyond the third-post-Minkowskian order following progress in the spinless case \cite{Bern:2021dqo, Bern:2021yeh, Dlapa:2021npj, Dlapa:2021vgp}.

\textbf{Acknowledgements:} 
We thank Harald Ita for collaboration during the initial phase of the project.
We thank Jan Steinhoff and Justin Vines for their expertise and timely help in checking the consistency of our Hamiltonian with known post-Newtonian results.
We thank Andres Luna, Chia-Hsien Shen, Dimitrios Kosmopoulos, Fei Teng, Justin Vines,
Radu Roiban, and Zvi Bern for insightful discussions and/or comments on the
manuscript.
The work of F.F.C.\ and M.K.\ is supported in part by the U.S. Department of
Energy under grant DE-SC0010102.  
G.L.\ is supported in part by the National Natural Science Foundation of
China Grants No.~11935013. G.L.\ thanks the Higgs Centre for Theoretical
Physics at the University of Edinburgh for the Visiting Researcher Scheme.
M.Z.’s work is supported in part by the U.K. Royal Society through Grant
URF\textbackslash R1\textbackslash 20109.
M.S.R.\ thanks the Higgs Centre for Theoretical Physics at the University of Edinburgh for hospitality and
the Mani L.\ Bhaumik Institute for generous support.

The computing for this project was performed on the HPC cluster at the Research
Computing Center at the Florida State University (FSU). For the purpose of open access,
the author have applied a Creative Commons Attribution (CC BY) license
to any Author Accepted Manuscript version arising from this submission.

\appendix
\onecolumngrid

\section{Supplementary Material}
\subsection{Full Theory Scattering Amplitudes}
\label{app:amplitudes}
Here we summarize the form factor decomposition of the $2\to 2$ scattering 
amplitudes in Eq.~(2) of the main text in conventional
dimensional regularization.
\subsubsection{Tree Level}
We decompose the tree-level scattering amplitude according to
\begin{equation}
 \mathcal{M}^{(0)} =\left(\frac{\kappa}{2}\right)^2 \frac{1}{\bq^2} \sum_{n=1}^4 M_n^{(0)}~T_n + \mathcal{O}(|\bq|^0)\;,
\end{equation}
with
\begin{equation}
  M_1^{(0)} = 2\mA^2\mS^2\left(\frac{1}{1-\ep} - 2\sigma^2\right)\;, \qquad 
  M_2^{(0)} = \frac{2\mS^2}{1-\ep}\;,\qquad
  M_3^{(0)} = 2\;, \qquad
  M_4^{(0)} = -4\mA\mS\sigma\;.
\end{equation}
%
\subsubsection{One Loop}
The one-loop scattering amplitude is given in the form of
\begin{equation}
 \mathcal{M}^{(1)} = \left(\frac{\kappa}{2}\right)^4 \frac{1}{\bq^2} \sum_{n=1}^4\sum_{k=0}^2 M_n^{(1,k)}~|\bq|^{k}~T_n + \mathcal{O}(|\bq|)\;.
\end{equation}
The super-classical terms are given by
\begin{align}
 M_1^{(1,0)} &= - \frac{2\mA^3\mS^3~f_2}{\sqrt{\sigma^2-1}}\left(\frac{1}{1-\ep} - 2\sigma^2\right)^2\;, \\
 M_2^{(1,0)} &= - \frac{\mA\mS^3~f_2}{(1-2\ep)\sqrt{\sigma^2-1}}\left[ \frac{2-3\ep}{(1-\ep)^2} - 4\sigma^2(1-\ep)\right]\;, \\
 M_3^{(1,0)} &=  \frac{\mA\mS~f_2}{\sqrt{\sigma^2-1}}\left[\frac{1-2\ep}{2(\sigma^2-1)(1-\ep)^2} +4\sigma^2 - \frac{2\ep}{(1-\ep)(1-2\ep)}\right]\;, \\
 M_4^{(1,0)} &=  \frac{4\mA^2\mS^2\sigma~f_2}{\sqrt{\sigma^2-1}}\left[\frac{1}{1-\ep} - 2\sigma^2\right]\;.
\end{align}
The classical contributions read
\begin{align}
 M_1^{(1,1)} &= \frac{\mA^2\mS^2(\mA+\mS)~f_1}{4(\sigma^2-1)}\frac{1-2\ep}{(1-\ep)^2}\Big( -3 + (3-4\ep)\sigma^2(6 - (5-4\ep)\sigma^2)\Big)\;, \\
 M_2^{(1,1)} &= \frac{\mS^2~f_1}{32(\sigma^2-1)}\left[ \frac{(5-4\ep)(1-12\ep+4\ep^2)}{(1-\ep)^2}\mA\sigma^4 - \frac{4(5-2\ep)(1-2\ep)}{(1-\ep)^2}\mS\right. \nonumber \\
 &+\frac{4(1 - \ep) (5 - 2\ep) (1 - 2\ep) (3 - 4\ep)}{(1-\ep)^3}\mS\sigma^2 -\frac{27 -\ep (67 - 4 \ep (14 - 3 \ep))}{(1-\ep)^3}\mA \nonumber \\
 &\left. +\frac{(78 - 2 \ep (147 - 2 \ep (113 - 2 \ep (35 - 6 \ep)))}{(1-\ep)^3}\mA\sigma^2 \right]\;, \\
 M_3^{(1,1)} &= \frac{f_1}{16(\sigma-1)^2}\left[\frac{2-4\ep(3-4\ep)}{(1-\ep)^2}\mS - \frac{8(1-2\ep)(3-4\ep)}{1-\ep}\mS\sigma^2 
 + \frac{2(1-2\ep)(5-4\ep)(3-4\ep)}{(1-\ep)^2}\mS\sigma^4\right. \nonumber \\
 &\left. + \frac{9-10\ep+8\ep^2}{(1-\ep)^2}\mA - \frac{2(33-2\ep(45-4\ep(11-4\ep)))}{(1-\ep)^2}\mA\sigma^2 + \frac{(5-4\ep)(13-2\ep(15-8\ep))}{(1-\ep)^2}\mA\sigma^4\right]\;, \\
M_4^{(1,1)} &= \frac{\mA\mS\sigma~f_1}{4(\sigma^2-1)}\frac{1-2\ep}{1-\ep}\left[ \mA(12 -4(5-4\ep)\sigma^2) + \mS\left(\frac{3(3-4\ep)}{1-\ep} -
\frac{(5-4\ep)(3-4\ep)}{1-\ep}\sigma^2\right)\right]\;.
\end{align}
And finally, the quantum corrections correspond to
\begin{align}
 M_1^{(1,2)} &= \frac{\mA\mS~f_2}{\sqrt{\sigma^2-1}}\left[ \frac{\ep}{2}\left(\frac{1}{1-\ep} -2\sigma^2\right)^2\frac{(\mA+\mS\sigma)^2}{\sigma^2-1}
 -\mS^2 \left(\frac{1+2\ep}{2(1-\ep)(1-2\ep)} - \frac{4\ep}{1-2\ep}\sigma^2 +2\ep\sigma^4\right)\right]\;, \\
 M_2^{(1,2)} &= \frac{\mS~f_2}{8(\sigma^2-1)^{3/2}}\frac{1}{(1-2\ep)(1-\ep)^2}\Bigg[ 
 -2(1-6\ep+3\ep^2)\frac{\mS^2}{\mA} + (8-44\ep+64\ep^2-40\ep^3+8\ep^4)\frac{\mS^2\sigma^2}{\mA} \nonumber \\
&-4(1-7\ep+7\ep^2)\mS\sigma + 16(1-\ep)^2(1-4\ep+\ep^2)\mS\sigma^3 -(1-4\ep+6\ep^2)\mA \nonumber \\
&+4\ep(1+3\ep-6\ep^2+2\ep^3)\mA\sigma^2 +8(1-\ep)^2(1-4\ep)\mA\sigma^4 \Bigg]\;, \\
 M_3^{(1,2)} &= \frac{f_2}{\mA\mS(\sigma^2-1)^{5/2}}\frac{1}{8(1-2\ep)}\Bigg[\frac{3}{1-\ep}\mA^2 - 4(6-\ep(7-4\ep))\mA^2\sigma^2 + 8(1-\ep)(3-2\ep)\mA^2\sigma^4 \nonumber \\
&- \frac{1+\ep}{(1-\ep)^2}\mS^2 - \frac{4-2\ep(13-2\ep(10-\ep(9-4\ep)))}{(1-\ep)^2}\mS^2\sigma^2 + \frac{8-16\ep(2-(2-\ep)\ep)}{1-\ep}\mS^2\sigma^4 \nonumber \\
&-\left(16 - \frac{6}{1-\ep}\right)\mA\mS\sigma + 8(3-4\ep)\ep\mA\mS\sigma^3 + 16(1-\ep)(1-2\ep)\mA\mS\sigma^5\Bigg]\;, \\
 M_4^{(1,2)} &= \frac{f_2}{(\sigma^2-1)^{3/2}}\Bigg[\mA^2\sigma -2(1-\ep)\mA^2\sigma^3 + \frac{1}{4(1-\ep)^2}\mA\mS - \frac{\ep}{1-\ep}\mA\mS\sigma^2 - (2-4\ep)\mA\mS\sigma^4 \nonumber \\
 &+\left(1+\frac{1}{2(1-\ep)} + \frac{1}{2(1-\ep)^2} - \frac{1}{1-2\ep}\right)\mS^2\sigma - \left(1-2\ep - \frac{1}{1-2\ep} + \frac{2}{1-\ep}\right)\mS^2\sigma^3\Bigg]\;.
\end{align}
Here, we have defined
\begin{equation}
 f_1 \equiv \frac{1}{4\pi}\left(\frac{4\pi\mu^2}{\bq^2}\right)^\ep \frac{\Gamma^2(\frac{1}{2}-\ep)\Gamma(\frac{1}{2}+\ep)}{2\sqrt{\pi}\Gamma(1-2\ep)}\;, 
 \qquad\qquad
 f_2 \equiv \frac{1}{4\pi}\left(\frac{4\pi\mu^2}{\bq^2}\right)^\ep \frac{i\Gamma^2(-\ep)\Gamma(1+\ep)}{4\Gamma(-2\ep)}\;.
\end{equation}
%
\subsubsection{Two Loops}
The two-loop scattering amplitude is given in the form of
\begin{equation}
 \mathcal{M}^{(2)} = -\left(\frac{\kappa}{2}\right)^6 \frac{1}{\bq^2} \mathcal{N} \sum_{n=1}^4\sum_{k=0}^2 M_n^{(2,k)}~|\bq|^{k}~T_n + \mathcal{O}(|\bq|)\;,
\end{equation}
where we keep a normalization factor of
\begin{equation}
\mathcal{N} =\left(\frac{\overline{\mu}^2}{\bq^2}\right)^{2\ep}= \left(\frac{\mu^2}{\bq^2}\right)^{2\ep}~\left(\frac{e^{\gamma_E}}{4\pi}\right)^{-2\ep}
\end{equation}
unexpanded.  Note that, at the $\mathcal{O}(1/|\bq|)$ order all master integrals are purely
imaginary and therefore do not contribute to the classical limit. Furthermore,
to extract the classical terms it is enough to know the $1/\ep$ poles of the
scattering amplitudes which are equal to $-1/2$ times the coefficients of $\log(\bq^2)$. With this, for the super-classical terms at $\mathcal{O}(1/\bq^2)$
we obtain
\begin{align}
 M_1^{(2,0)} &= \frac{\mA^4\mS^4(1-2\sigma^2)^2}{(8\pi)^2(\sigma^2-1)} \left[\frac{1-2\sigma^2}{\ep^2} +\frac{3}{\ep} \right] + \mathcal{O}(\ep^0)\;,\\
 M_2^{(2,0)} &= \frac{\mA^2\mS^4(1-2\sigma^2)}{(8\pi)^2(\sigma^2-1)} \left[\frac{1-2\sigma^2}{\ep^2} + \frac{4-2\sigma^2}{\ep}\right] + \mathcal{O}(\ep^0)\;,\\
 M_3^{(2,0)} &= \frac{\mA^2\mS^2}{(16\pi)^2(\sigma^2-1)^2} \left[\frac{(2\sigma^2-1)(1-8\sigma^2+8\sigma^4)}{\ep^2} + \frac{2(3-8\sigma^2)(\sigma^2-1)}{\ep}\right] + \mathcal{O}(\ep^0)\;,\\
 M_4^{(2,0)} &= -\frac{\mA^3\mS^3\sigma(1-2\sigma^2)}{(4\pi)^2(\sigma^2-1)}\left[\frac{1-2\sigma^2}{2\ep^2} + \frac{1}{\ep}\right] + \mathcal{O}(\ep^0)\;.
\end{align}
The super-classical terms at $\mathcal{O}(1/|\bq|)$ are given by
%
\begin{align}
 M_1^{(2,1)} &= -\mathrm{i}\pi \frac{3\mA^3\mS^3(\mA+\mS)}{(16\pi)^2\sqrt{\sigma^2-1}}\Bigg\{(1-7\sigma^2+10\sigma^4)\left[\frac{1}{\ep} - 2\log(2)\right] - 
 \frac{3+6\sigma^2-65\sigma^4+64\sigma^6}{3(\sigma^2-1)}\Bigg\} + \mathcal{O}(\ep)\;, \\
 M_2^{(2,1)} &= \frac{-\mathrm{i}\pi \mA\mS^3}{2(32\pi)^2(\sigma^2-1)^{3/2}}\Bigg\{ (2\sigma^2-1)\left(\mS 20(1-3\sigma^2) + \mA(27-78\sigma^2-5\sigma^4)\right)\left[\frac{1}{\ep}-2\log(2)\right] \nonumber \\
 &+ \frac{4}{3}\mS(-13-97\sigma^2+156\sigma^4) + \mA\frac{(-195+138\sigma^2+45\sigma^4+244\sigma^6)}{3}\Bigg\} + \mathcal{O}(\ep)\;, \\
 M_3^{(2,1)} &= \frac{\mathrm{i}\pi \mA\mS}{(32\pi)^2(\sigma^2-1)^{5/2}}\Bigg\{ \Big(\mA(2\sigma^2-1)(9-66\sigma^2+65\sigma^4) - \mS(2-28\sigma^2+78\sigma^4-60\sigma^6)\Big)
 \left[\frac{1}{\ep}-2\log(2)\right] \nonumber \\
 &+ \frac{2}{3}\Big(\mS(7-134\sigma^2+327\sigma^4-192\sigma^6) + \mA(-66+249\sigma^2-264\sigma^4+89\sigma^6)\Big)\Bigg\} + \mathcal{O}(\ep)\;, \\
 M_4^{(2,1)} &= \frac{-\mathrm{i}\pi\mA^2\mS^2\sigma}{(16\pi)^2(\sigma^2-1)^{3/2}}\Bigg\{(4\mA+3\mS)(3-11\sigma^2+10\sigma^4)\left[\frac{1}{\ep}-2\log(2)\right] \nonumber \\
 &\qquad\qquad\qquad\qquad+ \mS(3+47\sigma^2-64\sigma^4) + 4\mA(3+6\sigma^2-13\sigma^4)\Bigg\} + \mathcal{O}(\ep)\;.
\end{align}
The classical terms are
\begin{align}
 M_1^{(2,2)} &= \frac{1}{\ep^2}\frac{\mA^2\mS^4(1-2\sigma^2)}{(16\pi)^2(\sigma^2-1)} + \frac{1}{\ep}\Bigg[\frac{\mA^3\mS^3(3+12\sigma^2-4\sigma^4)}{(8\pi)^2\sqrt{\sigma^2-1}}\arccosh(\sigma) + \frac{\mA^4\mS^2 (-3 + 4\sigma^2) (1 - 8 \sigma^2 + 8\sigma^4)}{(16\pi)^2(\sigma^2-1)^2}\nonumber \\
&+\frac{\mA^2\mS^4 (-9 + 48 \sigma^2 - 78 \sigma^4 + 40 \sigma^6)}{(16\pi)^2(\sigma^2-1)^2} + 
   \frac{\mA^3\mS^3\sigma (-59 + 156 \sigma^2 - 162 \sigma^4 + 68 \sigma^6)}{6(8\pi)^2(\sigma^2-1)^2}\Bigg] + \mathcal{O}(\ep^0)\;, \\
 M_2^{(2,2)} &= \frac{1}{\ep^2}\Bigg[\frac{\mS^4 (1 - 2 \sigma^2) (1 - 4 \sigma^2)}{2(16\pi)^2(\sigma^2-1)^2} + 
 \frac{\mA\mS^3 \sigma (1 - 2 \sigma^2) (1 - 4 \sigma^2)}{(16\pi)^2(\sigma^2-1)^2} + 
 \frac{\mA^2\mS^2 (1 - 2 \sigma^2) (1 - 8 \sigma^4)}{(32\pi)^2(\sigma^2-1)^2} \Bigg] \nonumber \\
 &+ \frac{1}{\ep}\Bigg[\frac{\mA\mS^3 (3 + 24 \sigma^2 - 6 \sigma^4 + 4 \sigma^6)}{2(8\pi)^2(\sigma^2-1)^{5/2}}\arccosh(\sigma)
 + \frac{\mS^4 (-9 + 62 \sigma^2 - 72 \sigma^4)}{(32\pi)^2(\sigma^2-1)^2} \nonumber \\
 &- \frac{\mA \mS^3 \sigma (1059 - 278 \sigma^2 + 1064 \sigma^4)}{30(16\pi)^2(\sigma^2-1)^2}
 - \frac{\mA^2\mS^2 (7 - 60 \sigma^2 + 48 \sigma^4 + 32 \sigma^6)}{(32\pi)^2(\sigma^2-1)^2}\Bigg] + \mathcal{O}(\ep^0)\;, \\
 M_3^{(2,2)} &= \frac{1}{\ep^2}\Bigg[ \frac{\mA \mS \sigma (1 - 2 \sigma^2) (5 - 8 \sigma^4)}{2(16\pi)^2(\sigma^2-1)^3}
 - \frac{3\mA^2 (1 - 2 \sigma^2) (1 - 8 \sigma^2 + 8 \sigma^4)}{(32\pi)^2(\sigma^2-1)^3}
 + \frac{\mS^2 (1 + 2 (\sigma^2 - 8 \sigma^4 + 8 \sigma^6))}{(32\pi)^2(\sigma^2-1)^3}\Bigg] \nonumber \\
 &+ \frac{1}{\ep}\Bigg[\frac{\mA\mS (24 + 162 \sigma^2 + 35 \sigma^4 - 104 \sigma^6 + 8 \sigma^8)}{(16\pi)^2(\sigma^2-1)^{7/2}}\arccosh(\sigma)
 + \frac{\mS^2 (19 - 98 \sigma^2 + 200 \sigma^4 - 128 \sigma^6)}{2(32\pi)^2(\sigma^2-1)^3} \nonumber \\
 &- \frac{\mA\mS\sigma (7943 + 6986 \sigma^2 - 8516\sigma^4 + 1192\sigma^6)}{15(32\pi)^2(\sigma^2-1)^3}
 + \frac{\mA^2 (-21 + 2 \sigma^2 (7 + 64 \sigma^2 - 64 \sigma^4))}{2(32\pi)^2(\sigma^2-1)^3} \Bigg] + \mathcal{O}(\ep^0)\;, \\
 M_4^{(2,2)} &= \frac{1}{\ep^2}\Bigg[ - \frac{(\mA^3\mS+\mA\mS^3)\sigma (1 - 2 \sigma^2)^2}{2(8\pi)^2(\sigma^2-1)^2}
 - \frac{\mA^2\mS^2 (1 - 2 \sigma^2) (1 - 8 \sigma^4)}{2(16\pi)^2(\sigma^2-1)^2} \Bigg] \nonumber \\
 &+ \frac{1}{\ep}\Bigg[-\frac{\mA^2\mS^2 \sigma (-6 + \sigma^2) (1 + 2 \sigma^2)}{2(4\pi)^2(\sigma^2-1)^{3/2}}\arccosh(\sigma)
 + \frac{\mA\mS^3 \sigma (3 - 16 \sigma^2 + 16 \sigma^4)}{2(8\pi)^2(\sigma^2-1)^2} \nonumber \\
 &+ \frac{\mA^3 \mS (\sigma - 5 \sigma^3 + 5 \sigma^5)}{2(4\pi)^2(\sigma^2-1)^2}
 + \frac{\mA^2 \mS^2 (77 + 189 \sigma^2 - 360 \sigma^4 + 136 \sigma^6)}{3(16\pi)^2(\sigma^2-1)^2}\Bigg] + \mathcal{O}(\ep^0)\;.
\end{align}
%
\subsection{Polarization Tensors in Non-Relativistic EFT}
\label{app:tensors}
In this section we give the expansion of the relativistic tensors $T_n=\ep_{1,\mu}T_n^{\mu\nu}\ep_{4,\nu}^\star$
from Eq.~(2) of the main text in terms of operators $O_n$ defined by
rest-frame polarization vectors (see Eq.~(6) of the main text).
\begin{align}
 T_1 &= -O_1 + \frac{O_2}{\mA^2(\gamma_1+1)}\left(1 + \frac{\bq^2}{4\mA^2(\gamma_1+1)}\right)
 -\frac{O_3}{\mA^2(\gamma_1+1)}\left(1 + \frac{\bq^2}{8\mA^2(\gamma_1+1)}\right) 
 + \frac{O_4}{2\mA^4(\gamma_1+1)^2}\;, \\
 T_2 &= -\frac{O_2~\bq^2}{2\mA^2(\gamma_1+1)}\left(1 + \frac{\bq^2}{4\mA^2(\gamma_1+1)}\right) 
 + O_3\left(1+\frac{\bq^2}{4\mA^2(\gamma_1+1)}\right)^2 
 - \frac{O_4~\bq^2}{4\mA^4(\gamma_1+1)^2}\;, \\
 T_3 &= -\frac{O_2~\bq^2}{2\mA^2}\left(E(E-\mA) - \frac{(2E-\mA)\bq^2}{4\mA(\gamma_1+1)} 
       + \frac{\bq^2}{16\mA^2(\gamma_1+1)^2}\right) + 
         \frac{O_3~\bq^2}{4\mA^2}\left((E-\mA) - \frac{\bq^2}{4\mA(\gamma_1+1)}\right)^2 \nonumber \\
       &-\frac{O_4}{\mA^2}\left( E - \frac{\bq^2}{4\mA(\gamma_1+1)}\right)^2\;, \\
 T_4 &= -\frac{O_2}{\mA}\left(E + \frac{(E-\mA)\bq^2}{2\mA^2(\gamma_1+1)} - \frac{\bq^2}{8\mA^3(\gamma_1+1)^2}\right)
 + O_3\left( -1 + \frac{E}{\mA} + \frac{(E-2\mA)\bq^2}{4\mA^3(\gamma_1+1)} - \frac{\bq^2}{16\mA^4(\gamma_1+1)^2}\right) \nonumber \\
 &-\frac{O_4}{\mA^3(\gamma_1+1)}\left( E - \frac{\bq^2}{4\mA(\gamma_1+1)}\right)\;,
\end{align}
where $E = E_A + E_\phi$, $\gamma_1 = E_A/\mA$ and, for brevity, $O_n
\equiv \hat{\bep}_{1}^iO^{ij}_n(\bpb,\bq) \hat{\bep}_{4}^{\star j}$ (where the
$\hat{\bep}_i$ are the 3-dimensional rest frame polarization vectors). 
Note that these expressions are not truncated in $|\bq|$. 
%
\subsection{Post-Newtonian expanded coefficients \label{app:PN}}
The coefficients computed in the main text can directly be expanded in the
non-relativistic limit $|\bp|\ll 1$.  Expressions including up to
$\mathcal{O}(\bp^{10})$ are given in the ancillary files to this manuscript\footnote{We
include five computer files, including amplitudes computed in the full
theory [\texttt{FullTheory\_Amplitudes.m}], amplitudes computed from the EFT
[\texttt{EFT\_Amplitudes.m}], translations between relativistic and
non-relativistic tensors in Eqs.~(3) and (7) [\texttt{Tensor\_expansion.m}], and
the matching coefficients defined in Eq.~(8) [\texttt{Coefficients.m}] as well
as non-relativistic expansions thereof [\texttt{Coefficients\_Expanded.m}].}, here we only
list the first two orders.
{
\begin{align}
\tilde{c}_3^{(1)} ={}& {-}\frac{1}{4} m_A m_\phi \left(m_A^2{+}8 m_A m_\phi{+}m_\phi^2\right){-}\frac{\bp^2 \left(517 m_A^2 m_\phi^2{+}230 m_A^3 m_\phi{+}25 m_A^4{+}230 \
	m_A m_\phi^3{+}25 m_\phi^4\right)}{8 m_A m_\phi}{+}\mathcal{O}(\bp^4)\,,\label{eq:c1PN}\\
\tilde{c}_3^{(2)} ={}& {-}\frac{1490 m_A^3 m_\phi^2{+}1428 m_A^2 m_\phi^3{+}636 m_A^4 m_\phi{+}84 m_A^5{+}570 m_A m_\phi^4{+}75 m_\phi^5}{24 m_A (m_A{+}m_\phi)^2}\nonumber\\
&{-}\frac{\bp^2 \left(40225 m_A^4 m_\phi^2{+}34882 m_A^3 m_\phi^3{+}12034 m_A^2 m_\phi^4{+}19940 m_A^5 m_\phi{+}3384 m_A^6{+}478 m_A m_\phi^5{-}315 m_\phi^6\right)}{96 m_A^3 m_\phi (m_A{+}m_\phi)^2}{+}\mathcal{O}(\bp^4)\,,\\
\tilde{c}_3^{(3)} ={}& \frac{m_\phi \left(2025 m_A^2 m_\phi^2{+}1000 m_A^3 m_\phi{+}132 m_A^4{+}1381 m_A m_\phi^3{+}264 m_\phi^4\right)}{240 m_A (m_A{+}m_\phi)^2}\\
&{+}\frac{\bp^2 \left(515832 m_A^4 m_\phi^2{+}659844 m_A^3 m_\phi^3{+}339726 m_A^2 m_\phi^4{+}158568 m_A^5 m_\phi{+}14448 m_A^6{+}45280 m_A m_\phi^5{-}6027 m_\phi^6\right)}{6720 m_A^3 m_\phi (m_A{+}m_\phi)^2}{+}\mathcal{O}(\bp^4)\nonumber\,,\\
\tilde{c}_3^{(4)} ={}& \frac{119982 m_A^4 m_\phi^2{+}111716 m_A^3 m_\phi^3{+}33749 m_A^2 m_\phi^4{+}49536 m_A^5 m_\phi{+}5628 m_A^6{-}4973 m_A m_\phi^5{-}2891 m_\phi^6}{6720 m_A^3 m_\phi (m_A{+}m_\phi)^2}\nonumber\\
&{+}\frac{\bp^2 \left(965286 m_A^6 m_\phi^2{+}2559750 m_A^5 m_\phi^3{+}1043018 m_A^4 m_\phi^4{-}1657386 m_A^3 m_\phi^5{-}1397739 m_A^2 m_\phi^6\right)}{80640 m_A^5 m_\phi^3 (m_A{+}m_\phi)^2}\nonumber\\
&{+}\frac{\bp^2 \left({-}109202 m_A^7 m_\phi{-}45528 m_A^8{-}164033 m_A m_\phi^7{+}59997 m_\phi^8\right)}{80640 m_A^5 m_\phi^3 (m_A{+}m_\phi)^2}{+}\mathcal{O}(\bp^4)\,,\\
\tilde{c}_3^{(5)} ={}&\frac{m_\phi \left(1765 m_A^2 m_\phi^2{+}1520 m_A^3 m_\phi{+}432 m_A^4{+}901 m_A m_\phi^3{+}214 m_\phi^4\right)}{960 m_A (m_A{+}m_\phi)^2}\label{eq:c5PN}\\
&{+}\frac{\bp^2 \left(348602 m_A^4 m_\phi^2{+}432624 m_A^3 m_\phi^3{+}225346 m_A^2 m_\phi^4{+}106768 m_A^5 m_\phi{+}4648 m_A^6{+}33380 m_A m_\phi^5{-}5537 m_\phi^6\right)}{26880 m_A^3 m_\phi (m_A{+}m_\phi)^2}+\mathcal{O}(\bp^4)\,.\nonumber
\end{align}}


%


\begin{thebibliography}{192}%
	\makeatletter
	\providecommand \@ifxundefined [1]{%
		\@ifx{#1\undefined}
	}%
	\providecommand \@ifnum [1]{%
		\ifnum #1\expandafter \@firstoftwo
		\else \expandafter \@secondoftwo
		\fi
	}%
	\providecommand \@ifx [1]{%
		\ifx #1\expandafter \@firstoftwo
		\else \expandafter \@secondoftwo
		\fi
	}%
	\providecommand \natexlab [1]{#1}%
	\providecommand \enquote  [1]{``#1''}%
	\providecommand \bibnamefont  [1]{#1}%
	\providecommand \bibfnamefont [1]{#1}%
	\providecommand \citenamefont [1]{#1}%
	\providecommand \href@noop [0]{\@secondoftwo}%
	\providecommand \href [0]{\begingroup \@sanitize@url \@href}%
	\providecommand \@href[1]{\@@startlink{#1}\@@href}%
	\providecommand \@@href[1]{\endgroup#1\@@endlink}%
	\providecommand \@sanitize@url [0]{\catcode `\\12\catcode `\$12\catcode
		`\&12\catcode `\#12\catcode `\^12\catcode `\_12\catcode `\%12\relax}%
	\providecommand \@@startlink[1]{}%
	\providecommand \@@endlink[0]{}%
	\providecommand \url  [0]{\begingroup\@sanitize@url \@url }%
	\providecommand \@url [1]{\endgroup\@href {#1}{\urlprefix }}%
	\providecommand \urlprefix  [0]{URL }%
	\providecommand \Eprint [0]{\href }%
	\providecommand \doibase [0]{http://dx.doi.org/}%
	\providecommand \selectlanguage [0]{\@gobble}%
	\providecommand \bibinfo  [0]{\@secondoftwo}%
	\providecommand \bibfield  [0]{\@secondoftwo}%
	\providecommand \translation [1]{[#1]}%
	\providecommand \BibitemOpen [0]{}%
	\providecommand \bibitemStop [0]{}%
	\providecommand \bibitemNoStop [0]{.\EOS\space}%
	\providecommand \EOS [0]{\spacefactor3000\relax}%
	\providecommand \BibitemShut  [1]{\csname bibitem#1\endcsname}%
	\let\auto@bib@innerbib\@empty
	\bibitem [{\citenamefont {Abbott}\ \emph {et~al.}(2016)\citenamefont {Abbott}
		\emph {et~al.}}]{LIGOScientific:2016aoc}%
	\BibitemOpen
	\bibfield  {author} {\bibinfo {author} {\bibfnamefont {B.~P.}\ \bibnamefont
			{Abbott}} \emph {et~al.} (\bibinfo {collaboration} {LIGO Scientific,
			Virgo}),\ }\bibfield  {title} {\enquote {\bibinfo {title} {{Observation of
					Gravitational Waves from a Binary Black Hole Merger}},}\ }\href {\doibase
		10.1103/PhysRevLett.116.061102} {\bibfield  {journal} {\bibinfo  {journal}
			{Phys. Rev. Lett.}\ }\textbf {\bibinfo {volume} {116}},\ \bibinfo {pages}
		{061102} (\bibinfo {year} {2016})},\ \Eprint
	{http://arxiv.org/abs/1602.03837} {arXiv:1602.03837 [gr-qc]} \BibitemShut
	{NoStop}%
	\bibitem [{\citenamefont {Abbott}\ \emph {et~al.}(2021)\citenamefont {Abbott}
		\emph {et~al.}}]{LIGOScientific:2021djp}%
	\BibitemOpen
	\bibfield  {author} {\bibinfo {author} {\bibfnamefont {R.}~\bibnamefont
			{Abbott}} \emph {et~al.} (\bibinfo {collaboration} {LIGO Scientific, VIRGO,
			KAGRA}),\ }\bibfield  {title} {\enquote {\bibinfo {title} {{GWTC-3: Compact
					Binary Coalescences Observed by LIGO and Virgo During the Second Part of the
					Third Observing Run}},}\ }\href@noop {} {\  (\bibinfo {year} {2021})},\
	\Eprint {http://arxiv.org/abs/2111.03606} {arXiv:2111.03606 [gr-qc]}
	\BibitemShut {NoStop}%
	\bibitem [{\citenamefont {Punturo}\ \emph {et~al.}(2010)\citenamefont {Punturo}
		\emph {et~al.}}]{Punturo_2010}%
	\BibitemOpen
	\bibfield  {author} {\bibinfo {author} {\bibfnamefont {M}~\bibnamefont
			{Punturo}} \emph {et~al.},\ }\bibfield  {title} {\enquote {\bibinfo {title}
			{The einstein telescope: a third-generation gravitational wave
				observatory},}\ }\href {\doibase 10.1088/0264-9381/27/19/194002} {\bibfield
		{journal} {\bibinfo  {journal} {Classical and Quantum Gravity}\ }\textbf
		{\bibinfo {volume} {27}},\ \bibinfo {pages} {194002} (\bibinfo {year}
		{2010})}\BibitemShut {NoStop}%
	\bibitem [{\citenamefont {Reitze}\ \emph {et~al.}(2019)\citenamefont {Reitze}
		\emph {et~al.}}]{Reitze:2019iox}%
	\BibitemOpen
	\bibfield  {author} {\bibinfo {author} {\bibfnamefont {David}\ \bibnamefont
			{Reitze}} \emph {et~al.},\ }\bibfield  {title} {\enquote {\bibinfo {title}
			{{Cosmic Explorer: The U.S. Contribution to Gravitational-Wave Astronomy
					beyond LIGO}},}\ }\href@noop {} {\bibfield  {journal} {\bibinfo  {journal}
			{Bull. Am. Astron. Soc.}\ }\textbf {\bibinfo {volume} {51}},\ \bibinfo
		{pages} {035} (\bibinfo {year} {2019})},\ \Eprint
	{http://arxiv.org/abs/1907.04833} {arXiv:1907.04833 [astro-ph.IM]}
	\BibitemShut {NoStop}%
	\bibitem [{\citenamefont {Bertotti}(1956)}]{Bertotti:1956pxu}%
	\BibitemOpen
	\bibfield  {author} {\bibinfo {author} {\bibfnamefont {B.}~\bibnamefont
			{Bertotti}},\ }\bibfield  {title} {\enquote {\bibinfo {title} {{On
					gravitational motion}},}\ }\href {\doibase 10.1007/bf02746175} {\bibfield
		{journal} {\bibinfo  {journal} {Nuovo Cim.}\ }\textbf {\bibinfo {volume}
			{4}},\ \bibinfo {pages} {898--906} (\bibinfo {year} {1956})}\BibitemShut
	{NoStop}%
	\bibitem [{\citenamefont {Kerr}(1959)}]{Kerr:1959zlt}%
	\BibitemOpen
	\bibfield  {author} {\bibinfo {author} {\bibfnamefont {R.~P.}\ \bibnamefont
			{Kerr}},\ }\bibfield  {title} {\enquote {\bibinfo {title} {{The
					Lorentz-covariant approximation method in general relativity I}},}\ }\href
	{\doibase 10.1007/bf02732767} {\bibfield  {journal} {\bibinfo  {journal}
			{Nuovo Cim.}\ }\textbf {\bibinfo {volume} {13}},\ \bibinfo {pages} {469--491}
		(\bibinfo {year} {1959})}\BibitemShut {NoStop}%
	\bibitem [{\citenamefont {Bertotti}\ and\ \citenamefont
		{Plebanski}(1960)}]{Bertotti:1960wuq}%
	\BibitemOpen
	\bibfield  {author} {\bibinfo {author} {\bibfnamefont {B.}~\bibnamefont
			{Bertotti}}\ and\ \bibinfo {author} {\bibfnamefont {J.}~\bibnamefont
			{Plebanski}},\ }\bibfield  {title} {\enquote {\bibinfo {title} {{Theory of
					gravitational perturbations in the fast motion approximation}},}\ }\href
	{\doibase 10.1016/0003-4916(60)90132-9} {\bibfield  {journal} {\bibinfo
			{journal} {Annals Phys.}\ }\textbf {\bibinfo {volume} {11}},\ \bibinfo
		{pages} {169--200} (\bibinfo {year} {1960})}\BibitemShut {NoStop}%
	\bibitem [{\citenamefont {Portilla}(1979)}]{Portilla:1979xx}%
	\BibitemOpen
	\bibfield  {author} {\bibinfo {author} {\bibfnamefont {M.}~\bibnamefont
			{Portilla}},\ }\bibfield  {title} {\enquote {\bibinfo {title} {{Momentum and
					angular momentum of two gravitating particles}},}\ }\href {\doibase
		10.1088/0305-4470/12/7/025} {\bibfield  {journal} {\bibinfo  {journal} {J.
				Phys. A}\ }\textbf {\bibinfo {volume} {12}},\ \bibinfo {pages} {1075--1090}
		(\bibinfo {year} {1979})}\BibitemShut {NoStop}%
	\bibitem [{\citenamefont {Westpfahl}\ and\ \citenamefont
		{Goller}(1979)}]{Westpfahl:1979gu}%
	\BibitemOpen
	\bibfield  {author} {\bibinfo {author} {\bibfnamefont {K.}~\bibnamefont
			{Westpfahl}}\ and\ \bibinfo {author} {\bibfnamefont {M.}~\bibnamefont
			{Goller}},\ }\bibfield  {title} {\enquote {\bibinfo {title} {{Gravitational
					scattering of two relativistic particles in post-linear approximation}},}\
	}\href {\doibase 10.1007/BF02817047} {\bibfield  {journal} {\bibinfo
			{journal} {Lett. Nuovo Cim.}\ }\textbf {\bibinfo {volume} {26}},\ \bibinfo
		{pages} {573--576} (\bibinfo {year} {1979})}\BibitemShut {NoStop}%
	\bibitem [{\citenamefont {Portilla}(1980)}]{Portilla:1980uz}%
	\BibitemOpen
	\bibfield  {author} {\bibinfo {author} {\bibfnamefont {M.}~\bibnamefont
			{Portilla}},\ }\bibfield  {title} {\enquote {\bibinfo {title} {{Scattering of
					two gravitating particles: Classical approach}},}\ }\href {\doibase
		10.1088/0305-4470/13/12/017} {\bibfield  {journal} {\bibinfo  {journal} {J.
				Phys. A}\ }\textbf {\bibinfo {volume} {13}},\ \bibinfo {pages} {3677--3683}
		(\bibinfo {year} {1980})}\BibitemShut {NoStop}%
	\bibitem [{\citenamefont {Bel}\ \emph {et~al.}(1981)\citenamefont {Bel},
		\citenamefont {Damour}, \citenamefont {Deruelle}, \citenamefont {Ibanez},\
		and\ \citenamefont {Martin}}]{Bel:1981be}%
	\BibitemOpen
	\bibfield  {author} {\bibinfo {author} {\bibfnamefont {LLuis}\ \bibnamefont
			{Bel}}, \bibinfo {author} {\bibfnamefont {T.}~\bibnamefont {Damour}},
		\bibinfo {author} {\bibfnamefont {N.}~\bibnamefont {Deruelle}}, \bibinfo
		{author} {\bibfnamefont {J.}~\bibnamefont {Ibanez}}, \ and\ \bibinfo {author}
		{\bibfnamefont {J.}~\bibnamefont {Martin}},\ }\bibfield  {title} {\enquote
		{\bibinfo {title} {{Poincar\'e-invariant gravitational field and equations of
					motion of two pointlike objects: The postlinear approximation of general
					relativity}},}\ }\href {\doibase 10.1007/BF00756073} {\bibfield  {journal}
		{\bibinfo  {journal} {Gen. Rel. Grav.}\ }\textbf {\bibinfo {volume} {13}},\
		\bibinfo {pages} {963--1004} (\bibinfo {year} {1981})}\BibitemShut {NoStop}%
	\bibitem [{\citenamefont {Westpfahl}(1985)}]{Westpfahl:1985tsl}%
	\BibitemOpen
	\bibfield  {author} {\bibinfo {author} {\bibfnamefont {Konradin}\
			\bibnamefont {Westpfahl}},\ }\bibfield  {title} {\enquote {\bibinfo {title}
			{{High-Speed Scattering of Charged and Uncharged Particles in General
					Relativity}},}\ }\href {\doibase 10.1002/prop.2190330802} {\bibfield
		{journal} {\bibinfo  {journal} {Fortsch. Phys.}\ }\textbf {\bibinfo {volume}
			{33}},\ \bibinfo {pages} {417--493} (\bibinfo {year} {1985})}\BibitemShut
	{NoStop}%
	\bibitem [{\citenamefont {Ledvinka}\ \emph {et~al.}(2008)\citenamefont
		{Ledvinka}, \citenamefont {Schaefer},\ and\ \citenamefont
		{Bicak}}]{Ledvinka:2008tk}%
	\BibitemOpen
	\bibfield  {author} {\bibinfo {author} {\bibfnamefont {Tomas}\ \bibnamefont
			{Ledvinka}}, \bibinfo {author} {\bibfnamefont {Gerhard}\ \bibnamefont
			{Schaefer}}, \ and\ \bibinfo {author} {\bibfnamefont {Jiri}\ \bibnamefont
			{Bicak}},\ }\bibfield  {title} {\enquote {\bibinfo {title} {{Relativistic
					Closed-Form Hamiltonian for Many-Body Gravitating Systems in the
					Post-Minkowskian Approximation}},}\ }\href {\doibase
		10.1103/PhysRevLett.100.251101} {\bibfield  {journal} {\bibinfo  {journal}
			{Phys. Rev. Lett.}\ }\textbf {\bibinfo {volume} {100}},\ \bibinfo {pages}
		{251101} (\bibinfo {year} {2008})},\ \Eprint {http://arxiv.org/abs/0807.0214}
	{arXiv:0807.0214 [gr-qc]} \BibitemShut {NoStop}%
	\bibitem [{\citenamefont {Damour}(2016)}]{Damour:2016gwp}%
	\BibitemOpen
	\bibfield  {author} {\bibinfo {author} {\bibfnamefont {Thibault}\
			\bibnamefont {Damour}},\ }\bibfield  {title} {\enquote {\bibinfo {title}
			{{Gravitational scattering, post-Minkowskian approximation and Effective
					One-Body theory}},}\ }\href {\doibase 10.1103/PhysRevD.94.104015} {\bibfield
		{journal} {\bibinfo  {journal} {Phys. Rev. D}\ }\textbf {\bibinfo {volume}
			{94}},\ \bibinfo {pages} {104015} (\bibinfo {year} {2016})},\ \Eprint
	{http://arxiv.org/abs/1609.00354} {arXiv:1609.00354 [gr-qc]} \BibitemShut
	{NoStop}%
	\bibitem [{\citenamefont {Bjerrum-Bohr}\ \emph {et~al.}(2018)\citenamefont
		{Bjerrum-Bohr}, \citenamefont {Damgaard}, \citenamefont {Festuccia},
		\citenamefont {Plant\'e},\ and\ \citenamefont
		{Vanhove}}]{Bjerrum-Bohr:2018xdl}%
	\BibitemOpen
	\bibfield  {author} {\bibinfo {author} {\bibfnamefont {N.~E.~J.}\
			\bibnamefont {Bjerrum-Bohr}}, \bibinfo {author} {\bibfnamefont {Poul~H.}\
			\bibnamefont {Damgaard}}, \bibinfo {author} {\bibfnamefont {Guido}\
			\bibnamefont {Festuccia}}, \bibinfo {author} {\bibfnamefont {Ludovic}\
			\bibnamefont {Plant\'e}}, \ and\ \bibinfo {author} {\bibfnamefont {Pierre}\
			\bibnamefont {Vanhove}},\ }\bibfield  {title} {\enquote {\bibinfo {title}
			{{General Relativity from Scattering Amplitudes}},}\ }\href {\doibase
		10.1103/PhysRevLett.121.171601} {\bibfield  {journal} {\bibinfo  {journal}
			{Phys. Rev. Lett.}\ }\textbf {\bibinfo {volume} {121}},\ \bibinfo {pages}
		{171601} (\bibinfo {year} {2018})},\ \Eprint
	{http://arxiv.org/abs/1806.04920} {arXiv:1806.04920 [hep-th]} \BibitemShut
	{NoStop}%
	\bibitem [{\citenamefont {Cheung}\ \emph {et~al.}(2018)\citenamefont {Cheung},
		\citenamefont {Rothstein},\ and\ \citenamefont {Solon}}]{Cheung:2018wkq}%
	\BibitemOpen
	\bibfield  {author} {\bibinfo {author} {\bibfnamefont {Clifford}\
			\bibnamefont {Cheung}}, \bibinfo {author} {\bibfnamefont {Ira~Z.}\
			\bibnamefont {Rothstein}}, \ and\ \bibinfo {author} {\bibfnamefont
			{Mikhail~P.}\ \bibnamefont {Solon}},\ }\bibfield  {title} {\enquote {\bibinfo
			{title} {{From Scattering Amplitudes to Classical Potentials in the
					Post-Minkowskian Expansion}},}\ }\href {\doibase
		10.1103/PhysRevLett.121.251101} {\bibfield  {journal} {\bibinfo  {journal}
			{Phys. Rev. Lett.}\ }\textbf {\bibinfo {volume} {121}},\ \bibinfo {pages}
		{251101} (\bibinfo {year} {2018})},\ \Eprint
	{http://arxiv.org/abs/1808.02489} {arXiv:1808.02489 [hep-th]} \BibitemShut
	{NoStop}%
	\bibitem [{\citenamefont {Kosower}\ \emph {et~al.}(2019)\citenamefont
		{Kosower}, \citenamefont {Maybee},\ and\ \citenamefont
		{O'Connell}}]{Kosower:2018adc}%
	\BibitemOpen
	\bibfield  {author} {\bibinfo {author} {\bibfnamefont {David~A.}\
			\bibnamefont {Kosower}}, \bibinfo {author} {\bibfnamefont {Ben}\ \bibnamefont
			{Maybee}}, \ and\ \bibinfo {author} {\bibfnamefont {Donal}\ \bibnamefont
			{O'Connell}},\ }\bibfield  {title} {\enquote {\bibinfo {title} {{Amplitudes,
					Observables, and Classical Scattering}},}\ }\href {\doibase
		10.1007/JHEP02(2019)137} {\bibfield  {journal} {\bibinfo  {journal} {JHEP}\
		}\textbf {\bibinfo {volume} {02}},\ \bibinfo {pages} {137} (\bibinfo {year}
		{2019})},\ \Eprint {http://arxiv.org/abs/1811.10950} {arXiv:1811.10950
		[hep-th]} \BibitemShut {NoStop}%
	\bibitem [{\citenamefont {Bern}\ \emph
		{et~al.}(2019{\natexlab{a}})\citenamefont {Bern}, \citenamefont {Cheung},
		\citenamefont {Roiban}, \citenamefont {Shen}, \citenamefont {Solon},\ and\
		\citenamefont {Zeng}}]{Bern:2019nnu}%
	\BibitemOpen
	\bibfield  {author} {\bibinfo {author} {\bibfnamefont {Zvi}\ \bibnamefont
			{Bern}}, \bibinfo {author} {\bibfnamefont {Clifford}\ \bibnamefont {Cheung}},
		\bibinfo {author} {\bibfnamefont {Radu}\ \bibnamefont {Roiban}}, \bibinfo
		{author} {\bibfnamefont {Chia-Hsien}\ \bibnamefont {Shen}}, \bibinfo {author}
		{\bibfnamefont {Mikhail~P.}\ \bibnamefont {Solon}}, \ and\ \bibinfo {author}
		{\bibfnamefont {Mao}\ \bibnamefont {Zeng}},\ }\bibfield  {title} {\enquote
		{\bibinfo {title} {{Scattering Amplitudes and the Conservative Hamiltonian
					for Binary Systems at Third Post-Minkowskian Order}},}\ }\href {\doibase
		10.1103/PhysRevLett.122.201603} {\bibfield  {journal} {\bibinfo  {journal}
			{Phys. Rev. Lett.}\ }\textbf {\bibinfo {volume} {122}},\ \bibinfo {pages}
		{201603} (\bibinfo {year} {2019}{\natexlab{a}})},\ \Eprint
	{http://arxiv.org/abs/1901.04424} {arXiv:1901.04424 [hep-th]} \BibitemShut
	{NoStop}%
	\bibitem [{\citenamefont {Bern}\ \emph
		{et~al.}(2019{\natexlab{b}})\citenamefont {Bern}, \citenamefont {Cheung},
		\citenamefont {Roiban}, \citenamefont {Shen}, \citenamefont {Solon},\ and\
		\citenamefont {Zeng}}]{Bern:2019crd}%
	\BibitemOpen
	\bibfield  {author} {\bibinfo {author} {\bibfnamefont {Zvi}\ \bibnamefont
			{Bern}}, \bibinfo {author} {\bibfnamefont {Clifford}\ \bibnamefont {Cheung}},
		\bibinfo {author} {\bibfnamefont {Radu}\ \bibnamefont {Roiban}}, \bibinfo
		{author} {\bibfnamefont {Chia-Hsien}\ \bibnamefont {Shen}}, \bibinfo {author}
		{\bibfnamefont {Mikhail~P.}\ \bibnamefont {Solon}}, \ and\ \bibinfo {author}
		{\bibfnamefont {Mao}\ \bibnamefont {Zeng}},\ }\bibfield  {title} {\enquote
		{\bibinfo {title} {{Black Hole Binary Dynamics from the Double Copy and
					Effective Theory}},}\ }\href {\doibase 10.1007/JHEP10(2019)206} {\bibfield
		{journal} {\bibinfo  {journal} {JHEP}\ }\textbf {\bibinfo {volume} {10}},\
		\bibinfo {pages} {206} (\bibinfo {year} {2019}{\natexlab{b}})},\ \Eprint
	{http://arxiv.org/abs/1908.01493} {arXiv:1908.01493 [hep-th]} \BibitemShut
	{NoStop}%
	\bibitem [{\citenamefont {Di~Vecchia}\ \emph {et~al.}(2021)\citenamefont
		{Di~Vecchia}, \citenamefont {Heissenberg}, \citenamefont {Russo},\ and\
		\citenamefont {Veneziano}}]{DiVecchia:2021bdo}%
	\BibitemOpen
	\bibfield  {author} {\bibinfo {author} {\bibfnamefont {Paolo}\ \bibnamefont
			{Di~Vecchia}}, \bibinfo {author} {\bibfnamefont {Carlo}\ \bibnamefont
			{Heissenberg}}, \bibinfo {author} {\bibfnamefont {Rodolfo}\ \bibnamefont
			{Russo}}, \ and\ \bibinfo {author} {\bibfnamefont {Gabriele}\ \bibnamefont
			{Veneziano}},\ }\bibfield  {title} {\enquote {\bibinfo {title} {{The eikonal
					approach to gravitational scattering and radiation at $ \mathcal{O}
					$(G$^{3}$)}},}\ }\href {\doibase 10.1007/JHEP07(2021)169} {\bibfield
		{journal} {\bibinfo  {journal} {JHEP}\ }\textbf {\bibinfo {volume} {07}},\
		\bibinfo {pages} {169} (\bibinfo {year} {2021})},\ \Eprint
	{http://arxiv.org/abs/2104.03256} {arXiv:2104.03256 [hep-th]} \BibitemShut
	{NoStop}%
	\bibitem [{\citenamefont {Herrmann}\ \emph
		{et~al.}(2021{\natexlab{a}})\citenamefont {Herrmann}, \citenamefont
		{Parra-Martinez}, \citenamefont {Ruf},\ and\ \citenamefont
		{Zeng}}]{Herrmann:2021tct}%
	\BibitemOpen
	\bibfield  {author} {\bibinfo {author} {\bibfnamefont {Enrico}\ \bibnamefont
			{Herrmann}}, \bibinfo {author} {\bibfnamefont {Julio}\ \bibnamefont
			{Parra-Martinez}}, \bibinfo {author} {\bibfnamefont {Michael~S.}\
			\bibnamefont {Ruf}}, \ and\ \bibinfo {author} {\bibfnamefont {Mao}\
			\bibnamefont {Zeng}},\ }\bibfield  {title} {\enquote {\bibinfo {title}
			{{Radiative classical gravitational observables at $ \mathcal{O} $(G$^{3}$)
					from scattering amplitudes}},}\ }\href {\doibase 10.1007/JHEP10(2021)148}
	{\bibfield  {journal} {\bibinfo  {journal} {JHEP}\ }\textbf {\bibinfo
			{volume} {10}},\ \bibinfo {pages} {148} (\bibinfo {year}
		{2021}{\natexlab{a}})},\ \Eprint {http://arxiv.org/abs/2104.03957}
	{arXiv:2104.03957 [hep-th]} \BibitemShut {NoStop}%
	\bibitem [{\citenamefont {Bjerrum-Bohr}\ \emph {et~al.}(2021)\citenamefont
		{Bjerrum-Bohr}, \citenamefont {Damgaard}, \citenamefont {Plant\'e},\ and\
		\citenamefont {Vanhove}}]{Bjerrum-Bohr:2021din}%
	\BibitemOpen
	\bibfield  {author} {\bibinfo {author} {\bibfnamefont {N.~Emil~J.}\
			\bibnamefont {Bjerrum-Bohr}}, \bibinfo {author} {\bibfnamefont {Poul~H.}\
			\bibnamefont {Damgaard}}, \bibinfo {author} {\bibfnamefont {Ludovic}\
			\bibnamefont {Plant\'e}}, \ and\ \bibinfo {author} {\bibfnamefont {Pierre}\
			\bibnamefont {Vanhove}},\ }\bibfield  {title} {\enquote {\bibinfo {title}
			{{The amplitude for classical gravitational scattering at third
					Post-Minkowskian order}},}\ }\href {\doibase 10.1007/JHEP08(2021)172}
	{\bibfield  {journal} {\bibinfo  {journal} {JHEP}\ }\textbf {\bibinfo
			{volume} {08}},\ \bibinfo {pages} {172} (\bibinfo {year} {2021})},\ \Eprint
	{http://arxiv.org/abs/2105.05218} {arXiv:2105.05218 [hep-th]} \BibitemShut
	{NoStop}%
	\bibitem [{\citenamefont {Brandhuber}\ \emph {et~al.}(2021)\citenamefont
		{Brandhuber}, \citenamefont {Chen}, \citenamefont {Travaglini},\ and\
		\citenamefont {Wen}}]{Brandhuber:2021eyq}%
	\BibitemOpen
	\bibfield  {author} {\bibinfo {author} {\bibfnamefont {Andreas}\ \bibnamefont
			{Brandhuber}}, \bibinfo {author} {\bibfnamefont {Gang}\ \bibnamefont {Chen}},
		\bibinfo {author} {\bibfnamefont {Gabriele}\ \bibnamefont {Travaglini}}, \
		and\ \bibinfo {author} {\bibfnamefont {Congkao}\ \bibnamefont {Wen}},\
	}\bibfield  {title} {\enquote {\bibinfo {title} {{Classical gravitational
					scattering from a gauge-invariant double copy}},}\ }\href {\doibase
		10.1007/JHEP10(2021)118} {\bibfield  {journal} {\bibinfo  {journal} {JHEP}\
		}\textbf {\bibinfo {volume} {10}},\ \bibinfo {pages} {118} (\bibinfo {year}
		{2021})},\ \Eprint {http://arxiv.org/abs/2108.04216} {arXiv:2108.04216
		[hep-th]} \BibitemShut {NoStop}%
	\bibitem [{\citenamefont {Bern}\ \emph
		{et~al.}(2021{\natexlab{a}})\citenamefont {Bern}, \citenamefont
		{Parra-Martinez}, \citenamefont {Roiban}, \citenamefont {Ruf}, \citenamefont
		{Shen}, \citenamefont {Solon},\ and\ \citenamefont {Zeng}}]{Bern:2021dqo}%
	\BibitemOpen
	\bibfield  {author} {\bibinfo {author} {\bibfnamefont {Zvi}\ \bibnamefont
			{Bern}}, \bibinfo {author} {\bibfnamefont {Julio}\ \bibnamefont
			{Parra-Martinez}}, \bibinfo {author} {\bibfnamefont {Radu}\ \bibnamefont
			{Roiban}}, \bibinfo {author} {\bibfnamefont {Michael~S.}\ \bibnamefont
			{Ruf}}, \bibinfo {author} {\bibfnamefont {Chia-Hsien}\ \bibnamefont {Shen}},
		\bibinfo {author} {\bibfnamefont {Mikhail~P.}\ \bibnamefont {Solon}}, \ and\
		\bibinfo {author} {\bibfnamefont {Mao}\ \bibnamefont {Zeng}},\ }\bibfield
	{title} {\enquote {\bibinfo {title} {{Scattering Amplitudes and Conservative
					Binary Dynamics at ${\cal O}(G^4)$}},}\ }\href {\doibase
		10.1103/PhysRevLett.126.171601} {\bibfield  {journal} {\bibinfo  {journal}
			{Phys. Rev. Lett.}\ }\textbf {\bibinfo {volume} {126}},\ \bibinfo {pages}
		{171601} (\bibinfo {year} {2021}{\natexlab{a}})},\ \Eprint
	{http://arxiv.org/abs/2101.07254} {arXiv:2101.07254 [hep-th]} \BibitemShut
	{NoStop}%
	\bibitem [{\citenamefont {Bern}\ \emph
		{et~al.}(2022{\natexlab{a}})\citenamefont {Bern}, \citenamefont
		{Parra-Martinez}, \citenamefont {Roiban}, \citenamefont {Ruf}, \citenamefont
		{Shen}, \citenamefont {Solon},\ and\ \citenamefont {Zeng}}]{Bern:2021yeh}%
	\BibitemOpen
	\bibfield  {author} {\bibinfo {author} {\bibfnamefont {Zvi}\ \bibnamefont
			{Bern}}, \bibinfo {author} {\bibfnamefont {Julio}\ \bibnamefont
			{Parra-Martinez}}, \bibinfo {author} {\bibfnamefont {Radu}\ \bibnamefont
			{Roiban}}, \bibinfo {author} {\bibfnamefont {Michael~S.}\ \bibnamefont
			{Ruf}}, \bibinfo {author} {\bibfnamefont {Chia-Hsien}\ \bibnamefont {Shen}},
		\bibinfo {author} {\bibfnamefont {Mikhail~P.}\ \bibnamefont {Solon}}, \ and\
		\bibinfo {author} {\bibfnamefont {Mao}\ \bibnamefont {Zeng}},\ }\bibfield
	{title} {\enquote {\bibinfo {title} {{Scattering Amplitudes, the Tail Effect,
					and Conservative Binary Dynamics at O(G4)}},}\ }\href {\doibase
		10.1103/PhysRevLett.128.161103} {\bibfield  {journal} {\bibinfo  {journal}
			{Phys. Rev. Lett.}\ }\textbf {\bibinfo {volume} {128}},\ \bibinfo {pages}
		{161103} (\bibinfo {year} {2022}{\natexlab{a}})},\ \Eprint
	{http://arxiv.org/abs/2112.10750} {arXiv:2112.10750 [hep-th]} \BibitemShut
	{NoStop}%
	\bibitem [{\citenamefont {K\"alin}\ \emph
		{et~al.}(2020{\natexlab{a}})\citenamefont {K\"alin}, \citenamefont {Liu},\
		and\ \citenamefont {Porto}}]{Kalin:2020fhe}%
	\BibitemOpen
	\bibfield  {author} {\bibinfo {author} {\bibfnamefont {Gregor}\ \bibnamefont
			{K\"alin}}, \bibinfo {author} {\bibfnamefont {Zhengwen}\ \bibnamefont {Liu}},
		\ and\ \bibinfo {author} {\bibfnamefont {Rafael~A.}\ \bibnamefont {Porto}},\
	}\bibfield  {title} {\enquote {\bibinfo {title} {{Conservative Dynamics of
					Binary Systems to Third Post-Minkowskian Order from the Effective Field
					Theory Approach}},}\ }\href {\doibase 10.1103/PhysRevLett.125.261103}
	{\bibfield  {journal} {\bibinfo  {journal} {Phys. Rev. Lett.}\ }\textbf
		{\bibinfo {volume} {125}},\ \bibinfo {pages} {261103} (\bibinfo {year}
		{2020}{\natexlab{a}})},\ \Eprint {http://arxiv.org/abs/2007.04977}
	{arXiv:2007.04977 [hep-th]} \BibitemShut {NoStop}%
	\bibitem [{\citenamefont {Dlapa}\ \emph {et~al.}(2021)\citenamefont {Dlapa},
		\citenamefont {K\"alin}, \citenamefont {Liu},\ and\ \citenamefont
		{Porto}}]{Dlapa:2021npj}%
	\BibitemOpen
	\bibfield  {author} {\bibinfo {author} {\bibfnamefont {Christoph}\
			\bibnamefont {Dlapa}}, \bibinfo {author} {\bibfnamefont {Gregor}\
			\bibnamefont {K\"alin}}, \bibinfo {author} {\bibfnamefont {Zhengwen}\
			\bibnamefont {Liu}}, \ and\ \bibinfo {author} {\bibfnamefont {Rafael~A.}\
			\bibnamefont {Porto}},\ }\bibfield  {title} {\enquote {\bibinfo {title}
			{{Dynamics of Binary Systems to Fourth Post-Minkowskian Order from the
					Effective Field Theory Approach}},}\ }\href@noop {} {\  (\bibinfo {year}
		{2021})},\ \Eprint {http://arxiv.org/abs/2106.08276} {arXiv:2106.08276
		[hep-th]} \BibitemShut {NoStop}%
	\bibitem [{\citenamefont {Dlapa}\ \emph {et~al.}(2022)\citenamefont {Dlapa},
		\citenamefont {K\"alin}, \citenamefont {Liu},\ and\ \citenamefont
		{Porto}}]{Dlapa:2021vgp}%
	\BibitemOpen
	\bibfield  {author} {\bibinfo {author} {\bibfnamefont {Christoph}\
			\bibnamefont {Dlapa}}, \bibinfo {author} {\bibfnamefont {Gregor}\
			\bibnamefont {K\"alin}}, \bibinfo {author} {\bibfnamefont {Zhengwen}\
			\bibnamefont {Liu}}, \ and\ \bibinfo {author} {\bibfnamefont {Rafael~A.}\
			\bibnamefont {Porto}},\ }\bibfield  {title} {\enquote {\bibinfo {title}
			{{Conservative Dynamics of Binary Systems at Fourth Post-Minkowskian Order in
					the Large-Eccentricity Expansion}},}\ }\href {\doibase
		10.1103/PhysRevLett.128.161104} {\bibfield  {journal} {\bibinfo  {journal}
			{Phys. Rev. Lett.}\ }\textbf {\bibinfo {volume} {128}},\ \bibinfo {pages}
		{161104} (\bibinfo {year} {2022})},\ \Eprint
	{http://arxiv.org/abs/2112.11296} {arXiv:2112.11296 [hep-th]} \BibitemShut
	{NoStop}%
	\bibitem [{\citenamefont {Foffa}\ \emph {et~al.}(2019)\citenamefont {Foffa},
		\citenamefont {Mastrolia}, \citenamefont {Sturani}, \citenamefont {Sturm},\
		and\ \citenamefont {Torres~Bobadilla}}]{Foffa:2019hrb}%
	\BibitemOpen
	\bibfield  {author} {\bibinfo {author} {\bibfnamefont {Stefano}\ \bibnamefont
			{Foffa}}, \bibinfo {author} {\bibfnamefont {Pierpaolo}\ \bibnamefont
			{Mastrolia}}, \bibinfo {author} {\bibfnamefont {Riccardo}\ \bibnamefont
			{Sturani}}, \bibinfo {author} {\bibfnamefont {Christian}\ \bibnamefont
			{Sturm}}, \ and\ \bibinfo {author} {\bibfnamefont {William~J.}\ \bibnamefont
			{Torres~Bobadilla}},\ }\bibfield  {title} {\enquote {\bibinfo {title}
			{{Static two-body potential at fifth post-Newtonian order}},}\ }\href
	{\doibase 10.1103/PhysRevLett.122.241605} {\bibfield  {journal} {\bibinfo
			{journal} {Phys. Rev. Lett.}\ }\textbf {\bibinfo {volume} {122}},\ \bibinfo
		{pages} {241605} (\bibinfo {year} {2019})},\ \Eprint
	{http://arxiv.org/abs/1902.10571} {arXiv:1902.10571 [gr-qc]} \BibitemShut
	{NoStop}%
	\bibitem [{\citenamefont {Bl\"umlein}\ \emph
		{et~al.}(2020{\natexlab{a}})\citenamefont {Bl\"umlein}, \citenamefont
		{Maier},\ and\ \citenamefont {Marquard}}]{Blumlein:2019zku}%
	\BibitemOpen
	\bibfield  {author} {\bibinfo {author} {\bibfnamefont {J.}~\bibnamefont
			{Bl\"umlein}}, \bibinfo {author} {\bibfnamefont {A.}~\bibnamefont {Maier}}, \
		and\ \bibinfo {author} {\bibfnamefont {P.}~\bibnamefont {Marquard}},\
	}\bibfield  {title} {\enquote {\bibinfo {title} {{Five-Loop Static
					Contribution to the Gravitational Interaction Potential of Two Point
					Masses}},}\ }\href {\doibase 10.1016/j.physletb.2019.135100} {\bibfield
		{journal} {\bibinfo  {journal} {Phys. Lett. B}\ }\textbf {\bibinfo {volume}
			{800}},\ \bibinfo {pages} {135100} (\bibinfo {year} {2020}{\natexlab{a}})},\
	\Eprint {http://arxiv.org/abs/1902.11180} {arXiv:1902.11180 [gr-qc]}
	\BibitemShut {NoStop}%
	\bibitem [{\citenamefont {Bl\"umlein}\ \emph
		{et~al.}(2021{\natexlab{a}})\citenamefont {Bl\"umlein}, \citenamefont
		{Maier}, \citenamefont {Marquard},\ and\ \citenamefont
		{Sch\"afer}}]{Blumlein:2020pyo}%
	\BibitemOpen
	\bibfield  {author} {\bibinfo {author} {\bibfnamefont {J.}~\bibnamefont
			{Bl\"umlein}}, \bibinfo {author} {\bibfnamefont {A.}~\bibnamefont {Maier}},
		\bibinfo {author} {\bibfnamefont {P.}~\bibnamefont {Marquard}}, \ and\
		\bibinfo {author} {\bibfnamefont {G.}~\bibnamefont {Sch\"afer}},\ }\bibfield
	{title} {\enquote {\bibinfo {title} {{The fifth-order post-Newtonian
					Hamiltonian dynamics of two-body systems from an effective field theory
					approach: potential contributions}},}\ }\href {\doibase
		10.1016/j.nuclphysb.2021.115352} {\bibfield  {journal} {\bibinfo  {journal}
			{Nucl. Phys. B}\ }\textbf {\bibinfo {volume} {965}},\ \bibinfo {pages}
		{115352} (\bibinfo {year} {2021}{\natexlab{a}})},\ \Eprint
	{http://arxiv.org/abs/2010.13672} {arXiv:2010.13672 [gr-qc]} \BibitemShut
	{NoStop}%
	\bibitem [{\citenamefont {Bl\"umlein}\ \emph
		{et~al.}(2020{\natexlab{b}})\citenamefont {Bl\"umlein}, \citenamefont
		{Maier}, \citenamefont {Marquard},\ and\ \citenamefont
		{Sch\"afer}}]{Blumlein:2020pog}%
	\BibitemOpen
	\bibfield  {author} {\bibinfo {author} {\bibfnamefont {J.}~\bibnamefont
			{Bl\"umlein}}, \bibinfo {author} {\bibfnamefont {A.}~\bibnamefont {Maier}},
		\bibinfo {author} {\bibfnamefont {P.}~\bibnamefont {Marquard}}, \ and\
		\bibinfo {author} {\bibfnamefont {G.}~\bibnamefont {Sch\"afer}},\ }\bibfield
	{title} {\enquote {\bibinfo {title} {{Fourth post-Newtonian Hamiltonian
					dynamics of two-body systems from an effective field theory approach}},}\
	}\href {\doibase 10.1016/j.nuclphysb.2020.115041} {\bibfield  {journal}
		{\bibinfo  {journal} {Nucl. Phys. B}\ }\textbf {\bibinfo {volume} {955}},\
		\bibinfo {pages} {115041} (\bibinfo {year} {2020}{\natexlab{b}})},\ \Eprint
	{http://arxiv.org/abs/2003.01692} {arXiv:2003.01692 [gr-qc]} \BibitemShut
	{NoStop}%
	\bibitem [{\citenamefont {Bl\"umlein}\ \emph
		{et~al.}(2021{\natexlab{b}})\citenamefont {Bl\"umlein}, \citenamefont
		{Maier}, \citenamefont {Marquard},\ and\ \citenamefont
		{Sch\"afer}}]{Blumlein:2021txe}%
	\BibitemOpen
	\bibfield  {author} {\bibinfo {author} {\bibfnamefont {J.}~\bibnamefont
			{Bl\"umlein}}, \bibinfo {author} {\bibfnamefont {A.}~\bibnamefont {Maier}},
		\bibinfo {author} {\bibfnamefont {P.}~\bibnamefont {Marquard}}, \ and\
		\bibinfo {author} {\bibfnamefont {G.}~\bibnamefont {Sch\"afer}},\ }\bibfield
	{title} {\enquote {\bibinfo {title} {{The fifth-order post-Newtonian
					Hamiltonian dynamics of two-body systems from an effective field theory
					approach}},}\ }\href@noop {} {\  (\bibinfo {year} {2021}{\natexlab{b}})},\
	\Eprint {http://arxiv.org/abs/2110.13822} {arXiv:2110.13822 [gr-qc]}
	\BibitemShut {NoStop}%
	\bibitem [{\citenamefont {Bl\"umlein}\ \emph
		{et~al.}(2020{\natexlab{c}})\citenamefont {Bl\"umlein}, \citenamefont
		{Maier}, \citenamefont {Marquard},\ and\ \citenamefont
		{Sch\"afer}}]{Blumlein:2020znm}%
	\BibitemOpen
	\bibfield  {author} {\bibinfo {author} {\bibfnamefont {J.}~\bibnamefont
			{Bl\"umlein}}, \bibinfo {author} {\bibfnamefont {A.}~\bibnamefont {Maier}},
		\bibinfo {author} {\bibfnamefont {P.}~\bibnamefont {Marquard}}, \ and\
		\bibinfo {author} {\bibfnamefont {G.}~\bibnamefont {Sch\"afer}},\ }\bibfield
	{title} {\enquote {\bibinfo {title} {{Testing binary dynamics in gravity at
					the sixth post-Newtonian level}},}\ }\href {\doibase
		10.1016/j.physletb.2020.135496} {\bibfield  {journal} {\bibinfo  {journal}
			{Phys. Lett. B}\ }\textbf {\bibinfo {volume} {807}},\ \bibinfo {pages}
		{135496} (\bibinfo {year} {2020}{\natexlab{c}})},\ \Eprint
	{http://arxiv.org/abs/2003.07145} {arXiv:2003.07145 [gr-qc]} \BibitemShut
	{NoStop}%
	\bibitem [{\citenamefont {Bl\"umlein}\ \emph
		{et~al.}(2021{\natexlab{c}})\citenamefont {Bl\"umlein}, \citenamefont
		{Maier}, \citenamefont {Marquard},\ and\ \citenamefont
		{Sch\"afer}}]{Blumlein:2021txj}%
	\BibitemOpen
	\bibfield  {author} {\bibinfo {author} {\bibfnamefont {J.}~\bibnamefont
			{Bl\"umlein}}, \bibinfo {author} {\bibfnamefont {A.}~\bibnamefont {Maier}},
		\bibinfo {author} {\bibfnamefont {P.}~\bibnamefont {Marquard}}, \ and\
		\bibinfo {author} {\bibfnamefont {G.}~\bibnamefont {Sch\"afer}},\ }\bibfield
	{title} {\enquote {\bibinfo {title} {{The 6th post-Newtonian potential terms
					at $O(G_N^4)$}},}\ }\href {\doibase 10.1016/j.physletb.2021.136260}
	{\bibfield  {journal} {\bibinfo  {journal} {Phys. Lett. B}\ }\textbf
		{\bibinfo {volume} {816}},\ \bibinfo {pages} {136260} (\bibinfo {year}
		{2021}{\natexlab{c}})},\ \Eprint {http://arxiv.org/abs/2101.08630}
	{arXiv:2101.08630 [gr-qc]} \BibitemShut {NoStop}%
	\bibitem [{\citenamefont {Bini}\ \emph
		{et~al.}(2020{\natexlab{a}})\citenamefont {Bini}, \citenamefont {Damour},\
		and\ \citenamefont {Geralico}}]{Bini:2020wpo}%
	\BibitemOpen
	\bibfield  {author} {\bibinfo {author} {\bibfnamefont {Donato}\ \bibnamefont
			{Bini}}, \bibinfo {author} {\bibfnamefont {Thibault}\ \bibnamefont {Damour}},
		\ and\ \bibinfo {author} {\bibfnamefont {Andrea}\ \bibnamefont {Geralico}},\
	}\bibfield  {title} {\enquote {\bibinfo {title} {{Binary dynamics at the
					fifth and fifth-and-a-half post-Newtonian orders}},}\ }\href {\doibase
		10.1103/PhysRevD.102.024062} {\bibfield  {journal} {\bibinfo  {journal}
			{Phys. Rev. D}\ }\textbf {\bibinfo {volume} {102}},\ \bibinfo {pages}
		{024062} (\bibinfo {year} {2020}{\natexlab{a}})},\ \Eprint
	{http://arxiv.org/abs/2003.11891} {arXiv:2003.11891 [gr-qc]} \BibitemShut
	{NoStop}%
	\bibitem [{\citenamefont {Bini}\ \emph
		{et~al.}(2020{\natexlab{b}})\citenamefont {Bini}, \citenamefont {Damour},\
		and\ \citenamefont {Geralico}}]{Bini:2020nsb}%
	\BibitemOpen
	\bibfield  {author} {\bibinfo {author} {\bibfnamefont {Donato}\ \bibnamefont
			{Bini}}, \bibinfo {author} {\bibfnamefont {Thibault}\ \bibnamefont {Damour}},
		\ and\ \bibinfo {author} {\bibfnamefont {Andrea}\ \bibnamefont {Geralico}},\
	}\bibfield  {title} {\enquote {\bibinfo {title} {{Sixth post-Newtonian
					local-in-time dynamics of binary systems}},}\ }\href {\doibase
		10.1103/PhysRevD.102.024061} {\bibfield  {journal} {\bibinfo  {journal}
			{Phys. Rev. D}\ }\textbf {\bibinfo {volume} {102}},\ \bibinfo {pages}
		{024061} (\bibinfo {year} {2020}{\natexlab{b}})},\ \Eprint
	{http://arxiv.org/abs/2004.05407} {arXiv:2004.05407 [gr-qc]} \BibitemShut
	{NoStop}%
	\bibitem [{\citenamefont {Bini}\ \emph
		{et~al.}(2020{\natexlab{c}})\citenamefont {Bini}, \citenamefont {Damour},
		\citenamefont {Geralico}, \citenamefont {Laporta},\ and\ \citenamefont
		{Mastrolia}}]{Bini:2020uiq}%
	\BibitemOpen
	\bibfield  {author} {\bibinfo {author} {\bibfnamefont {Donato}\ \bibnamefont
			{Bini}}, \bibinfo {author} {\bibfnamefont {Thibault}\ \bibnamefont {Damour}},
		\bibinfo {author} {\bibfnamefont {Andrea}\ \bibnamefont {Geralico}}, \bibinfo
		{author} {\bibfnamefont {Stefano}\ \bibnamefont {Laporta}}, \ and\ \bibinfo
		{author} {\bibfnamefont {Pierpaolo}\ \bibnamefont {Mastrolia}},\ }\bibfield
	{title} {\enquote {\bibinfo {title} {{Gravitational dynamics at $O(G^6)$:
					perturbative gravitational scattering meets experimental mathematics}},}\
	}\href@noop {} {\  (\bibinfo {year} {2020}{\natexlab{c}})},\ \Eprint
	{http://arxiv.org/abs/2008.09389} {arXiv:2008.09389 [gr-qc]} \BibitemShut
	{NoStop}%
	\bibitem [{\citenamefont {Bini}\ \emph {et~al.}(2021)\citenamefont {Bini},
		\citenamefont {Damour}, \citenamefont {Geralico}, \citenamefont {Laporta},\
		and\ \citenamefont {Mastrolia}}]{Bini:2020rzn}%
	\BibitemOpen
	\bibfield  {author} {\bibinfo {author} {\bibfnamefont {Donato}\ \bibnamefont
			{Bini}}, \bibinfo {author} {\bibfnamefont {Thibault}\ \bibnamefont {Damour}},
		\bibinfo {author} {\bibfnamefont {Andrea}\ \bibnamefont {Geralico}}, \bibinfo
		{author} {\bibfnamefont {Stefano}\ \bibnamefont {Laporta}}, \ and\ \bibinfo
		{author} {\bibfnamefont {Pierpaolo}\ \bibnamefont {Mastrolia}},\ }\bibfield
	{title} {\enquote {\bibinfo {title} {{Gravitational scattering at the seventh
					order in $G$: nonlocal contribution at the sixth post-Newtonian accuracy}},}\
	}\href {\doibase 10.1103/PhysRevD.103.044038} {\bibfield  {journal} {\bibinfo
			{journal} {Phys. Rev. D}\ }\textbf {\bibinfo {volume} {103}},\ \bibinfo
		{pages} {044038} (\bibinfo {year} {2021})},\ \Eprint
	{http://arxiv.org/abs/2012.12918} {arXiv:2012.12918 [gr-qc]} \BibitemShut
	{NoStop}%
	\bibitem [{\citenamefont {Buonanno}\ \emph {et~al.}(2009)\citenamefont
		{Buonanno}, \citenamefont {Iyer}, \citenamefont {Ochsner}, \citenamefont
		{Pan},\ and\ \citenamefont {Sathyaprakash}}]{Buonanno:2009zt}%
	\BibitemOpen
	\bibfield  {author} {\bibinfo {author} {\bibfnamefont {Alessandra}\
			\bibnamefont {Buonanno}}, \bibinfo {author} {\bibfnamefont {Bala}\
			\bibnamefont {Iyer}}, \bibinfo {author} {\bibfnamefont {Evan}\ \bibnamefont
			{Ochsner}}, \bibinfo {author} {\bibfnamefont {Yi}~\bibnamefont {Pan}}, \ and\
		\bibinfo {author} {\bibfnamefont {B.~S.}\ \bibnamefont {Sathyaprakash}},\
	}\bibfield  {title} {\enquote {\bibinfo {title} {{Comparison of
					post-Newtonian templates for compact binary inspiral signals in
					gravitational-wave detectors}},}\ }\href {\doibase
		10.1103/PhysRevD.80.084043} {\bibfield  {journal} {\bibinfo  {journal} {Phys.
				Rev. D}\ }\textbf {\bibinfo {volume} {80}},\ \bibinfo {pages} {084043}
		(\bibinfo {year} {2009})},\ \Eprint {http://arxiv.org/abs/0907.0700}
	{arXiv:0907.0700 [gr-qc]} \BibitemShut {NoStop}%
	\bibitem [{\citenamefont {Brown}\ \emph {et~al.}(2012)\citenamefont {Brown},
		\citenamefont {Harry}, \citenamefont {Lundgren},\ and\ \citenamefont
		{Nitz}}]{Brown:2012qf}%
	\BibitemOpen
	\bibfield  {author} {\bibinfo {author} {\bibfnamefont {Duncan~A.}\
			\bibnamefont {Brown}}, \bibinfo {author} {\bibfnamefont {Ian}\ \bibnamefont
			{Harry}}, \bibinfo {author} {\bibfnamefont {Andrew}\ \bibnamefont
			{Lundgren}}, \ and\ \bibinfo {author} {\bibfnamefont {Alexander~H.}\
			\bibnamefont {Nitz}},\ }\bibfield  {title} {\enquote {\bibinfo {title}
			{{Detecting binary neutron star systems with spin in advanced
					gravitational-wave detectors}},}\ }\href {\doibase
		10.1103/PhysRevD.86.084017} {\bibfield  {journal} {\bibinfo  {journal} {Phys.
				Rev. D}\ }\textbf {\bibinfo {volume} {86}},\ \bibinfo {pages} {084017}
		(\bibinfo {year} {2012})},\ \Eprint {http://arxiv.org/abs/1207.6406}
	{arXiv:1207.6406 [gr-qc]} \BibitemShut {NoStop}%
	\bibitem [{\citenamefont {Baird}\ \emph {et~al.}(2013)\citenamefont {Baird},
		\citenamefont {Fairhurst}, \citenamefont {Hannam},\ and\ \citenamefont
		{Murphy}}]{Baird:2012cu}%
	\BibitemOpen
	\bibfield  {author} {\bibinfo {author} {\bibfnamefont {Emily}\ \bibnamefont
			{Baird}}, \bibinfo {author} {\bibfnamefont {Stephen}\ \bibnamefont
			{Fairhurst}}, \bibinfo {author} {\bibfnamefont {Mark}\ \bibnamefont
			{Hannam}}, \ and\ \bibinfo {author} {\bibfnamefont {Patricia}\ \bibnamefont
			{Murphy}},\ }\bibfield  {title} {\enquote {\bibinfo {title} {{Degeneracy
					between mass and spin in black-hole-binary waveforms}},}\ }\href {\doibase
		10.1103/PhysRevD.87.024035} {\bibfield  {journal} {\bibinfo  {journal} {Phys.
				Rev. D}\ }\textbf {\bibinfo {volume} {87}},\ \bibinfo {pages} {024035}
		(\bibinfo {year} {2013})},\ \Eprint {http://arxiv.org/abs/1211.0546}
	{arXiv:1211.0546 [gr-qc]} \BibitemShut {NoStop}%
	\bibitem [{\citenamefont {Krishnendu}\ and\ \citenamefont
		{Ohme}(2022)}]{Krishnendu:2021cyi}%
	\BibitemOpen
	\bibfield  {author} {\bibinfo {author} {\bibfnamefont {N.~V.}\ \bibnamefont
			{Krishnendu}}\ and\ \bibinfo {author} {\bibfnamefont {Frank}\ \bibnamefont
			{Ohme}},\ }\bibfield  {title} {\enquote {\bibinfo {title} {{Interplay of
					spin-precession and higher harmonics in the parameter estimation of binary
					black holes}},}\ }\href {\doibase 10.1103/PhysRevD.105.064012} {\bibfield
		{journal} {\bibinfo  {journal} {Phys. Rev. D}\ }\textbf {\bibinfo {volume}
			{105}},\ \bibinfo {pages} {064012} (\bibinfo {year} {2022})},\ \Eprint
	{http://arxiv.org/abs/2110.00766} {arXiv:2110.00766 [gr-qc]} \BibitemShut
	{NoStop}%
	\bibitem [{\citenamefont {Barker}\ and\ \citenamefont
		{O'Connell}(1970)}]{Barker:1970zr}%
	\BibitemOpen
	\bibfield  {author} {\bibinfo {author} {\bibfnamefont {B.~M.}\ \bibnamefont
			{Barker}}\ and\ \bibinfo {author} {\bibfnamefont {R.~F.}\ \bibnamefont
			{O'Connell}},\ }\bibfield  {title} {\enquote {\bibinfo {title} {{Derivation
					of the equations of motion of a gyroscope from the quantum theory of
					gravitation}},}\ }\href {\doibase 10.1103/PhysRevD.2.1428} {\bibfield
		{journal} {\bibinfo  {journal} {Phys. Rev. D}\ }\textbf {\bibinfo {volume}
			{2}},\ \bibinfo {pages} {1428--1435} (\bibinfo {year} {1970})}\BibitemShut
	{NoStop}%
	\bibitem [{\citenamefont {Barker}\ and\ \citenamefont
		{O'Connell}(1975)}]{Barker:1975ae}%
	\BibitemOpen
	\bibfield  {author} {\bibinfo {author} {\bibfnamefont {B.~M.}\ \bibnamefont
			{Barker}}\ and\ \bibinfo {author} {\bibfnamefont {R.~F.}\ \bibnamefont
			{O'Connell}},\ }\bibfield  {title} {\enquote {\bibinfo {title}
			{{Gravitational Two-Body Problem with Arbitrary Masses, Spins, and Quadrupole
					Moments}},}\ }\href {\doibase 10.1103/PhysRevD.12.329} {\bibfield  {journal}
		{\bibinfo  {journal} {Phys. Rev. D}\ }\textbf {\bibinfo {volume} {12}},\
		\bibinfo {pages} {329--335} (\bibinfo {year} {1975})}\BibitemShut {NoStop}%
	\bibitem [{\citenamefont {Kidder}\ \emph {et~al.}(1993)\citenamefont {Kidder},
		\citenamefont {Will},\ and\ \citenamefont {Wiseman}}]{Kidder:1992fr}%
	\BibitemOpen
	\bibfield  {author} {\bibinfo {author} {\bibfnamefont {Lawrence~E.}\
			\bibnamefont {Kidder}}, \bibinfo {author} {\bibfnamefont {Clifford~M.}\
			\bibnamefont {Will}}, \ and\ \bibinfo {author} {\bibfnamefont {Alan~G.}\
			\bibnamefont {Wiseman}},\ }\bibfield  {title} {\enquote {\bibinfo {title}
			{{Spin effects in the inspiral of coalescing compact binaries}},}\ }\href
	{\doibase 10.1103/PhysRevD.47.R4183} {\bibfield  {journal} {\bibinfo
			{journal} {Phys. Rev. D}\ }\textbf {\bibinfo {volume} {47}},\ \bibinfo
		{pages} {R4183--R4187} (\bibinfo {year} {1993})},\ \Eprint
	{http://arxiv.org/abs/gr-qc/9211025} {arXiv:gr-qc/9211025} \BibitemShut
	{NoStop}%
	\bibitem [{\citenamefont {Kidder}(1995)}]{Kidder:1995zr}%
	\BibitemOpen
	\bibfield  {author} {\bibinfo {author} {\bibfnamefont {Lawrence~E.}\
			\bibnamefont {Kidder}},\ }\bibfield  {title} {\enquote {\bibinfo {title}
			{{Coalescing binary systems of compact objects to postNewtonian 5/2 order. 5.
					Spin effects}},}\ }\href {\doibase 10.1103/PhysRevD.52.821} {\bibfield
		{journal} {\bibinfo  {journal} {Phys. Rev. D}\ }\textbf {\bibinfo {volume}
			{52}},\ \bibinfo {pages} {821--847} (\bibinfo {year} {1995})},\ \Eprint
	{http://arxiv.org/abs/gr-qc/9506022} {arXiv:gr-qc/9506022} \BibitemShut
	{NoStop}%
	\bibitem [{\citenamefont {Tagoshi}\ \emph {et~al.}(2001)\citenamefont
		{Tagoshi}, \citenamefont {Ohashi},\ and\ \citenamefont
		{Owen}}]{Tagoshi:2000zg}%
	\BibitemOpen
	\bibfield  {author} {\bibinfo {author} {\bibfnamefont {Hideyuki}\
			\bibnamefont {Tagoshi}}, \bibinfo {author} {\bibfnamefont {Akira}\
			\bibnamefont {Ohashi}}, \ and\ \bibinfo {author} {\bibfnamefont
			{Benjamin~J.}\ \bibnamefont {Owen}},\ }\bibfield  {title} {\enquote {\bibinfo
			{title} {{Gravitational field and equations of motion of spinning compact
					binaries to 2.5 postNewtonian order}},}\ }\href {\doibase
		10.1103/PhysRevD.63.044006} {\bibfield  {journal} {\bibinfo  {journal} {Phys.
				Rev. D}\ }\textbf {\bibinfo {volume} {63}},\ \bibinfo {pages} {044006}
		(\bibinfo {year} {2001})},\ \Eprint {http://arxiv.org/abs/gr-qc/0010014}
	{arXiv:gr-qc/0010014} \BibitemShut {NoStop}%
	\bibitem [{\citenamefont {Faye}\ \emph {et~al.}(2006)\citenamefont {Faye},
		\citenamefont {Blanchet},\ and\ \citenamefont {Buonanno}}]{Faye:2006gx}%
	\BibitemOpen
	\bibfield  {author} {\bibinfo {author} {\bibfnamefont {Guillaume}\
			\bibnamefont {Faye}}, \bibinfo {author} {\bibfnamefont {Luc}\ \bibnamefont
			{Blanchet}}, \ and\ \bibinfo {author} {\bibfnamefont {Alessandra}\
			\bibnamefont {Buonanno}},\ }\bibfield  {title} {\enquote {\bibinfo {title}
			{{Higher-order spin effects in the dynamics of compact binaries. I. Equations
					of motion}},}\ }\href {\doibase 10.1103/PhysRevD.74.104033} {\bibfield
		{journal} {\bibinfo  {journal} {Phys. Rev. D}\ }\textbf {\bibinfo {volume}
			{74}},\ \bibinfo {pages} {104033} (\bibinfo {year} {2006})},\ \Eprint
	{http://arxiv.org/abs/gr-qc/0605139} {arXiv:gr-qc/0605139} \BibitemShut
	{NoStop}%
	\bibitem [{\citenamefont {Blanchet}\ \emph {et~al.}(2006)\citenamefont
		{Blanchet}, \citenamefont {Buonanno},\ and\ \citenamefont
		{Faye}}]{Blanchet:2006gy}%
	\BibitemOpen
	\bibfield  {author} {\bibinfo {author} {\bibfnamefont {Luc}\ \bibnamefont
			{Blanchet}}, \bibinfo {author} {\bibfnamefont {Alessandra}\ \bibnamefont
			{Buonanno}}, \ and\ \bibinfo {author} {\bibfnamefont {Guillaume}\
			\bibnamefont {Faye}},\ }\bibfield  {title} {\enquote {\bibinfo {title}
			{{Higher-order spin effects in the dynamics of compact binaries. II.
					Radiation field}},}\ }\href {\doibase 10.1103/PhysRevD.81.089901} {\bibfield
		{journal} {\bibinfo  {journal} {Phys. Rev. D}\ }\textbf {\bibinfo {volume}
			{74}},\ \bibinfo {pages} {104034} (\bibinfo {year} {2006})},\ \bibinfo {note}
	{[Erratum: Phys.Rev.D 75, 049903 (2007), Erratum: Phys.Rev.D 81, 089901
		(2010)]},\ \Eprint {http://arxiv.org/abs/gr-qc/0605140} {arXiv:gr-qc/0605140}
	\BibitemShut {NoStop}%
	\bibitem [{\citenamefont {Damour}\ \emph {et~al.}(2008)\citenamefont {Damour},
		\citenamefont {Jaranowski},\ and\ \citenamefont {Schaefer}}]{Damour:2007nc}%
	\BibitemOpen
	\bibfield  {author} {\bibinfo {author} {\bibfnamefont {Thibault}\
			\bibnamefont {Damour}}, \bibinfo {author} {\bibfnamefont {Piotr}\
			\bibnamefont {Jaranowski}}, \ and\ \bibinfo {author} {\bibfnamefont
			{Gerhard}\ \bibnamefont {Schaefer}},\ }\bibfield  {title} {\enquote {\bibinfo
			{title} {{Hamiltonian of two spinning compact bodies with next-to-leading
					order gravitational spin-orbit coupling}},}\ }\href {\doibase
		10.1103/PhysRevD.77.064032} {\bibfield  {journal} {\bibinfo  {journal} {Phys.
				Rev. D}\ }\textbf {\bibinfo {volume} {77}},\ \bibinfo {pages} {064032}
		(\bibinfo {year} {2008})},\ \Eprint {http://arxiv.org/abs/0711.1048}
	{arXiv:0711.1048 [gr-qc]} \BibitemShut {NoStop}%
	\bibitem [{\citenamefont {Steinhoff}\ \emph
		{et~al.}(2008{\natexlab{a}})\citenamefont {Steinhoff}, \citenamefont
		{Hergt},\ and\ \citenamefont {Schaefer}}]{Steinhoff:2007mb}%
	\BibitemOpen
	\bibfield  {author} {\bibinfo {author} {\bibfnamefont {Jan}\ \bibnamefont
			{Steinhoff}}, \bibinfo {author} {\bibfnamefont {Steven}\ \bibnamefont
			{Hergt}}, \ and\ \bibinfo {author} {\bibfnamefont {Gerhard}\ \bibnamefont
			{Schaefer}},\ }\bibfield  {title} {\enquote {\bibinfo {title} {{On the
					next-to-leading order gravitational spin(1)-spin(2) dynamics}},}\ }\href
	{\doibase 10.1103/PhysRevD.77.081501} {\bibfield  {journal} {\bibinfo
			{journal} {Phys. Rev. D}\ }\textbf {\bibinfo {volume} {77}},\ \bibinfo
		{pages} {081501} (\bibinfo {year} {2008}{\natexlab{a}})},\ \Eprint
	{http://arxiv.org/abs/0712.1716} {arXiv:0712.1716 [gr-qc]} \BibitemShut
	{NoStop}%
	\bibitem [{\citenamefont {Steinhoff}\ \emph
		{et~al.}(2008{\natexlab{b}})\citenamefont {Steinhoff}, \citenamefont
		{Schaefer},\ and\ \citenamefont {Hergt}}]{Steinhoff:2008zr}%
	\BibitemOpen
	\bibfield  {author} {\bibinfo {author} {\bibfnamefont {Jan}\ \bibnamefont
			{Steinhoff}}, \bibinfo {author} {\bibfnamefont {Gerhard}\ \bibnamefont
			{Schaefer}}, \ and\ \bibinfo {author} {\bibfnamefont {Steven}\ \bibnamefont
			{Hergt}},\ }\bibfield  {title} {\enquote {\bibinfo {title} {{ADM canonical
					formalism for gravitating spinning objects}},}\ }\href {\doibase
		10.1103/PhysRevD.77.104018} {\bibfield  {journal} {\bibinfo  {journal} {Phys.
				Rev. D}\ }\textbf {\bibinfo {volume} {77}},\ \bibinfo {pages} {104018}
		(\bibinfo {year} {2008}{\natexlab{b}})},\ \Eprint
	{http://arxiv.org/abs/0805.3136} {arXiv:0805.3136 [gr-qc]} \BibitemShut
	{NoStop}%
	\bibitem [{\citenamefont {Steinhoff}\ \emph
		{et~al.}(2008{\natexlab{c}})\citenamefont {Steinhoff}, \citenamefont
		{Hergt},\ and\ \citenamefont {Schaefer}}]{Steinhoff:2008ji}%
	\BibitemOpen
	\bibfield  {author} {\bibinfo {author} {\bibfnamefont {Jan}\ \bibnamefont
			{Steinhoff}}, \bibinfo {author} {\bibfnamefont {Steven}\ \bibnamefont
			{Hergt}}, \ and\ \bibinfo {author} {\bibfnamefont {Gerhard}\ \bibnamefont
			{Schaefer}},\ }\bibfield  {title} {\enquote {\bibinfo {title} {{Spin-squared
					Hamiltonian of next-to-leading order gravitational interaction}},}\ }\href
	{\doibase 10.1103/PhysRevD.78.101503} {\bibfield  {journal} {\bibinfo
			{journal} {Phys. Rev. D}\ }\textbf {\bibinfo {volume} {78}},\ \bibinfo
		{pages} {101503} (\bibinfo {year} {2008}{\natexlab{c}})},\ \Eprint
	{http://arxiv.org/abs/0809.2200} {arXiv:0809.2200 [gr-qc]} \BibitemShut
	{NoStop}%
	\bibitem [{\citenamefont {Marsat}\ \emph
		{et~al.}(2013{\natexlab{a}})\citenamefont {Marsat}, \citenamefont {Bohe},
		\citenamefont {Faye},\ and\ \citenamefont {Blanchet}}]{Marsat:2012fn}%
	\BibitemOpen
	\bibfield  {author} {\bibinfo {author} {\bibfnamefont {Sylvain}\ \bibnamefont
			{Marsat}}, \bibinfo {author} {\bibfnamefont {Alejandro}\ \bibnamefont
			{Bohe}}, \bibinfo {author} {\bibfnamefont {Guillaume}\ \bibnamefont {Faye}},
		\ and\ \bibinfo {author} {\bibfnamefont {Luc}\ \bibnamefont {Blanchet}},\
	}\bibfield  {title} {\enquote {\bibinfo {title} {{Next-to-next-to-leading
					order spin-orbit effects in the equations of motion of compact binary
					systems}},}\ }\href {\doibase 10.1088/0264-9381/30/5/055007} {\bibfield
		{journal} {\bibinfo  {journal} {Class. Quant. Grav.}\ }\textbf {\bibinfo
			{volume} {30}},\ \bibinfo {pages} {055007} (\bibinfo {year}
		{2013}{\natexlab{a}})},\ \Eprint {http://arxiv.org/abs/1210.4143}
	{arXiv:1210.4143 [gr-qc]} \BibitemShut {NoStop}%
	\bibitem [{\citenamefont {Hergt}\ \emph {et~al.}(2010)\citenamefont {Hergt},
		\citenamefont {Steinhoff},\ and\ \citenamefont {Schaefer}}]{Hergt:2010pa}%
	\BibitemOpen
	\bibfield  {author} {\bibinfo {author} {\bibfnamefont {Steven}\ \bibnamefont
			{Hergt}}, \bibinfo {author} {\bibfnamefont {Jan}\ \bibnamefont {Steinhoff}},
		\ and\ \bibinfo {author} {\bibfnamefont {Gerhard}\ \bibnamefont {Schaefer}},\
	}\bibfield  {title} {\enquote {\bibinfo {title} {{Reduced Hamiltonian for
					next-to-leading order Spin-Squared Dynamics of General Compact Binaries}},}\
	}\href {\doibase 10.1088/0264-9381/27/13/135007} {\bibfield  {journal}
		{\bibinfo  {journal} {Class. Quant. Grav.}\ }\textbf {\bibinfo {volume}
			{27}},\ \bibinfo {pages} {135007} (\bibinfo {year} {2010})},\ \Eprint
	{http://arxiv.org/abs/1002.2093} {arXiv:1002.2093 [gr-qc]} \BibitemShut
	{NoStop}%
	\bibitem [{\citenamefont {Hergt}\ \emph {et~al.}(2014)\citenamefont {Hergt},
		\citenamefont {Steinhoff},\ and\ \citenamefont {Schaefer}}]{Hergt:2012zx}%
	\BibitemOpen
	\bibfield  {author} {\bibinfo {author} {\bibfnamefont {Steven}\ \bibnamefont
			{Hergt}}, \bibinfo {author} {\bibfnamefont {Jan}\ \bibnamefont {Steinhoff}},
		\ and\ \bibinfo {author} {\bibfnamefont {Gerhard}\ \bibnamefont {Schaefer}},\
	}\bibfield  {title} {\enquote {\bibinfo {title} {{On the comparison of
					results regarding the post-Newtonian approximate treatment of the dynamics of
					extended spinning compact binaries}},}\ }\href {\doibase
		10.1088/1742-6596/484/1/012018} {\bibfield  {journal} {\bibinfo  {journal}
			{J. Phys. Conf. Ser.}\ }\textbf {\bibinfo {volume} {484}},\ \bibinfo {pages}
		{012018} (\bibinfo {year} {2014})},\ \Eprint {http://arxiv.org/abs/1205.4530}
	{arXiv:1205.4530 [gr-qc]} \BibitemShut {NoStop}%
	\bibitem [{\citenamefont {Bohe}\ \emph {et~al.}(2013)\citenamefont {Bohe},
		\citenamefont {Marsat}, \citenamefont {Faye},\ and\ \citenamefont
		{Blanchet}}]{Bohe:2012mr}%
	\BibitemOpen
	\bibfield  {author} {\bibinfo {author} {\bibfnamefont {Alejandro}\
			\bibnamefont {Bohe}}, \bibinfo {author} {\bibfnamefont {Sylvain}\
			\bibnamefont {Marsat}}, \bibinfo {author} {\bibfnamefont {Guillaume}\
			\bibnamefont {Faye}}, \ and\ \bibinfo {author} {\bibfnamefont {Luc}\
			\bibnamefont {Blanchet}},\ }\bibfield  {title} {\enquote {\bibinfo {title}
			{{Next-to-next-to-leading order spin-orbit effects in the near-zone metric
					and precession equations of compact binaries}},}\ }\href {\doibase
		10.1088/0264-9381/30/7/075017} {\bibfield  {journal} {\bibinfo  {journal}
			{Class. Quant. Grav.}\ }\textbf {\bibinfo {volume} {30}},\ \bibinfo {pages}
		{075017} (\bibinfo {year} {2013})},\ \Eprint {http://arxiv.org/abs/1212.5520}
	{arXiv:1212.5520 [gr-qc]} \BibitemShut {NoStop}%
	\bibitem [{\citenamefont {Hartung}\ \emph {et~al.}(2013)\citenamefont
		{Hartung}, \citenamefont {Steinhoff},\ and\ \citenamefont
		{Schafer}}]{Hartung:2013dza}%
	\BibitemOpen
	\bibfield  {author} {\bibinfo {author} {\bibfnamefont {Johannes}\
			\bibnamefont {Hartung}}, \bibinfo {author} {\bibfnamefont {Jan}\ \bibnamefont
			{Steinhoff}}, \ and\ \bibinfo {author} {\bibfnamefont {Gerhard}\ \bibnamefont
			{Schafer}},\ }\bibfield  {title} {\enquote {\bibinfo {title}
			{{Next-to-next-to-leading order post-Newtonian linear-in-spin binary
					Hamiltonians}},}\ }\href {\doibase 10.1002/andp.201200271} {\bibfield
		{journal} {\bibinfo  {journal} {Annalen Phys.}\ }\textbf {\bibinfo {volume}
			{525}},\ \bibinfo {pages} {359--394} (\bibinfo {year} {2013})},\ \Eprint
	{http://arxiv.org/abs/1302.6723} {arXiv:1302.6723 [gr-qc]} \BibitemShut
	{NoStop}%
	\bibitem [{\citenamefont {Marsat}\ \emph
		{et~al.}(2013{\natexlab{b}})\citenamefont {Marsat}, \citenamefont {Blanchet},
		\citenamefont {Bohe},\ and\ \citenamefont {Faye}}]{Marsat:2013wwa}%
	\BibitemOpen
	\bibfield  {author} {\bibinfo {author} {\bibfnamefont {Sylvain}\ \bibnamefont
			{Marsat}}, \bibinfo {author} {\bibfnamefont {Luc}\ \bibnamefont {Blanchet}},
		\bibinfo {author} {\bibfnamefont {Alejandro}\ \bibnamefont {Bohe}}, \ and\
		\bibinfo {author} {\bibfnamefont {Guillaume}\ \bibnamefont {Faye}},\
	}\bibfield  {title} {\enquote {\bibinfo {title} {{Gravitational waves from
					spinning compact object binaries: New post-Newtonian results}},}\ \
	}(\bibinfo {year} {2013})\ \Eprint {http://arxiv.org/abs/1312.5375}
	{arXiv:1312.5375 [gr-qc]} \BibitemShut {NoStop}%
	\bibitem [{\citenamefont {Boh\'e}\ \emph {et~al.}(2015)\citenamefont {Boh\'e},
		\citenamefont {Faye}, \citenamefont {Marsat},\ and\ \citenamefont
		{Porter}}]{Bohe:2015ana}%
	\BibitemOpen
	\bibfield  {author} {\bibinfo {author} {\bibfnamefont {Alejandro}\
			\bibnamefont {Boh\'e}}, \bibinfo {author} {\bibfnamefont {Guillaume}\
			\bibnamefont {Faye}}, \bibinfo {author} {\bibfnamefont {Sylvain}\
			\bibnamefont {Marsat}}, \ and\ \bibinfo {author} {\bibfnamefont {Edward~K.}\
			\bibnamefont {Porter}},\ }\bibfield  {title} {\enquote {\bibinfo {title}
			{{Quadratic-in-spin effects in the orbital dynamics and gravitational-wave
					energy flux of compact binaries at the 3PN order}},}\ }\href {\doibase
		10.1088/0264-9381/32/19/195010} {\bibfield  {journal} {\bibinfo  {journal}
			{Class. Quant. Grav.}\ }\textbf {\bibinfo {volume} {32}},\ \bibinfo {pages}
		{195010} (\bibinfo {year} {2015})},\ \Eprint
	{http://arxiv.org/abs/1501.01529} {arXiv:1501.01529 [gr-qc]} \BibitemShut
	{NoStop}%
	\bibitem [{\citenamefont {Bini}\ \emph {et~al.}(2017)\citenamefont {Bini},
		\citenamefont {Geralico},\ and\ \citenamefont {Vines}}]{Bini:2017pee}%
	\BibitemOpen
	\bibfield  {author} {\bibinfo {author} {\bibfnamefont {Donato}\ \bibnamefont
			{Bini}}, \bibinfo {author} {\bibfnamefont {Andrea}\ \bibnamefont {Geralico}},
		\ and\ \bibinfo {author} {\bibfnamefont {Justin}\ \bibnamefont {Vines}},\
	}\bibfield  {title} {\enquote {\bibinfo {title} {{Hyperbolic scattering of
					spinning particles by a Kerr black hole}},}\ }\href {\doibase
		10.1103/PhysRevD.96.084044} {\bibfield  {journal} {\bibinfo  {journal} {Phys.
				Rev. D}\ }\textbf {\bibinfo {volume} {96}},\ \bibinfo {pages} {084044}
		(\bibinfo {year} {2017})},\ \Eprint {http://arxiv.org/abs/1707.09814}
	{arXiv:1707.09814 [gr-qc]} \BibitemShut {NoStop}%
	\bibitem [{\citenamefont {Siemonsen}\ \emph {et~al.}(2018)\citenamefont
		{Siemonsen}, \citenamefont {Steinhoff},\ and\ \citenamefont
		{Vines}}]{Siemonsen:2017yux}%
	\BibitemOpen
	\bibfield  {author} {\bibinfo {author} {\bibfnamefont {Nils}\ \bibnamefont
			{Siemonsen}}, \bibinfo {author} {\bibfnamefont {Jan}\ \bibnamefont
			{Steinhoff}}, \ and\ \bibinfo {author} {\bibfnamefont {Justin}\ \bibnamefont
			{Vines}},\ }\bibfield  {title} {\enquote {\bibinfo {title} {{Gravitational
					waves from spinning binary black holes at the leading post-Newtonian orders
					at all orders in spin}},}\ }\href {\doibase 10.1103/PhysRevD.97.124046}
	{\bibfield  {journal} {\bibinfo  {journal} {Phys. Rev. D}\ }\textbf {\bibinfo
			{volume} {97}},\ \bibinfo {pages} {124046} (\bibinfo {year} {2018})},\
	\Eprint {http://arxiv.org/abs/1712.08603} {arXiv:1712.08603 [gr-qc]}
	\BibitemShut {NoStop}%
	\bibitem [{\citenamefont {Porto}(2006)}]{Porto:2005ac}%
	\BibitemOpen
	\bibfield  {author} {\bibinfo {author} {\bibfnamefont {Rafael~A.}\
			\bibnamefont {Porto}},\ }\bibfield  {title} {\enquote {\bibinfo {title}
			{{Post-Newtonian corrections to the motion of spinning bodies in NRGR}},}\
	}\href {\doibase 10.1103/PhysRevD.73.104031} {\bibfield  {journal} {\bibinfo
			{journal} {Phys. Rev. D}\ }\textbf {\bibinfo {volume} {73}},\ \bibinfo
		{pages} {104031} (\bibinfo {year} {2006})},\ \Eprint
	{http://arxiv.org/abs/gr-qc/0511061} {arXiv:gr-qc/0511061} \BibitemShut
	{NoStop}%
	\bibitem [{\citenamefont {Porto}\ and\ \citenamefont
		{Rothstein}(2006)}]{Porto:2006bt}%
	\BibitemOpen
	\bibfield  {author} {\bibinfo {author} {\bibfnamefont {Rafael~A.}\
			\bibnamefont {Porto}}\ and\ \bibinfo {author} {\bibfnamefont {Ira~Z.}\
			\bibnamefont {Rothstein}},\ }\bibfield  {title} {\enquote {\bibinfo {title}
			{{The Hyperfine Einstein-Infeld-Hoffmann potential}},}\ }\href {\doibase
		10.1103/PhysRevLett.97.021101} {\bibfield  {journal} {\bibinfo  {journal}
			{Phys. Rev. Lett.}\ }\textbf {\bibinfo {volume} {97}},\ \bibinfo {pages}
		{021101} (\bibinfo {year} {2006})},\ \Eprint
	{http://arxiv.org/abs/gr-qc/0604099} {arXiv:gr-qc/0604099} \BibitemShut
	{NoStop}%
	\bibitem [{\citenamefont {Porto}\ and\ \citenamefont
		{Rothstein}(2007)}]{Porto:2007tt}%
	\BibitemOpen
	\bibfield  {author} {\bibinfo {author} {\bibfnamefont {Rafael~A.}\
			\bibnamefont {Porto}}\ and\ \bibinfo {author} {\bibfnamefont {Ira~Z.}\
			\bibnamefont {Rothstein}},\ }\bibfield  {title} {\enquote {\bibinfo {title}
			{{Comment on `On the next-to-leading order gravitational spin(1) - spin(2)
					dynamics' by J. Steinhoff et al}},}\ }\href@noop {} {\  (\bibinfo {year}
		{2007})},\ \Eprint {http://arxiv.org/abs/0712.2032} {arXiv:0712.2032 [gr-qc]}
	\BibitemShut {NoStop}%
	\bibitem [{\citenamefont {Porto}\ and\ \citenamefont
		{Rothstein}(2008{\natexlab{a}})}]{Porto:2008tb}%
	\BibitemOpen
	\bibfield  {author} {\bibinfo {author} {\bibfnamefont {Rafael~A.}\
			\bibnamefont {Porto}}\ and\ \bibinfo {author} {\bibfnamefont {Ira~Z.}\
			\bibnamefont {Rothstein}},\ }\bibfield  {title} {\enquote {\bibinfo {title}
			{{Spin(1)Spin(2) Effects in the Motion of Inspiralling Compact Binaries at
					Third Order in the Post-Newtonian Expansion}},}\ }\href {\doibase
		10.1103/PhysRevD.78.044012} {\bibfield  {journal} {\bibinfo  {journal} {Phys.
				Rev. D}\ }\textbf {\bibinfo {volume} {78}},\ \bibinfo {pages} {044012}
		(\bibinfo {year} {2008}{\natexlab{a}})},\ \bibinfo {note} {[Erratum:
		Phys.Rev.D 81, 029904 (2010)]},\ \Eprint {http://arxiv.org/abs/0802.0720}
	{arXiv:0802.0720 [gr-qc]} \BibitemShut {NoStop}%
	\bibitem [{\citenamefont {Levi}(2010{\natexlab{a}})}]{Levi:2008nh}%
	\BibitemOpen
	\bibfield  {author} {\bibinfo {author} {\bibfnamefont {Michele}\ \bibnamefont
			{Levi}},\ }\bibfield  {title} {\enquote {\bibinfo {title} {{Next to Leading
					Order gravitational Spin1-Spin2 coupling with Kaluza-Klein reduction}},}\
	}\href {\doibase 10.1103/PhysRevD.82.064029} {\bibfield  {journal} {\bibinfo
			{journal} {Phys. Rev. D}\ }\textbf {\bibinfo {volume} {82}},\ \bibinfo
		{pages} {064029} (\bibinfo {year} {2010}{\natexlab{a}})},\ \Eprint
	{http://arxiv.org/abs/0802.1508} {arXiv:0802.1508 [gr-qc]} \BibitemShut
	{NoStop}%
	\bibitem [{\citenamefont {Porto}\ and\ \citenamefont
		{Rothstein}(2008{\natexlab{b}})}]{Porto:2008jj}%
	\BibitemOpen
	\bibfield  {author} {\bibinfo {author} {\bibfnamefont {Rafael~A}\
			\bibnamefont {Porto}}\ and\ \bibinfo {author} {\bibfnamefont {Ira~Z.}\
			\bibnamefont {Rothstein}},\ }\bibfield  {title} {\enquote {\bibinfo {title}
			{{Next to Leading Order Spin(1)Spin(1) Effects in the Motion of Inspiralling
					Compact Binaries}},}\ }\href {\doibase 10.1103/PhysRevD.78.044013} {\bibfield
		{journal} {\bibinfo  {journal} {Phys. Rev. D}\ }\textbf {\bibinfo {volume}
			{78}},\ \bibinfo {pages} {044013} (\bibinfo {year} {2008}{\natexlab{b}})},\
	\bibinfo {note} {[Erratum: Phys.Rev.D 81, 029905 (2010)]},\ \Eprint
	{http://arxiv.org/abs/0804.0260} {arXiv:0804.0260 [gr-qc]} \BibitemShut
	{NoStop}%
	\bibitem [{\citenamefont {Porto}(2010)}]{Porto:2010tr}%
	\BibitemOpen
	\bibfield  {author} {\bibinfo {author} {\bibfnamefont {Rafael~A.}\
			\bibnamefont {Porto}},\ }\bibfield  {title} {\enquote {\bibinfo {title}
			{{Next to leading order spin-orbit effects in the motion of inspiralling
					compact binaries}},}\ }\href {\doibase 10.1088/0264-9381/27/20/205001}
	{\bibfield  {journal} {\bibinfo  {journal} {Class. Quant. Grav.}\ }\textbf
		{\bibinfo {volume} {27}},\ \bibinfo {pages} {205001} (\bibinfo {year}
		{2010})},\ \Eprint {http://arxiv.org/abs/1005.5730} {arXiv:1005.5730 [gr-qc]}
	\BibitemShut {NoStop}%
	\bibitem [{\citenamefont {Levi}(2010{\natexlab{b}})}]{Levi:2010zu}%
	\BibitemOpen
	\bibfield  {author} {\bibinfo {author} {\bibfnamefont {Michele}\ \bibnamefont
			{Levi}},\ }\bibfield  {title} {\enquote {\bibinfo {title} {{Next to Leading
					Order gravitational Spin-Orbit coupling in an Effective Field Theory
					approach}},}\ }\href {\doibase 10.1103/PhysRevD.82.104004} {\bibfield
		{journal} {\bibinfo  {journal} {Phys. Rev. D}\ }\textbf {\bibinfo {volume}
			{82}},\ \bibinfo {pages} {104004} (\bibinfo {year} {2010}{\natexlab{b}})},\
	\Eprint {http://arxiv.org/abs/1006.4139} {arXiv:1006.4139 [gr-qc]}
	\BibitemShut {NoStop}%
	\bibitem [{\citenamefont {Porto}\ \emph {et~al.}(2011)\citenamefont {Porto},
		\citenamefont {Ross},\ and\ \citenamefont {Rothstein}}]{Porto:2010zg}%
	\BibitemOpen
	\bibfield  {author} {\bibinfo {author} {\bibfnamefont {Rafael~A.}\
			\bibnamefont {Porto}}, \bibinfo {author} {\bibfnamefont {Andreas}\
			\bibnamefont {Ross}}, \ and\ \bibinfo {author} {\bibfnamefont {Ira~Z.}\
			\bibnamefont {Rothstein}},\ }\bibfield  {title} {\enquote {\bibinfo {title}
			{{Spin induced multipole moments for the gravitational wave flux from binary
					inspirals to third Post-Newtonian order}},}\ }\href {\doibase
		10.1088/1475-7516/2011/03/009} {\bibfield  {journal} {\bibinfo  {journal}
			{JCAP}\ }\textbf {\bibinfo {volume} {03}},\ \bibinfo {pages} {009} (\bibinfo
		{year} {2011})},\ \Eprint {http://arxiv.org/abs/1007.1312} {arXiv:1007.1312
		[gr-qc]} \BibitemShut {NoStop}%
	\bibitem [{\citenamefont {Hartung}\ and\ \citenamefont
		{Steinhoff}(2011)}]{Hartung:2011ea}%
	\BibitemOpen
	\bibfield  {author} {\bibinfo {author} {\bibfnamefont {Johannes}\
			\bibnamefont {Hartung}}\ and\ \bibinfo {author} {\bibfnamefont {Jan}\
			\bibnamefont {Steinhoff}},\ }\bibfield  {title} {\enquote {\bibinfo {title}
			{{Next-to-next-to-leading order post-Newtonian spin(1)-spin(2) Hamiltonian
					for self-gravitating binaries}},}\ }\href {\doibase 10.1002/andp.201100163}
	{\bibfield  {journal} {\bibinfo  {journal} {Annalen Phys.}\ }\textbf
		{\bibinfo {volume} {523}},\ \bibinfo {pages} {919--924} (\bibinfo {year}
		{2011})},\ \Eprint {http://arxiv.org/abs/1107.4294} {arXiv:1107.4294 [gr-qc]}
	\BibitemShut {NoStop}%
	\bibitem [{\citenamefont {Levi}(2012)}]{Levi:2011eq}%
	\BibitemOpen
	\bibfield  {author} {\bibinfo {author} {\bibfnamefont {Michele}\ \bibnamefont
			{Levi}},\ }\bibfield  {title} {\enquote {\bibinfo {title} {{Binary dynamics
					from spin1-spin2 coupling at fourth post-Newtonian order}},}\ }\href
	{\doibase 10.1103/PhysRevD.85.064043} {\bibfield  {journal} {\bibinfo
			{journal} {Phys. Rev. D}\ }\textbf {\bibinfo {volume} {85}},\ \bibinfo
		{pages} {064043} (\bibinfo {year} {2012})},\ \Eprint
	{http://arxiv.org/abs/1107.4322} {arXiv:1107.4322 [gr-qc]} \BibitemShut
	{NoStop}%
	\bibitem [{\citenamefont {Porto}\ \emph {et~al.}(2012)\citenamefont {Porto},
		\citenamefont {Ross},\ and\ \citenamefont {Rothstein}}]{Porto:2012as}%
	\BibitemOpen
	\bibfield  {author} {\bibinfo {author} {\bibfnamefont {Rafael~A.}\
			\bibnamefont {Porto}}, \bibinfo {author} {\bibfnamefont {Andreas}\
			\bibnamefont {Ross}}, \ and\ \bibinfo {author} {\bibfnamefont {Ira~Z.}\
			\bibnamefont {Rothstein}},\ }\bibfield  {title} {\enquote {\bibinfo {title}
			{{Spin induced multipole moments for the gravitational wave amplitude from
					binary inspirals to 2.5 Post-Newtonian order}},}\ }\href {\doibase
		10.1088/1475-7516/2012/09/028} {\bibfield  {journal} {\bibinfo  {journal}
			{JCAP}\ }\textbf {\bibinfo {volume} {09}},\ \bibinfo {pages} {028} (\bibinfo
		{year} {2012})},\ \Eprint {http://arxiv.org/abs/1203.2962} {arXiv:1203.2962
		[gr-qc]} \BibitemShut {NoStop}%
	\bibitem [{\citenamefont {Levi}\ and\ \citenamefont
		{Steinhoff}(2014)}]{Levi:2014sba}%
	\BibitemOpen
	\bibfield  {author} {\bibinfo {author} {\bibfnamefont {Michele}\ \bibnamefont
			{Levi}}\ and\ \bibinfo {author} {\bibfnamefont {Jan}\ \bibnamefont
			{Steinhoff}},\ }\bibfield  {title} {\enquote {\bibinfo {title} {{Equivalence
					of ADM Hamiltonian and Effective Field Theory approaches at
					next-to-next-to-leading order spin1-spin2 coupling of binary inspirals}},}\
	}\href {\doibase 10.1088/1475-7516/2014/12/003} {\bibfield  {journal}
		{\bibinfo  {journal} {JCAP}\ }\textbf {\bibinfo {volume} {12}},\ \bibinfo
		{pages} {003} (\bibinfo {year} {2014})},\ \Eprint
	{http://arxiv.org/abs/1408.5762} {arXiv:1408.5762 [gr-qc]} \BibitemShut
	{NoStop}%
	\bibitem [{\citenamefont {Levi}\ and\ \citenamefont
		{Steinhoff}(2015{\natexlab{a}})}]{Levi:2014gsa}%
	\BibitemOpen
	\bibfield  {author} {\bibinfo {author} {\bibfnamefont {Michele}\ \bibnamefont
			{Levi}}\ and\ \bibinfo {author} {\bibfnamefont {Jan}\ \bibnamefont
			{Steinhoff}},\ }\bibfield  {title} {\enquote {\bibinfo {title} {{Leading
					order finite size effects with spins for inspiralling compact binaries}},}\
	}\href {\doibase 10.1007/JHEP06(2015)059} {\bibfield  {journal} {\bibinfo
			{journal} {JHEP}\ }\textbf {\bibinfo {volume} {06}},\ \bibinfo {pages} {059}
		(\bibinfo {year} {2015}{\natexlab{a}})},\ \Eprint
	{http://arxiv.org/abs/1410.2601} {arXiv:1410.2601 [gr-qc]} \BibitemShut
	{NoStop}%
	\bibitem [{\citenamefont {Levi}\ and\ \citenamefont
		{Steinhoff}(2015{\natexlab{b}})}]{Levi:2015msa}%
	\BibitemOpen
	\bibfield  {author} {\bibinfo {author} {\bibfnamefont {Michele}\ \bibnamefont
			{Levi}}\ and\ \bibinfo {author} {\bibfnamefont {Jan}\ \bibnamefont
			{Steinhoff}},\ }\bibfield  {title} {\enquote {\bibinfo {title} {{Spinning
					gravitating objects in the effective field theory in the post-Newtonian
					scheme}},}\ }\href {\doibase 10.1007/JHEP09(2015)219} {\bibfield  {journal}
		{\bibinfo  {journal} {JHEP}\ }\textbf {\bibinfo {volume} {09}},\ \bibinfo
		{pages} {219} (\bibinfo {year} {2015}{\natexlab{b}})},\ \Eprint
	{http://arxiv.org/abs/1501.04956} {arXiv:1501.04956 [gr-qc]} \BibitemShut
	{NoStop}%
	\bibitem [{\citenamefont {Levi}\ and\ \citenamefont
		{Steinhoff}(2016{\natexlab{a}})}]{Levi:2015uxa}%
	\BibitemOpen
	\bibfield  {author} {\bibinfo {author} {\bibfnamefont {Michele}\ \bibnamefont
			{Levi}}\ and\ \bibinfo {author} {\bibfnamefont {Jan}\ \bibnamefont
			{Steinhoff}},\ }\bibfield  {title} {\enquote {\bibinfo {title}
			{{Next-to-next-to-leading order gravitational spin-orbit coupling via the
					effective field theory for spinning objects in the post-Newtonian scheme}},}\
	}\href {\doibase 10.1088/1475-7516/2016/01/011} {\bibfield  {journal}
		{\bibinfo  {journal} {JCAP}\ }\textbf {\bibinfo {volume} {01}},\ \bibinfo
		{pages} {011} (\bibinfo {year} {2016}{\natexlab{a}})},\ \Eprint
	{http://arxiv.org/abs/1506.05056} {arXiv:1506.05056 [gr-qc]} \BibitemShut
	{NoStop}%
	\bibitem [{\citenamefont {Levi}\ and\ \citenamefont
		{Steinhoff}(2016{\natexlab{b}})}]{Levi:2015ixa}%
	\BibitemOpen
	\bibfield  {author} {\bibinfo {author} {\bibfnamefont {Michele}\ \bibnamefont
			{Levi}}\ and\ \bibinfo {author} {\bibfnamefont {Jan}\ \bibnamefont
			{Steinhoff}},\ }\bibfield  {title} {\enquote {\bibinfo {title}
			{{Next-to-next-to-leading order gravitational spin-squared potential via the
					effective field theory for spinning objects in the post-Newtonian scheme}},}\
	}\href {\doibase 10.1088/1475-7516/2016/01/008} {\bibfield  {journal}
		{\bibinfo  {journal} {JCAP}\ }\textbf {\bibinfo {volume} {01}},\ \bibinfo
		{pages} {008} (\bibinfo {year} {2016}{\natexlab{b}})},\ \Eprint
	{http://arxiv.org/abs/1506.05794} {arXiv:1506.05794 [gr-qc]} \BibitemShut
	{NoStop}%
	\bibitem [{\citenamefont {Levi}\ and\ \citenamefont
		{Steinhoff}(2021)}]{Levi:2016ofk}%
	\BibitemOpen
	\bibfield  {author} {\bibinfo {author} {\bibfnamefont {Michele}\ \bibnamefont
			{Levi}}\ and\ \bibinfo {author} {\bibfnamefont {Jan}\ \bibnamefont
			{Steinhoff}},\ }\bibfield  {title} {\enquote {\bibinfo {title} {{Complete
					conservative dynamics for inspiralling compact binaries with spins at the
					fourth post-Newtonian order}},}\ }\href {\doibase
		10.1088/1475-7516/2021/09/029} {\bibfield  {journal} {\bibinfo  {journal}
			{JCAP}\ }\textbf {\bibinfo {volume} {09}},\ \bibinfo {pages} {029} (\bibinfo
		{year} {2021})},\ \Eprint {http://arxiv.org/abs/1607.04252} {arXiv:1607.04252
		[gr-qc]} \BibitemShut {NoStop}%
	\bibitem [{\citenamefont {Maia}\ \emph
		{et~al.}(2017{\natexlab{a}})\citenamefont {Maia}, \citenamefont {Galley},
		\citenamefont {Leibovich},\ and\ \citenamefont {Porto}}]{Maia:2017gxn}%
	\BibitemOpen
	\bibfield  {author} {\bibinfo {author} {\bibfnamefont {Natalia~T.}\
			\bibnamefont {Maia}}, \bibinfo {author} {\bibfnamefont {Chad~R.}\
			\bibnamefont {Galley}}, \bibinfo {author} {\bibfnamefont {Adam~K.}\
			\bibnamefont {Leibovich}}, \ and\ \bibinfo {author} {\bibfnamefont
			{Rafael~A.}\ \bibnamefont {Porto}},\ }\bibfield  {title} {\enquote {\bibinfo
			{title} {{Radiation reaction for spinning bodies in effective field theory I:
					Spin-orbit effects}},}\ }\href {\doibase 10.1103/PhysRevD.96.084064}
	{\bibfield  {journal} {\bibinfo  {journal} {Phys. Rev. D}\ }\textbf {\bibinfo
			{volume} {96}},\ \bibinfo {pages} {084064} (\bibinfo {year}
		{2017}{\natexlab{a}})},\ \Eprint {http://arxiv.org/abs/1705.07934}
	{arXiv:1705.07934 [gr-qc]} \BibitemShut {NoStop}%
	\bibitem [{\citenamefont {Maia}\ \emph
		{et~al.}(2017{\natexlab{b}})\citenamefont {Maia}, \citenamefont {Galley},
		\citenamefont {Leibovich},\ and\ \citenamefont {Porto}}]{Maia:2017yok}%
	\BibitemOpen
	\bibfield  {author} {\bibinfo {author} {\bibfnamefont {Natalia~T.}\
			\bibnamefont {Maia}}, \bibinfo {author} {\bibfnamefont {Chad~R.}\
			\bibnamefont {Galley}}, \bibinfo {author} {\bibfnamefont {Adam~K.}\
			\bibnamefont {Leibovich}}, \ and\ \bibinfo {author} {\bibfnamefont
			{Rafael~A.}\ \bibnamefont {Porto}},\ }\bibfield  {title} {\enquote {\bibinfo
			{title} {{Radiation reaction for spinning bodies in effective field theory
					II: Spin-spin effects}},}\ }\href {\doibase 10.1103/PhysRevD.96.084065}
	{\bibfield  {journal} {\bibinfo  {journal} {Phys. Rev. D}\ }\textbf {\bibinfo
			{volume} {96}},\ \bibinfo {pages} {084065} (\bibinfo {year}
		{2017}{\natexlab{b}})},\ \Eprint {http://arxiv.org/abs/1705.07938}
	{arXiv:1705.07938 [gr-qc]} \BibitemShut {NoStop}%
	\bibitem [{\citenamefont {Levi}\ \emph
		{et~al.}(2021{\natexlab{a}})\citenamefont {Levi}, \citenamefont
		{Mougiakakos},\ and\ \citenamefont {Vieira}}]{Levi:2019kgk}%
	\BibitemOpen
	\bibfield  {author} {\bibinfo {author} {\bibfnamefont {Mich\`ele}\
			\bibnamefont {Levi}}, \bibinfo {author} {\bibfnamefont {Stavros}\
			\bibnamefont {Mougiakakos}}, \ and\ \bibinfo {author} {\bibfnamefont
			{Mariana}\ \bibnamefont {Vieira}},\ }\bibfield  {title} {\enquote {\bibinfo
			{title} {{Gravitational cubic-in-spin interaction at the next-to-leading
					post-Newtonian order}},}\ }\href {\doibase 10.1007/JHEP01(2021)036}
	{\bibfield  {journal} {\bibinfo  {journal} {JHEP}\ }\textbf {\bibinfo
			{volume} {01}},\ \bibinfo {pages} {036} (\bibinfo {year}
		{2021}{\natexlab{a}})},\ \Eprint {http://arxiv.org/abs/1912.06276}
	{arXiv:1912.06276 [hep-th]} \BibitemShut {NoStop}%
	\bibitem [{\citenamefont {Levi}\ \emph
		{et~al.}(2021{\natexlab{b}})\citenamefont {Levi}, \citenamefont {Mcleod},\
		and\ \citenamefont {Von~Hippel}}]{Levi:2020kvb}%
	\BibitemOpen
	\bibfield  {author} {\bibinfo {author} {\bibfnamefont {Mich\`ele}\
			\bibnamefont {Levi}}, \bibinfo {author} {\bibfnamefont {Andrew~J.}\
			\bibnamefont {Mcleod}}, \ and\ \bibinfo {author} {\bibfnamefont {Matthew}\
			\bibnamefont {Von~Hippel}},\ }\bibfield  {title} {\enquote {\bibinfo {title}
			{{N$^{3}$LO gravitational spin-orbit coupling at order G$^{4}$}},}\ }\href
	{\doibase 10.1007/JHEP07(2021)115} {\bibfield  {journal} {\bibinfo  {journal}
			{JHEP}\ }\textbf {\bibinfo {volume} {07}},\ \bibinfo {pages} {115} (\bibinfo
		{year} {2021}{\natexlab{b}})},\ \Eprint {http://arxiv.org/abs/2003.02827}
	{arXiv:2003.02827 [hep-th]} \BibitemShut {NoStop}%
	\bibitem [{\citenamefont {Levi}\ \emph
		{et~al.}(2021{\natexlab{c}})\citenamefont {Levi}, \citenamefont {Mcleod},\
		and\ \citenamefont {Von~Hippel}}]{Levi:2020uwu}%
	\BibitemOpen
	\bibfield  {author} {\bibinfo {author} {\bibfnamefont {Mich\`ele}\
			\bibnamefont {Levi}}, \bibinfo {author} {\bibfnamefont {Andrew~J.}\
			\bibnamefont {Mcleod}}, \ and\ \bibinfo {author} {\bibfnamefont {Matthew}\
			\bibnamefont {Von~Hippel}},\ }\bibfield  {title} {\enquote {\bibinfo {title}
			{{N$^{3}$LO gravitational quadratic-in-spin interactions at G$^{4}$}},}\
	}\href {\doibase 10.1007/JHEP07(2021)116} {\bibfield  {journal} {\bibinfo
			{journal} {JHEP}\ }\textbf {\bibinfo {volume} {07}},\ \bibinfo {pages} {116}
		(\bibinfo {year} {2021}{\natexlab{c}})},\ \Eprint
	{http://arxiv.org/abs/2003.07890} {arXiv:2003.07890 [hep-th]} \BibitemShut
	{NoStop}%
	\bibitem [{\citenamefont {Antonelli}\ \emph
		{et~al.}(2020{\natexlab{a}})\citenamefont {Antonelli}, \citenamefont
		{Kavanagh}, \citenamefont {Khalil}, \citenamefont {Steinhoff},\ and\
		\citenamefont {Vines}}]{Antonelli:2020aeb}%
	\BibitemOpen
	\bibfield  {author} {\bibinfo {author} {\bibfnamefont {Andrea}\ \bibnamefont
			{Antonelli}}, \bibinfo {author} {\bibfnamefont {Chris}\ \bibnamefont
			{Kavanagh}}, \bibinfo {author} {\bibfnamefont {Mohammed}\ \bibnamefont
			{Khalil}}, \bibinfo {author} {\bibfnamefont {Jan}\ \bibnamefont {Steinhoff}},
		\ and\ \bibinfo {author} {\bibfnamefont {Justin}\ \bibnamefont {Vines}},\
	}\bibfield  {title} {\enquote {\bibinfo {title} {{Gravitational spin-orbit
					coupling through third-subleading post-Newtonian order: from first-order
					self-force to arbitrary mass ratios}},}\ }\href {\doibase
		10.1103/PhysRevLett.125.011103} {\bibfield  {journal} {\bibinfo  {journal}
			{Phys. Rev. Lett.}\ }\textbf {\bibinfo {volume} {125}},\ \bibinfo {pages}
		{011103} (\bibinfo {year} {2020}{\natexlab{a}})},\ \Eprint
	{http://arxiv.org/abs/2003.11391} {arXiv:2003.11391 [gr-qc]} \BibitemShut
	{NoStop}%
	\bibitem [{\citenamefont {Levi}\ and\ \citenamefont
		{Teng}(2021)}]{Levi:2020lfn}%
	\BibitemOpen
	\bibfield  {author} {\bibinfo {author} {\bibfnamefont {Mich\`ele}\
			\bibnamefont {Levi}}\ and\ \bibinfo {author} {\bibfnamefont {Fei}\
			\bibnamefont {Teng}},\ }\bibfield  {title} {\enquote {\bibinfo {title} {{NLO
					gravitational quartic-in-spin interaction}},}\ }\href {\doibase
		10.1007/JHEP01(2021)066} {\bibfield  {journal} {\bibinfo  {journal} {JHEP}\
		}\textbf {\bibinfo {volume} {01}},\ \bibinfo {pages} {066} (\bibinfo {year}
		{2021})},\ \Eprint {http://arxiv.org/abs/2008.12280} {arXiv:2008.12280
		[hep-th]} \BibitemShut {NoStop}%
	\bibitem [{\citenamefont {Antonelli}\ \emph
		{et~al.}(2020{\natexlab{b}})\citenamefont {Antonelli}, \citenamefont
		{Kavanagh}, \citenamefont {Khalil}, \citenamefont {Steinhoff},\ and\
		\citenamefont {Vines}}]{Antonelli:2020ybz}%
	\BibitemOpen
	\bibfield  {author} {\bibinfo {author} {\bibfnamefont {Andrea}\ \bibnamefont
			{Antonelli}}, \bibinfo {author} {\bibfnamefont {Chris}\ \bibnamefont
			{Kavanagh}}, \bibinfo {author} {\bibfnamefont {Mohammed}\ \bibnamefont
			{Khalil}}, \bibinfo {author} {\bibfnamefont {Jan}\ \bibnamefont {Steinhoff}},
		\ and\ \bibinfo {author} {\bibfnamefont {Justin}\ \bibnamefont {Vines}},\
	}\bibfield  {title} {\enquote {\bibinfo {title} {{Gravitational spin-orbit
					and aligned spin$_1$-spin$_2$ couplings through third-subleading
					post-Newtonian orders}},}\ }\href {\doibase 10.1103/PhysRevD.102.124024}
	{\bibfield  {journal} {\bibinfo  {journal} {Phys. Rev. D}\ }\textbf {\bibinfo
			{volume} {102}},\ \bibinfo {pages} {124024} (\bibinfo {year}
		{2020}{\natexlab{b}})},\ \Eprint {http://arxiv.org/abs/2010.02018}
	{arXiv:2010.02018 [gr-qc]} \BibitemShut {NoStop}%
	\bibitem [{\citenamefont {Goldberger}\ \emph {et~al.}(2021)\citenamefont
		{Goldberger}, \citenamefont {Li},\ and\ \citenamefont
		{Rothstein}}]{Goldberger:2020fot}%
	\BibitemOpen
	\bibfield  {author} {\bibinfo {author} {\bibfnamefont {Walter~D.}\
			\bibnamefont {Goldberger}}, \bibinfo {author} {\bibfnamefont {Jingping}\
			\bibnamefont {Li}}, \ and\ \bibinfo {author} {\bibfnamefont {Ira~Z.}\
			\bibnamefont {Rothstein}},\ }\bibfield  {title} {\enquote {\bibinfo {title}
			{{Non-conservative effects on spinning black holes from world-line effective
					field theory}},}\ }\href {\doibase 10.1007/JHEP06(2021)053} {\bibfield
		{journal} {\bibinfo  {journal} {JHEP}\ }\textbf {\bibinfo {volume} {06}},\
		\bibinfo {pages} {053} (\bibinfo {year} {2021})},\ \Eprint
	{http://arxiv.org/abs/2012.14869} {arXiv:2012.14869 [hep-th]} \BibitemShut
	{NoStop}%
	\bibitem [{\citenamefont {Liu}\ \emph {et~al.}(2021)\citenamefont {Liu},
		\citenamefont {Porto},\ and\ \citenamefont {Yang}}]{Liu:2021zxr}%
	\BibitemOpen
	\bibfield  {author} {\bibinfo {author} {\bibfnamefont {Zhengwen}\
			\bibnamefont {Liu}}, \bibinfo {author} {\bibfnamefont {Rafael~A.}\
			\bibnamefont {Porto}}, \ and\ \bibinfo {author} {\bibfnamefont {Zixin}\
			\bibnamefont {Yang}},\ }\bibfield  {title} {\enquote {\bibinfo {title} {{Spin
					Effects in the Effective Field Theory Approach to Post-Minkowskian
					Conservative Dynamics}},}\ }\href {\doibase 10.1007/JHEP06(2021)012}
	{\bibfield  {journal} {\bibinfo  {journal} {JHEP}\ }\textbf {\bibinfo
			{volume} {06}},\ \bibinfo {pages} {012} (\bibinfo {year} {2021})},\ \Eprint
	{http://arxiv.org/abs/2102.10059} {arXiv:2102.10059 [hep-th]} \BibitemShut
	{NoStop}%
	\bibitem [{\citenamefont {Cho}\ \emph {et~al.}(2021)\citenamefont {Cho},
		\citenamefont {Pardo},\ and\ \citenamefont {Porto}}]{Cho:2021mqw}%
	\BibitemOpen
	\bibfield  {author} {\bibinfo {author} {\bibfnamefont {Gihyuk}\ \bibnamefont
			{Cho}}, \bibinfo {author} {\bibfnamefont {Brian}\ \bibnamefont {Pardo}}, \
		and\ \bibinfo {author} {\bibfnamefont {Rafael~A.}\ \bibnamefont {Porto}},\
	}\bibfield  {title} {\enquote {\bibinfo {title} {{Gravitational radiation
					from inspiralling compact objects: Spin-spin effects completed at the
					next-to-leading post-Newtonian order}},}\ }\href {\doibase
		10.1103/PhysRevD.104.024037} {\bibfield  {journal} {\bibinfo  {journal}
			{Phys. Rev. D}\ }\textbf {\bibinfo {volume} {104}},\ \bibinfo {pages}
		{024037} (\bibinfo {year} {2021})},\ \Eprint
	{http://arxiv.org/abs/2103.14612} {arXiv:2103.14612 [gr-qc]} \BibitemShut
	{NoStop}%
	\bibitem [{\citenamefont {Kim}\ \emph {et~al.}(2021)\citenamefont {Kim},
		\citenamefont {Levi},\ and\ \citenamefont {Yin}}]{Kim:2021rfj}%
	\BibitemOpen
	\bibfield  {author} {\bibinfo {author} {\bibfnamefont {Jung-Wook}\
			\bibnamefont {Kim}}, \bibinfo {author} {\bibfnamefont {Mich\`ele}\
			\bibnamefont {Levi}}, \ and\ \bibinfo {author} {\bibfnamefont {Zhewei}\
			\bibnamefont {Yin}},\ }\bibfield  {title} {\enquote {\bibinfo {title}
			{{Quadratic-in-spin interactions at fifth post-Newtonian order probe new
					physics}},}\ }\href@noop {} {\  (\bibinfo {year} {2021})},\ \Eprint
	{http://arxiv.org/abs/2112.01509} {arXiv:2112.01509 [hep-th]} \BibitemShut
	{NoStop}%
	\bibitem [{\citenamefont {Cho}\ \emph {et~al.}(2022{\natexlab{a}})\citenamefont
		{Cho}, \citenamefont {Porto},\ and\ \citenamefont {Yang}}]{Cho:2022syn}%
	\BibitemOpen
	\bibfield  {author} {\bibinfo {author} {\bibfnamefont {Gihyuk}\ \bibnamefont
			{Cho}}, \bibinfo {author} {\bibfnamefont {Rafael~A.}\ \bibnamefont {Porto}},
		\ and\ \bibinfo {author} {\bibfnamefont {Zixin}\ \bibnamefont {Yang}},\
	}\bibfield  {title} {\enquote {\bibinfo {title} {{Gravitational radiation
					from inspiralling compact objects: Spin effects to fourth Post-Newtonian
					order}},}\ }\href@noop {} {\  (\bibinfo {year} {2022}{\natexlab{a}})},\
	\Eprint {http://arxiv.org/abs/2201.05138} {arXiv:2201.05138 [gr-qc]}
	\BibitemShut {NoStop}%
	\bibitem [{\citenamefont {Bini}\ and\ \citenamefont
		{Damour}(2017)}]{Bini:2017xzy}%
	\BibitemOpen
	\bibfield  {author} {\bibinfo {author} {\bibfnamefont {Donato}\ \bibnamefont
			{Bini}}\ and\ \bibinfo {author} {\bibfnamefont {Thibault}\ \bibnamefont
			{Damour}},\ }\bibfield  {title} {\enquote {\bibinfo {title} {{Gravitational
					spin-orbit coupling in binary systems, post-Minkowskian approximation and
					effective one-body theory}},}\ }\href {\doibase 10.1103/PhysRevD.96.104038}
	{\bibfield  {journal} {\bibinfo  {journal} {Phys. Rev. D}\ }\textbf {\bibinfo
			{volume} {96}},\ \bibinfo {pages} {104038} (\bibinfo {year} {2017})},\
	\Eprint {http://arxiv.org/abs/1709.00590} {arXiv:1709.00590 [gr-qc]}
	\BibitemShut {NoStop}%
	\bibitem [{\citenamefont {Bini}\ and\ \citenamefont
		{Damour}(2018)}]{Bini:2018ywr}%
	\BibitemOpen
	\bibfield  {author} {\bibinfo {author} {\bibfnamefont {Donato}\ \bibnamefont
			{Bini}}\ and\ \bibinfo {author} {\bibfnamefont {Thibault}\ \bibnamefont
			{Damour}},\ }\bibfield  {title} {\enquote {\bibinfo {title} {{Gravitational
					spin-orbit coupling in binary systems at the second post-Minkowskian
					approximation}},}\ }\href {\doibase 10.1103/PhysRevD.98.044036} {\bibfield
		{journal} {\bibinfo  {journal} {Phys. Rev. D}\ }\textbf {\bibinfo {volume}
			{98}},\ \bibinfo {pages} {044036} (\bibinfo {year} {2018})},\ \Eprint
	{http://arxiv.org/abs/1805.10809} {arXiv:1805.10809 [gr-qc]} \BibitemShut
	{NoStop}%
	\bibitem [{\citenamefont {Vines}(2018)}]{Vines:2017hyw}%
	\BibitemOpen
	\bibfield  {author} {\bibinfo {author} {\bibfnamefont {Justin}\ \bibnamefont
			{Vines}},\ }\bibfield  {title} {\enquote {\bibinfo {title} {{Scattering of
					two spinning black holes in post-Minkowskian gravity, to all orders in spin,
					and effective-one-body mappings}},}\ }\href {\doibase
		10.1088/1361-6382/aaa3a8} {\bibfield  {journal} {\bibinfo  {journal} {Class.
				Quant. Grav.}\ }\textbf {\bibinfo {volume} {35}},\ \bibinfo {pages} {084002}
		(\bibinfo {year} {2018})},\ \Eprint {http://arxiv.org/abs/1709.06016}
	{arXiv:1709.06016 [gr-qc]} \BibitemShut {NoStop}%
	\bibitem [{\citenamefont {Vines}\ \emph {et~al.}(2019)\citenamefont {Vines},
		\citenamefont {Steinhoff},\ and\ \citenamefont {Buonanno}}]{Vines:2018gqi}%
	\BibitemOpen
	\bibfield  {author} {\bibinfo {author} {\bibfnamefont {Justin}\ \bibnamefont
			{Vines}}, \bibinfo {author} {\bibfnamefont {Jan}\ \bibnamefont {Steinhoff}},
		\ and\ \bibinfo {author} {\bibfnamefont {Alessandra}\ \bibnamefont
			{Buonanno}},\ }\bibfield  {title} {\enquote {\bibinfo {title}
			{{Spinning-black-hole scattering and the test-black-hole limit at second
					post-Minkowskian order}},}\ }\href {\doibase 10.1103/PhysRevD.99.064054}
	{\bibfield  {journal} {\bibinfo  {journal} {Phys. Rev. D}\ }\textbf {\bibinfo
			{volume} {99}},\ \bibinfo {pages} {064054} (\bibinfo {year} {2019})},\
	\Eprint {http://arxiv.org/abs/1812.00956} {arXiv:1812.00956 [gr-qc]}
	\BibitemShut {NoStop}%
	\bibitem [{\citenamefont {Guevara}(2019)}]{Guevara:2017csg}%
	\BibitemOpen
	\bibfield  {author} {\bibinfo {author} {\bibfnamefont {Alfredo}\ \bibnamefont
			{Guevara}},\ }\bibfield  {title} {\enquote {\bibinfo {title} {{Holomorphic
					Classical Limit for Spin Effects in Gravitational and Electromagnetic
					Scattering}},}\ }\href {\doibase 10.1007/JHEP04(2019)033} {\bibfield
		{journal} {\bibinfo  {journal} {JHEP}\ }\textbf {\bibinfo {volume} {04}},\
		\bibinfo {pages} {033} (\bibinfo {year} {2019})},\ \Eprint
	{http://arxiv.org/abs/1706.02314} {arXiv:1706.02314 [hep-th]} \BibitemShut
	{NoStop}%
	\bibitem [{\citenamefont {Guevara}\ \emph
		{et~al.}(2019{\natexlab{a}})\citenamefont {Guevara}, \citenamefont
		{Ochirov},\ and\ \citenamefont {Vines}}]{Guevara:2018wpp}%
	\BibitemOpen
	\bibfield  {author} {\bibinfo {author} {\bibfnamefont {Alfredo}\ \bibnamefont
			{Guevara}}, \bibinfo {author} {\bibfnamefont {Alexander}\ \bibnamefont
			{Ochirov}}, \ and\ \bibinfo {author} {\bibfnamefont {Justin}\ \bibnamefont
			{Vines}},\ }\bibfield  {title} {\enquote {\bibinfo {title} {{Scattering of
					Spinning Black Holes from Exponentiated Soft Factors}},}\ }\href {\doibase
		10.1007/JHEP09(2019)056} {\bibfield  {journal} {\bibinfo  {journal} {JHEP}\
		}\textbf {\bibinfo {volume} {09}},\ \bibinfo {pages} {056} (\bibinfo {year}
		{2019}{\natexlab{a}})},\ \Eprint {http://arxiv.org/abs/1812.06895}
	{arXiv:1812.06895 [hep-th]} \BibitemShut {NoStop}%
	\bibitem [{\citenamefont {Chung}\ \emph {et~al.}(2019)\citenamefont {Chung},
		\citenamefont {Huang}, \citenamefont {Kim},\ and\ \citenamefont
		{Lee}}]{Chung:2018kqs}%
	\BibitemOpen
	\bibfield  {author} {\bibinfo {author} {\bibfnamefont {Ming-Zhi}\
			\bibnamefont {Chung}}, \bibinfo {author} {\bibfnamefont {Yu-Tin}\
			\bibnamefont {Huang}}, \bibinfo {author} {\bibfnamefont {Jung-Wook}\
			\bibnamefont {Kim}}, \ and\ \bibinfo {author} {\bibfnamefont {Sangmin}\
			\bibnamefont {Lee}},\ }\bibfield  {title} {\enquote {\bibinfo {title} {{The
					simplest massive S-matrix: from minimal coupling to Black Holes}},}\ }\href
	{\doibase 10.1007/JHEP04(2019)156} {\bibfield  {journal} {\bibinfo  {journal}
			{JHEP}\ }\textbf {\bibinfo {volume} {04}},\ \bibinfo {pages} {156} (\bibinfo
		{year} {2019})},\ \Eprint {http://arxiv.org/abs/1812.08752} {arXiv:1812.08752
		[hep-th]} \BibitemShut {NoStop}%
	\bibitem [{\citenamefont {Arkani-Hamed}\ \emph {et~al.}(2020)\citenamefont
		{Arkani-Hamed}, \citenamefont {Huang},\ and\ \citenamefont
		{O'Connell}}]{Arkani-Hamed:2019ymq}%
	\BibitemOpen
	\bibfield  {author} {\bibinfo {author} {\bibfnamefont {Nima}\ \bibnamefont
			{Arkani-Hamed}}, \bibinfo {author} {\bibfnamefont {Yu-tin}\ \bibnamefont
			{Huang}}, \ and\ \bibinfo {author} {\bibfnamefont {Donal}\ \bibnamefont
			{O'Connell}},\ }\bibfield  {title} {\enquote {\bibinfo {title} {{Kerr black
					holes as elementary particles}},}\ }\href {\doibase 10.1007/JHEP01(2020)046}
	{\bibfield  {journal} {\bibinfo  {journal} {JHEP}\ }\textbf {\bibinfo
			{volume} {01}},\ \bibinfo {pages} {046} (\bibinfo {year} {2020})},\ \Eprint
	{http://arxiv.org/abs/1906.10100} {arXiv:1906.10100 [hep-th]} \BibitemShut
	{NoStop}%
	\bibitem [{\citenamefont {Guevara}\ \emph
		{et~al.}(2019{\natexlab{b}})\citenamefont {Guevara}, \citenamefont
		{Ochirov},\ and\ \citenamefont {Vines}}]{Guevara:2019fsj}%
	\BibitemOpen
	\bibfield  {author} {\bibinfo {author} {\bibfnamefont {Alfredo}\ \bibnamefont
			{Guevara}}, \bibinfo {author} {\bibfnamefont {Alexander}\ \bibnamefont
			{Ochirov}}, \ and\ \bibinfo {author} {\bibfnamefont {Justin}\ \bibnamefont
			{Vines}},\ }\bibfield  {title} {\enquote {\bibinfo {title} {{Black-hole
					scattering with general spin directions from minimal-coupling amplitudes}},}\
	}\href {\doibase 10.1103/PhysRevD.100.104024} {\bibfield  {journal} {\bibinfo
			{journal} {Phys. Rev. D}\ }\textbf {\bibinfo {volume} {100}},\ \bibinfo
		{pages} {104024} (\bibinfo {year} {2019}{\natexlab{b}})},\ \Eprint
	{http://arxiv.org/abs/1906.10071} {arXiv:1906.10071 [hep-th]} \BibitemShut
	{NoStop}%
	\bibitem [{\citenamefont {Chung}\ \emph
		{et~al.}(2020{\natexlab{a}})\citenamefont {Chung}, \citenamefont {Huang},\
		and\ \citenamefont {Kim}}]{Chung:2019duq}%
	\BibitemOpen
	\bibfield  {author} {\bibinfo {author} {\bibfnamefont {Ming-Zhi}\
			\bibnamefont {Chung}}, \bibinfo {author} {\bibfnamefont {Yu-Tin}\
			\bibnamefont {Huang}}, \ and\ \bibinfo {author} {\bibfnamefont {Jung-Wook}\
			\bibnamefont {Kim}},\ }\bibfield  {title} {\enquote {\bibinfo {title}
			{{Classical potential for general spinning bodies}},}\ }\href {\doibase
		10.1007/JHEP09(2020)074} {\bibfield  {journal} {\bibinfo  {journal} {JHEP}\
		}\textbf {\bibinfo {volume} {09}},\ \bibinfo {pages} {074} (\bibinfo {year}
		{2020}{\natexlab{a}})},\ \Eprint {http://arxiv.org/abs/1908.08463}
	{arXiv:1908.08463 [hep-th]} \BibitemShut {NoStop}%
	\bibitem [{\citenamefont {Damgaard}\ \emph {et~al.}(2019)\citenamefont
		{Damgaard}, \citenamefont {Haddad},\ and\ \citenamefont
		{Helset}}]{Damgaard:2019lfh}%
	\BibitemOpen
	\bibfield  {author} {\bibinfo {author} {\bibfnamefont {Poul~H.}\ \bibnamefont
			{Damgaard}}, \bibinfo {author} {\bibfnamefont {Kays}\ \bibnamefont {Haddad}},
		\ and\ \bibinfo {author} {\bibfnamefont {Andreas}\ \bibnamefont {Helset}},\
	}\bibfield  {title} {\enquote {\bibinfo {title} {{Heavy Black Hole Effective
					Theory}},}\ }\href {\doibase 10.1007/JHEP11(2019)070} {\bibfield  {journal}
		{\bibinfo  {journal} {JHEP}\ }\textbf {\bibinfo {volume} {11}},\ \bibinfo
		{pages} {070} (\bibinfo {year} {2019})},\ \Eprint
	{http://arxiv.org/abs/1908.10308} {arXiv:1908.10308 [hep-ph]} \BibitemShut
	{NoStop}%
	\bibitem [{\citenamefont {Aoude}\ \emph {et~al.}(2020)\citenamefont {Aoude},
		\citenamefont {Haddad},\ and\ \citenamefont {Helset}}]{Aoude:2020onz}%
	\BibitemOpen
	\bibfield  {author} {\bibinfo {author} {\bibfnamefont {Rafael}\ \bibnamefont
			{Aoude}}, \bibinfo {author} {\bibfnamefont {Kays}\ \bibnamefont {Haddad}}, \
		and\ \bibinfo {author} {\bibfnamefont {Andreas}\ \bibnamefont {Helset}},\
	}\bibfield  {title} {\enquote {\bibinfo {title} {{On-shell heavy particle
					effective theories}},}\ }\href {\doibase 10.1007/JHEP05(2020)051} {\bibfield
		{journal} {\bibinfo  {journal} {JHEP}\ }\textbf {\bibinfo {volume} {05}},\
		\bibinfo {pages} {051} (\bibinfo {year} {2020})},\ \Eprint
	{http://arxiv.org/abs/2001.09164} {arXiv:2001.09164 [hep-th]} \BibitemShut
	{NoStop}%
	\bibitem [{\citenamefont {Chung}\ \emph
		{et~al.}(2020{\natexlab{b}})\citenamefont {Chung}, \citenamefont {Huang},
		\citenamefont {Kim},\ and\ \citenamefont {Lee}}]{Chung:2020rrz}%
	\BibitemOpen
	\bibfield  {author} {\bibinfo {author} {\bibfnamefont {Ming-Zhi}\
			\bibnamefont {Chung}}, \bibinfo {author} {\bibfnamefont {Yu-tin}\
			\bibnamefont {Huang}}, \bibinfo {author} {\bibfnamefont {Jung-Wook}\
			\bibnamefont {Kim}}, \ and\ \bibinfo {author} {\bibfnamefont {Sangmin}\
			\bibnamefont {Lee}},\ }\bibfield  {title} {\enquote {\bibinfo {title}
			{{Complete Hamiltonian for spinning binary systems at first post-Minkowskian
					order}},}\ }\href {\doibase 10.1007/JHEP05(2020)105} {\bibfield  {journal}
		{\bibinfo  {journal} {JHEP}\ }\textbf {\bibinfo {volume} {05}},\ \bibinfo
		{pages} {105} (\bibinfo {year} {2020}{\natexlab{b}})},\ \Eprint
	{http://arxiv.org/abs/2003.06600} {arXiv:2003.06600 [hep-th]} \BibitemShut
	{NoStop}%
	\bibitem [{\citenamefont {Guevara}\ \emph {et~al.}(2021)\citenamefont
		{Guevara}, \citenamefont {Maybee}, \citenamefont {Ochirov}, \citenamefont
		{O'connell},\ and\ \citenamefont {Vines}}]{Guevara:2020xjx}%
	\BibitemOpen
	\bibfield  {author} {\bibinfo {author} {\bibfnamefont {Alfredo}\ \bibnamefont
			{Guevara}}, \bibinfo {author} {\bibfnamefont {Ben}\ \bibnamefont {Maybee}},
		\bibinfo {author} {\bibfnamefont {Alexander}\ \bibnamefont {Ochirov}},
		\bibinfo {author} {\bibfnamefont {Donal}\ \bibnamefont {O'connell}}, \ and\
		\bibinfo {author} {\bibfnamefont {Justin}\ \bibnamefont {Vines}},\ }\bibfield
	{title} {\enquote {\bibinfo {title} {{A worldsheet for Kerr}},}\ }\href
	{\doibase 10.1007/JHEP03(2021)201} {\bibfield  {journal} {\bibinfo  {journal}
			{JHEP}\ }\textbf {\bibinfo {volume} {03}},\ \bibinfo {pages} {201} (\bibinfo
		{year} {2021})},\ \Eprint {http://arxiv.org/abs/2012.11570} {arXiv:2012.11570
		[hep-th]} \BibitemShut {NoStop}%
	\bibitem [{\citenamefont {Bern}\ \emph
		{et~al.}(2021{\natexlab{b}})\citenamefont {Bern}, \citenamefont {Luna},
		\citenamefont {Roiban}, \citenamefont {Shen},\ and\ \citenamefont
		{Zeng}}]{Bern:2020buy}%
	\BibitemOpen
	\bibfield  {author} {\bibinfo {author} {\bibfnamefont {Zvi}\ \bibnamefont
			{Bern}}, \bibinfo {author} {\bibfnamefont {Andres}\ \bibnamefont {Luna}},
		\bibinfo {author} {\bibfnamefont {Radu}\ \bibnamefont {Roiban}}, \bibinfo
		{author} {\bibfnamefont {Chia-Hsien}\ \bibnamefont {Shen}}, \ and\ \bibinfo
		{author} {\bibfnamefont {Mao}\ \bibnamefont {Zeng}},\ }\bibfield  {title}
	{\enquote {\bibinfo {title} {{Spinning black hole binary dynamics, scattering
					amplitudes, and effective field theory}},}\ }\href {\doibase
		10.1103/PhysRevD.104.065014} {\bibfield  {journal} {\bibinfo  {journal}
			{Phys. Rev. D}\ }\textbf {\bibinfo {volume} {104}},\ \bibinfo {pages}
		{065014} (\bibinfo {year} {2021}{\natexlab{b}})},\ \Eprint
	{http://arxiv.org/abs/2005.03071} {arXiv:2005.03071 [hep-th]} \BibitemShut
	{NoStop}%
	\bibitem [{\citenamefont {Kosmopoulos}\ and\ \citenamefont
		{Luna}(2021)}]{Kosmopoulos:2021zoq}%
	\BibitemOpen
	\bibfield  {author} {\bibinfo {author} {\bibfnamefont {Dimitrios}\
			\bibnamefont {Kosmopoulos}}\ and\ \bibinfo {author} {\bibfnamefont {Andres}\
			\bibnamefont {Luna}},\ }\bibfield  {title} {\enquote {\bibinfo {title}
			{{Quadratic-in-spin Hamiltonian at $ \mathcal{O} $(G$^{2}$) from scattering
					amplitudes}},}\ }\href {\doibase 10.1007/JHEP07(2021)037} {\bibfield
		{journal} {\bibinfo  {journal} {JHEP}\ }\textbf {\bibinfo {volume} {07}},\
		\bibinfo {pages} {037} (\bibinfo {year} {2021})},\ \Eprint
	{http://arxiv.org/abs/2102.10137} {arXiv:2102.10137 [hep-th]} \BibitemShut
	{NoStop}%
	\bibitem [{\citenamefont {Chen}\ \emph {et~al.}(2021)\citenamefont {Chen},
		\citenamefont {Chung}, \citenamefont {Huang},\ and\ \citenamefont
		{Kim}}]{Chen:2021qkk}%
	\BibitemOpen
	\bibfield  {author} {\bibinfo {author} {\bibfnamefont {Wei-Ming}\
			\bibnamefont {Chen}}, \bibinfo {author} {\bibfnamefont {Ming-Zhi}\
			\bibnamefont {Chung}}, \bibinfo {author} {\bibfnamefont {Yu-tin}\
			\bibnamefont {Huang}}, \ and\ \bibinfo {author} {\bibfnamefont {Jung-Wook}\
			\bibnamefont {Kim}},\ }\bibfield  {title} {\enquote {\bibinfo {title} {{The
					2PM Hamiltonian for binary Kerr to quartic in spin}},}\ }\href@noop {} {\
		(\bibinfo {year} {2021})},\ \Eprint {http://arxiv.org/abs/2111.13639}
	{arXiv:2111.13639 [hep-th]} \BibitemShut {NoStop}%
	\bibitem [{\citenamefont {Bern}\ \emph
		{et~al.}(2022{\natexlab{b}})\citenamefont {Bern}, \citenamefont
		{Kosmopoulos}, \citenamefont {Luna}, \citenamefont {Roiban},\ and\
		\citenamefont {Teng}}]{Bern:2022kto}%
	\BibitemOpen
	\bibfield  {author} {\bibinfo {author} {\bibfnamefont {Zvi}\ \bibnamefont
			{Bern}}, \bibinfo {author} {\bibfnamefont {Dimitrios}\ \bibnamefont
			{Kosmopoulos}}, \bibinfo {author} {\bibfnamefont {Andr\'es}\ \bibnamefont
			{Luna}}, \bibinfo {author} {\bibfnamefont {Radu}\ \bibnamefont {Roiban}}, \
		and\ \bibinfo {author} {\bibfnamefont {Fei}\ \bibnamefont {Teng}},\
	}\bibfield  {title} {\enquote {\bibinfo {title} {{Binary Dynamics Through the
					Fifth Power of Spin at $\mathcal{O}(G^2)$}},}\ }\href@noop {} {\  (\bibinfo
		{year} {2022}{\natexlab{b}})},\ \Eprint {http://arxiv.org/abs/2203.06202}
	{arXiv:2203.06202 [hep-th]} \BibitemShut {NoStop}%
	\bibitem [{\citenamefont {Aoude}\ \emph
		{et~al.}(2022{\natexlab{a}})\citenamefont {Aoude}, \citenamefont {Haddad},\
		and\ \citenamefont {Helset}}]{Aoude:2022trd}%
	\BibitemOpen
	\bibfield  {author} {\bibinfo {author} {\bibfnamefont {Rafael}\ \bibnamefont
			{Aoude}}, \bibinfo {author} {\bibfnamefont {Kays}\ \bibnamefont {Haddad}}, \
		and\ \bibinfo {author} {\bibfnamefont {Andreas}\ \bibnamefont {Helset}},\
	}\bibfield  {title} {\enquote {\bibinfo {title} {{Searching for Kerr in the
					2PM amplitude}},}\ }\href@noop {} {\  (\bibinfo {year}
		{2022}{\natexlab{a}})},\ \Eprint {http://arxiv.org/abs/2203.06197}
	{arXiv:2203.06197 [hep-th]} \BibitemShut {NoStop}%
	\bibitem [{\citenamefont {Aoude}\ \emph
		{et~al.}(2022{\natexlab{b}})\citenamefont {Aoude}, \citenamefont {Haddad},\
		and\ \citenamefont {Helset}}]{Aoude:2022thd}%
	\BibitemOpen
	\bibfield  {author} {\bibinfo {author} {\bibfnamefont {Rafael}\ \bibnamefont
			{Aoude}}, \bibinfo {author} {\bibfnamefont {Kays}\ \bibnamefont {Haddad}}, \
		and\ \bibinfo {author} {\bibfnamefont {Andreas}\ \bibnamefont {Helset}},\
	}\bibfield  {title} {\enquote {\bibinfo {title} {{Classical gravitational
					spinning-spinless scattering at $\mathcal{O}(G^{2} S^{\infty})$}},}\
	}\href@noop {} {\  (\bibinfo {year} {2022}{\natexlab{b}})},\ \Eprint
	{http://arxiv.org/abs/2205.02809} {arXiv:2205.02809 [hep-th]} \BibitemShut
	{NoStop}%
	\bibitem [{\citenamefont {Mogull}\ \emph {et~al.}(2021)\citenamefont {Mogull},
		\citenamefont {Plefka},\ and\ \citenamefont {Steinhoff}}]{Mogull:2020sak}%
	\BibitemOpen
	\bibfield  {author} {\bibinfo {author} {\bibfnamefont {Gustav}\ \bibnamefont
			{Mogull}}, \bibinfo {author} {\bibfnamefont {Jan}\ \bibnamefont {Plefka}}, \
		and\ \bibinfo {author} {\bibfnamefont {Jan}\ \bibnamefont {Steinhoff}},\
	}\bibfield  {title} {\enquote {\bibinfo {title} {{Classical black hole
					scattering from a worldline quantum field theory}},}\ }\href {\doibase
		10.1007/JHEP02(2021)048} {\bibfield  {journal} {\bibinfo  {journal} {JHEP}\
		}\textbf {\bibinfo {volume} {02}},\ \bibinfo {pages} {048} (\bibinfo {year}
		{2021})},\ \Eprint {http://arxiv.org/abs/2010.02865} {arXiv:2010.02865
		[hep-th]} \BibitemShut {NoStop}%
	\bibitem [{\citenamefont {Jakobsen}\ \emph {et~al.}(2021)\citenamefont
		{Jakobsen}, \citenamefont {Mogull}, \citenamefont {Plefka},\ and\
		\citenamefont {Steinhoff}}]{Jakobsen:2021smu}%
	\BibitemOpen
	\bibfield  {author} {\bibinfo {author} {\bibfnamefont {Gustav~Uhre}\
			\bibnamefont {Jakobsen}}, \bibinfo {author} {\bibfnamefont {Gustav}\
			\bibnamefont {Mogull}}, \bibinfo {author} {\bibfnamefont {Jan}\ \bibnamefont
			{Plefka}}, \ and\ \bibinfo {author} {\bibfnamefont {Jan}\ \bibnamefont
			{Steinhoff}},\ }\bibfield  {title} {\enquote {\bibinfo {title} {{Classical
					Gravitational Bremsstrahlung from a Worldline Quantum Field Theory}},}\
	}\href {\doibase 10.1103/PhysRevLett.126.201103} {\bibfield  {journal}
		{\bibinfo  {journal} {Phys. Rev. Lett.}\ }\textbf {\bibinfo {volume} {126}},\
		\bibinfo {pages} {201103} (\bibinfo {year} {2021})},\ \Eprint
	{http://arxiv.org/abs/2101.12688} {arXiv:2101.12688 [gr-qc]} \BibitemShut
	{NoStop}%
	\bibitem [{\citenamefont {Jakobsen}\ \emph
		{et~al.}(2022{\natexlab{a}})\citenamefont {Jakobsen}, \citenamefont {Mogull},
		\citenamefont {Plefka},\ and\ \citenamefont {Steinhoff}}]{Jakobsen:2021lvp}%
	\BibitemOpen
	\bibfield  {author} {\bibinfo {author} {\bibfnamefont {Gustav~Uhre}\
			\bibnamefont {Jakobsen}}, \bibinfo {author} {\bibfnamefont {Gustav}\
			\bibnamefont {Mogull}}, \bibinfo {author} {\bibfnamefont {Jan}\ \bibnamefont
			{Plefka}}, \ and\ \bibinfo {author} {\bibfnamefont {Jan}\ \bibnamefont
			{Steinhoff}},\ }\bibfield  {title} {\enquote {\bibinfo {title}
			{{Gravitational Bremsstrahlung and Hidden Supersymmetry of Spinning
					Bodies}},}\ }\href {\doibase 10.1103/PhysRevLett.128.011101} {\bibfield
		{journal} {\bibinfo  {journal} {Phys. Rev. Lett.}\ }\textbf {\bibinfo
			{volume} {128}},\ \bibinfo {pages} {011101} (\bibinfo {year}
		{2022}{\natexlab{a}})},\ \Eprint {http://arxiv.org/abs/2106.10256}
	{arXiv:2106.10256 [hep-th]} \BibitemShut {NoStop}%
	\bibitem [{\citenamefont {Jakobsen}\ \emph
		{et~al.}(2022{\natexlab{b}})\citenamefont {Jakobsen}, \citenamefont {Mogull},
		\citenamefont {Plefka},\ and\ \citenamefont {Steinhoff}}]{Jakobsen:2021zvh}%
	\BibitemOpen
	\bibfield  {author} {\bibinfo {author} {\bibfnamefont {Gustav~Uhre}\
			\bibnamefont {Jakobsen}}, \bibinfo {author} {\bibfnamefont {Gustav}\
			\bibnamefont {Mogull}}, \bibinfo {author} {\bibfnamefont {Jan}\ \bibnamefont
			{Plefka}}, \ and\ \bibinfo {author} {\bibfnamefont {Jan}\ \bibnamefont
			{Steinhoff}},\ }\bibfield  {title} {\enquote {\bibinfo {title} {{SUSY in the
					sky with gravitons}},}\ }\href {\doibase 10.1007/JHEP01(2022)027} {\bibfield
		{journal} {\bibinfo  {journal} {JHEP}\ }\textbf {\bibinfo {volume} {01}},\
		\bibinfo {pages} {027} (\bibinfo {year} {2022}{\natexlab{b}})},\ \Eprint
	{http://arxiv.org/abs/2109.04465} {arXiv:2109.04465 [hep-th]} \BibitemShut
	{NoStop}%
	\bibitem [{\citenamefont {Jakobsen}\ and\ \citenamefont
		{Mogull}(2022)}]{Jakobsen:2022fcj}%
	\BibitemOpen
	\bibfield  {author} {\bibinfo {author} {\bibfnamefont {Gustav~Uhre}\
			\bibnamefont {Jakobsen}}\ and\ \bibinfo {author} {\bibfnamefont {Gustav}\
			\bibnamefont {Mogull}},\ }\bibfield  {title} {\enquote {\bibinfo {title}
			{{Conservative and Radiative Dynamics of Spinning Bodies at Third
					Post-Minkowskian Order Using Worldline Quantum Field Theory}},}\ }\href
	{\doibase 10.1103/PhysRevLett.128.141102} {\bibfield  {journal} {\bibinfo
			{journal} {Phys. Rev. Lett.}\ }\textbf {\bibinfo {volume} {128}},\ \bibinfo
		{pages} {141102} (\bibinfo {year} {2022})},\ \Eprint
	{http://arxiv.org/abs/2201.07778} {arXiv:2201.07778 [hep-th]} \BibitemShut
	{NoStop}%
	\bibitem [{\citenamefont {Alessio}\ and\ \citenamefont
		{Di~Vecchia}(2022)}]{Alessio:2022kwv}%
	\BibitemOpen
	\bibfield  {author} {\bibinfo {author} {\bibfnamefont {Francesco}\
			\bibnamefont {Alessio}}\ and\ \bibinfo {author} {\bibfnamefont {Paolo}\
			\bibnamefont {Di~Vecchia}},\ }\bibfield  {title} {\enquote {\bibinfo {title}
			{{Radiation reaction for spinning black-hole scattering}},}\ }\href@noop {}
	{\  (\bibinfo {year} {2022})},\ \Eprint {http://arxiv.org/abs/2203.13272}
	{arXiv:2203.13272 [hep-th]} \BibitemShut {NoStop}%
	\bibitem [{\citenamefont {Cristofoli}\ \emph {et~al.}(2021)\citenamefont
		{Cristofoli}, \citenamefont {Gonzo}, \citenamefont {Moynihan}, \citenamefont
		{O'Connell}, \citenamefont {Ross}, \citenamefont {Sergola},\ and\
		\citenamefont {White}}]{Cristofoli:2021jas}%
	\BibitemOpen
	\bibfield  {author} {\bibinfo {author} {\bibfnamefont {Andrea}\ \bibnamefont
			{Cristofoli}}, \bibinfo {author} {\bibfnamefont {Riccardo}\ \bibnamefont
			{Gonzo}}, \bibinfo {author} {\bibfnamefont {Nathan}\ \bibnamefont
			{Moynihan}}, \bibinfo {author} {\bibfnamefont {Donal}\ \bibnamefont
			{O'Connell}}, \bibinfo {author} {\bibfnamefont {Alasdair}\ \bibnamefont
			{Ross}}, \bibinfo {author} {\bibfnamefont {Matteo}\ \bibnamefont {Sergola}},
		\ and\ \bibinfo {author} {\bibfnamefont {Chris~D.}\ \bibnamefont {White}},\
	}\bibfield  {title} {\enquote {\bibinfo {title} {{The Uncertainty Principle
					and Classical Amplitudes}},}\ }\href@noop {} {\  (\bibinfo {year} {2021})},\
	\Eprint {http://arxiv.org/abs/2112.07556} {arXiv:2112.07556 [hep-th]}
	\BibitemShut {NoStop}%
	\bibitem [{\citenamefont {Damour}\ and\ \citenamefont
		{Esposito-Farese}(1996)}]{Damour:1995kt}%
	\BibitemOpen
	\bibfield  {author} {\bibinfo {author} {\bibfnamefont {Thibault}\
			\bibnamefont {Damour}}\ and\ \bibinfo {author} {\bibfnamefont {Gilles}\
			\bibnamefont {Esposito-Farese}},\ }\bibfield  {title} {\enquote {\bibinfo
			{title} {{Testing gravity to second postNewtonian order: A Field theory
					approach}},}\ }\href {\doibase 10.1103/PhysRevD.53.5541} {\bibfield
		{journal} {\bibinfo  {journal} {Phys. Rev. D}\ }\textbf {\bibinfo {volume}
			{53}},\ \bibinfo {pages} {5541--5578} (\bibinfo {year} {1996})},\ \Eprint
	{http://arxiv.org/abs/gr-qc/9506063} {arXiv:gr-qc/9506063} \BibitemShut
	{NoStop}%
	\bibitem [{\citenamefont {Vaidya}(2015)}]{Vaidya:2014kza}%
	\BibitemOpen
	\bibfield  {author} {\bibinfo {author} {\bibfnamefont {Varun}\ \bibnamefont
			{Vaidya}},\ }\bibfield  {title} {\enquote {\bibinfo {title} {{Gravitational
					spin Hamiltonians from the S matrix}},}\ }\href {\doibase
		10.1103/PhysRevD.91.024017} {\bibfield  {journal} {\bibinfo  {journal} {Phys.
				Rev. D}\ }\textbf {\bibinfo {volume} {91}},\ \bibinfo {pages} {024017}
		(\bibinfo {year} {2015})},\ \Eprint {http://arxiv.org/abs/1410.5348}
	{arXiv:1410.5348 [hep-th]} \BibitemShut {NoStop}%
	\bibitem [{\citenamefont {Bern}\ \emph {et~al.}(1994)\citenamefont {Bern},
		\citenamefont {Dixon}, \citenamefont {Dunbar},\ and\ \citenamefont
		{Kosower}}]{Bern:1994zx}%
	\BibitemOpen
	\bibfield  {author} {\bibinfo {author} {\bibfnamefont {Zvi}\ \bibnamefont
			{Bern}}, \bibinfo {author} {\bibfnamefont {Lance~J.}\ \bibnamefont {Dixon}},
		\bibinfo {author} {\bibfnamefont {David~C.}\ \bibnamefont {Dunbar}}, \ and\
		\bibinfo {author} {\bibfnamefont {David~A.}\ \bibnamefont {Kosower}},\
	}\bibfield  {title} {\enquote {\bibinfo {title} {{One loop n point gauge
					theory amplitudes, unitarity and collinear limits}},}\ }\href {\doibase
		10.1016/0550-3213(94)90179-1} {\bibfield  {journal} {\bibinfo  {journal}
			{Nucl. Phys. B}\ }\textbf {\bibinfo {volume} {425}},\ \bibinfo {pages}
		{217--260} (\bibinfo {year} {1994})},\ \Eprint
	{http://arxiv.org/abs/hep-ph/9403226} {arXiv:hep-ph/9403226} \BibitemShut
	{NoStop}%
	\bibitem [{\citenamefont {Bern}\ \emph {et~al.}(1995)\citenamefont {Bern},
		\citenamefont {Dixon}, \citenamefont {Dunbar},\ and\ \citenamefont
		{Kosower}}]{Bern:1994cg}%
	\BibitemOpen
	\bibfield  {author} {\bibinfo {author} {\bibfnamefont {Zvi}\ \bibnamefont
			{Bern}}, \bibinfo {author} {\bibfnamefont {Lance~J.}\ \bibnamefont {Dixon}},
		\bibinfo {author} {\bibfnamefont {David~C.}\ \bibnamefont {Dunbar}}, \ and\
		\bibinfo {author} {\bibfnamefont {David~A.}\ \bibnamefont {Kosower}},\
	}\bibfield  {title} {\enquote {\bibinfo {title} {{Fusing gauge theory tree
					amplitudes into loop amplitudes}},}\ }\href {\doibase
		10.1016/0550-3213(94)00488-Z} {\bibfield  {journal} {\bibinfo  {journal}
			{Nucl. Phys. B}\ }\textbf {\bibinfo {volume} {435}},\ \bibinfo {pages}
		{59--101} (\bibinfo {year} {1995})},\ \Eprint
	{http://arxiv.org/abs/hep-ph/9409265} {arXiv:hep-ph/9409265} \BibitemShut
	{NoStop}%
	\bibitem [{\citenamefont {Bern}\ and\ \citenamefont
		{Morgan}(1996)}]{Bern:1995db}%
	\BibitemOpen
	\bibfield  {author} {\bibinfo {author} {\bibfnamefont {Z.}~\bibnamefont
			{Bern}}\ and\ \bibinfo {author} {\bibfnamefont {A.~G.}\ \bibnamefont
			{Morgan}},\ }\bibfield  {title} {\enquote {\bibinfo {title} {{Massive loop
					amplitudes from unitarity}},}\ }\href {\doibase 10.1016/0550-3213(96)00078-8}
	{\bibfield  {journal} {\bibinfo  {journal} {Nucl. Phys. B}\ }\textbf
		{\bibinfo {volume} {467}},\ \bibinfo {pages} {479--509} (\bibinfo {year}
		{1996})},\ \Eprint {http://arxiv.org/abs/hep-ph/9511336}
	{arXiv:hep-ph/9511336} \BibitemShut {NoStop}%
	\bibitem [{\citenamefont {Bern}\ \emph {et~al.}(1998)\citenamefont {Bern},
		\citenamefont {Dixon},\ and\ \citenamefont {Kosower}}]{Bern:1997sc}%
	\BibitemOpen
	\bibfield  {author} {\bibinfo {author} {\bibfnamefont {Zvi}\ \bibnamefont
			{Bern}}, \bibinfo {author} {\bibfnamefont {Lance~J.}\ \bibnamefont {Dixon}},
		\ and\ \bibinfo {author} {\bibfnamefont {David~A.}\ \bibnamefont {Kosower}},\
	}\bibfield  {title} {\enquote {\bibinfo {title} {{One loop amplitudes for e+
					e- to four partons}},}\ }\href {\doibase 10.1016/S0550-3213(97)00703-7}
	{\bibfield  {journal} {\bibinfo  {journal} {Nucl. Phys. B}\ }\textbf
		{\bibinfo {volume} {513}},\ \bibinfo {pages} {3--86} (\bibinfo {year}
		{1998})},\ \Eprint {http://arxiv.org/abs/hep-ph/9708239}
	{arXiv:hep-ph/9708239} \BibitemShut {NoStop}%
	\bibitem [{\citenamefont {Britto}\ \emph {et~al.}(2005)\citenamefont {Britto},
		\citenamefont {Cachazo},\ and\ \citenamefont {Feng}}]{Britto:2004nc}%
	\BibitemOpen
	\bibfield  {author} {\bibinfo {author} {\bibfnamefont {Ruth}\ \bibnamefont
			{Britto}}, \bibinfo {author} {\bibfnamefont {Freddy}\ \bibnamefont
			{Cachazo}}, \ and\ \bibinfo {author} {\bibfnamefont {Bo}~\bibnamefont
			{Feng}},\ }\bibfield  {title} {\enquote {\bibinfo {title} {{Generalized
					unitarity and one-loop amplitudes in N=4 super-Yang-Mills}},}\ }\href
	{\doibase 10.1016/j.nuclphysb.2005.07.014} {\bibfield  {journal} {\bibinfo
			{journal} {Nucl. Phys. B}\ }\textbf {\bibinfo {volume} {725}},\ \bibinfo
		{pages} {275--305} (\bibinfo {year} {2005})},\ \Eprint
	{http://arxiv.org/abs/hep-th/0412103} {arXiv:hep-th/0412103} \BibitemShut
	{NoStop}%
	\bibitem [{\citenamefont {Ita}(2016)}]{Ita:2015tya}%
	\BibitemOpen
	\bibfield  {author} {\bibinfo {author} {\bibfnamefont {Harald}\ \bibnamefont
			{Ita}},\ }\bibfield  {title} {\enquote {\bibinfo {title} {{Two-loop Integrand
					Decomposition into Master Integrals and Surface Terms}},}\ }\href {\doibase
		10.1103/PhysRevD.94.116015} {\bibfield  {journal} {\bibinfo  {journal} {Phys.
				Rev. D}\ }\textbf {\bibinfo {volume} {94}},\ \bibinfo {pages} {116015}
		(\bibinfo {year} {2016})},\ \Eprint {http://arxiv.org/abs/1510.05626}
	{arXiv:1510.05626 [hep-th]} \BibitemShut {NoStop}%
	\bibitem [{\citenamefont {Abreu}\ \emph
		{et~al.}(2017{\natexlab{a}})\citenamefont {Abreu}, \citenamefont
		{Febres~Cordero}, \citenamefont {Ita}, \citenamefont {Jaquier},\ and\
		\citenamefont {Page}}]{Abreu:2017idw}%
	\BibitemOpen
	\bibfield  {author} {\bibinfo {author} {\bibfnamefont {S.}~\bibnamefont
			{Abreu}}, \bibinfo {author} {\bibfnamefont {F.}~\bibnamefont
			{Febres~Cordero}}, \bibinfo {author} {\bibfnamefont {H.}~\bibnamefont {Ita}},
		\bibinfo {author} {\bibfnamefont {M.}~\bibnamefont {Jaquier}}, \ and\
		\bibinfo {author} {\bibfnamefont {B.}~\bibnamefont {Page}},\ }\bibfield
	{title} {\enquote {\bibinfo {title} {{Subleading Poles in the Numerical
					Unitarity Method at Two Loops}},}\ }\href {\doibase
		10.1103/PhysRevD.95.096011} {\bibfield  {journal} {\bibinfo  {journal} {Phys.
				Rev. D}\ }\textbf {\bibinfo {volume} {95}},\ \bibinfo {pages} {096011}
		(\bibinfo {year} {2017}{\natexlab{a}})},\ \Eprint
	{http://arxiv.org/abs/1703.05255} {arXiv:1703.05255 [hep-ph]} \BibitemShut
	{NoStop}%
	\bibitem [{\citenamefont {Abreu}\ \emph
		{et~al.}(2017{\natexlab{b}})\citenamefont {Abreu}, \citenamefont
		{Febres~Cordero}, \citenamefont {Ita}, \citenamefont {Jaquier}, \citenamefont
		{Page},\ and\ \citenamefont {Zeng}}]{Abreu:2017xsl}%
	\BibitemOpen
	\bibfield  {author} {\bibinfo {author} {\bibfnamefont {S.}~\bibnamefont
			{Abreu}}, \bibinfo {author} {\bibfnamefont {F.}~\bibnamefont
			{Febres~Cordero}}, \bibinfo {author} {\bibfnamefont {H.}~\bibnamefont {Ita}},
		\bibinfo {author} {\bibfnamefont {M.}~\bibnamefont {Jaquier}}, \bibinfo
		{author} {\bibfnamefont {B.}~\bibnamefont {Page}}, \ and\ \bibinfo {author}
		{\bibfnamefont {M.}~\bibnamefont {Zeng}},\ }\bibfield  {title} {\enquote
		{\bibinfo {title} {{Two-Loop Four-Gluon Amplitudes from Numerical
					Unitarity}},}\ }\href {\doibase 10.1103/PhysRevLett.119.142001} {\bibfield
		{journal} {\bibinfo  {journal} {Phys. Rev. Lett.}\ }\textbf {\bibinfo
			{volume} {119}},\ \bibinfo {pages} {142001} (\bibinfo {year}
		{2017}{\natexlab{b}})},\ \Eprint {http://arxiv.org/abs/1703.05273}
	{arXiv:1703.05273 [hep-ph]} \BibitemShut {NoStop}%
	\bibitem [{\citenamefont {Abreu}\ \emph
		{et~al.}(2018{\natexlab{a}})\citenamefont {Abreu}, \citenamefont
		{Febres~Cordero}, \citenamefont {Ita}, \citenamefont {Page},\ and\
		\citenamefont {Zeng}}]{Abreu:2017hqn}%
	\BibitemOpen
	\bibfield  {author} {\bibinfo {author} {\bibfnamefont {Samuel}\ \bibnamefont
			{Abreu}}, \bibinfo {author} {\bibfnamefont {Fernando}\ \bibnamefont
			{Febres~Cordero}}, \bibinfo {author} {\bibfnamefont {Harald}\ \bibnamefont
			{Ita}}, \bibinfo {author} {\bibfnamefont {Ben}\ \bibnamefont {Page}}, \ and\
		\bibinfo {author} {\bibfnamefont {Mao}\ \bibnamefont {Zeng}},\ }\bibfield
	{title} {\enquote {\bibinfo {title} {{Planar Two-Loop Five-Gluon Amplitudes
					from Numerical Unitarity}},}\ }\href {\doibase 10.1103/PhysRevD.97.116014}
	{\bibfield  {journal} {\bibinfo  {journal} {Phys. Rev. D}\ }\textbf {\bibinfo
			{volume} {97}},\ \bibinfo {pages} {116014} (\bibinfo {year}
		{2018}{\natexlab{a}})},\ \Eprint {http://arxiv.org/abs/1712.03946}
	{arXiv:1712.03946 [hep-ph]} \BibitemShut {NoStop}%
	\bibitem [{\citenamefont {Abreu}\ \emph
		{et~al.}(2018{\natexlab{b}})\citenamefont {Abreu}, \citenamefont
		{Febres~Cordero}, \citenamefont {Ita}, \citenamefont {Page},\ and\
		\citenamefont {Sotnikov}}]{Abreu:2018jgq}%
	\BibitemOpen
	\bibfield  {author} {\bibinfo {author} {\bibfnamefont {S.}~\bibnamefont
			{Abreu}}, \bibinfo {author} {\bibfnamefont {F.}~\bibnamefont
			{Febres~Cordero}}, \bibinfo {author} {\bibfnamefont {H.}~\bibnamefont {Ita}},
		\bibinfo {author} {\bibfnamefont {B.}~\bibnamefont {Page}}, \ and\ \bibinfo
		{author} {\bibfnamefont {V.}~\bibnamefont {Sotnikov}},\ }\bibfield  {title}
	{\enquote {\bibinfo {title} {{Planar Two-Loop Five-Parton Amplitudes from
					Numerical Unitarity}},}\ }\href {\doibase 10.1007/JHEP11(2018)116} {\bibfield
		{journal} {\bibinfo  {journal} {JHEP}\ }\textbf {\bibinfo {volume} {11}},\
		\bibinfo {pages} {116} (\bibinfo {year} {2018}{\natexlab{b}})},\ \Eprint
	{http://arxiv.org/abs/1809.09067} {arXiv:1809.09067 [hep-ph]} \BibitemShut
	{NoStop}%
	\bibitem [{\citenamefont {Neill}\ and\ \citenamefont
		{Rothstein}(2013)}]{Neill:2013wsa}%
	\BibitemOpen
	\bibfield  {author} {\bibinfo {author} {\bibfnamefont {Duff}\ \bibnamefont
			{Neill}}\ and\ \bibinfo {author} {\bibfnamefont {Ira~Z.}\ \bibnamefont
			{Rothstein}},\ }\bibfield  {title} {\enquote {\bibinfo {title} {{Classical
					Space-Times from the S Matrix}},}\ }\href {\doibase
		10.1016/j.nuclphysb.2013.09.007} {\bibfield  {journal} {\bibinfo  {journal}
			{Nucl. Phys. B}\ }\textbf {\bibinfo {volume} {877}},\ \bibinfo {pages}
		{177--189} (\bibinfo {year} {2013})},\ \Eprint
	{http://arxiv.org/abs/1304.7263} {arXiv:1304.7263 [hep-th]} \BibitemShut
	{NoStop}%
	\bibitem [{\citenamefont {Maybee}\ \emph {et~al.}(2019)\citenamefont {Maybee},
		\citenamefont {O'Connell},\ and\ \citenamefont {Vines}}]{Maybee:2019jus}%
	\BibitemOpen
	\bibfield  {author} {\bibinfo {author} {\bibfnamefont {Ben}\ \bibnamefont
			{Maybee}}, \bibinfo {author} {\bibfnamefont {Donal}\ \bibnamefont
			{O'Connell}}, \ and\ \bibinfo {author} {\bibfnamefont {Justin}\ \bibnamefont
			{Vines}},\ }\bibfield  {title} {\enquote {\bibinfo {title} {{Observables and
					amplitudes for spinning particles and black holes}},}\ }\href {\doibase
		10.1007/JHEP12(2019)156} {\bibfield  {journal} {\bibinfo  {journal} {JHEP}\
		}\textbf {\bibinfo {volume} {12}},\ \bibinfo {pages} {156} (\bibinfo {year}
		{2019})},\ \Eprint {http://arxiv.org/abs/1906.09260} {arXiv:1906.09260
		[hep-th]} \BibitemShut {NoStop}%
	\bibitem [{\citenamefont {Landshoff}\ and\ \citenamefont
		{Polkinghorne}(1969)}]{Landshoff:1969yyn}%
	\BibitemOpen
	\bibfield  {author} {\bibinfo {author} {\bibfnamefont {P.~V.}\ \bibnamefont
			{Landshoff}}\ and\ \bibinfo {author} {\bibfnamefont {J.~C.}\ \bibnamefont
			{Polkinghorne}},\ }\bibfield  {title} {\enquote {\bibinfo {title}
			{{Iterations of regge cuts}},}\ }\href {\doibase 10.1103/PhysRev.181.1989}
	{\bibfield  {journal} {\bibinfo  {journal} {Phys. Rev.}\ }\textbf {\bibinfo
			{volume} {181}},\ \bibinfo {pages} {1989--1995} (\bibinfo {year}
		{1969})}\BibitemShut {NoStop}%
	\bibitem [{\citenamefont {Parra-Martinez}\ \emph {et~al.}(2020)\citenamefont
		{Parra-Martinez}, \citenamefont {Ruf},\ and\ \citenamefont
		{Zeng}}]{Parra-Martinez:2020dzs}%
	\BibitemOpen
	\bibfield  {author} {\bibinfo {author} {\bibfnamefont {Julio}\ \bibnamefont
			{Parra-Martinez}}, \bibinfo {author} {\bibfnamefont {Michael~S.}\
			\bibnamefont {Ruf}}, \ and\ \bibinfo {author} {\bibfnamefont {Mao}\
			\bibnamefont {Zeng}},\ }\bibfield  {title} {\enquote {\bibinfo {title}
			{{Extremal black hole scattering at $\mathcal{O}(G^3)$: graviton dominance,
					eikonal exponentiation, and differential equations}},}\ }\href {\doibase
		10.1007/JHEP11(2020)023} {\bibfield  {journal} {\bibinfo  {journal} {JHEP}\
		}\textbf {\bibinfo {volume} {11}},\ \bibinfo {pages} {023} (\bibinfo {year}
		{2020})},\ \Eprint {http://arxiv.org/abs/2005.04236} {arXiv:2005.04236
		[hep-th]} \BibitemShut {NoStop}%
	\bibitem [{\citenamefont {Abreu}\ \emph {et~al.}(2021)\citenamefont {Abreu},
		\citenamefont {Dormans}, \citenamefont {Febres~Cordero}, \citenamefont {Ita},
		\citenamefont {Kraus}, \citenamefont {Page}, \citenamefont {Pascual},
		\citenamefont {Ruf},\ and\ \citenamefont {Sotnikov}}]{Abreu:2020xvt}%
	\BibitemOpen
	\bibfield  {author} {\bibinfo {author} {\bibfnamefont {S.}~\bibnamefont
			{Abreu}}, \bibinfo {author} {\bibfnamefont {J.}~\bibnamefont {Dormans}},
		\bibinfo {author} {\bibfnamefont {F.}~\bibnamefont {Febres~Cordero}},
		\bibinfo {author} {\bibfnamefont {H.}~\bibnamefont {Ita}}, \bibinfo {author}
		{\bibfnamefont {M.}~\bibnamefont {Kraus}}, \bibinfo {author} {\bibfnamefont
			{B.}~\bibnamefont {Page}}, \bibinfo {author} {\bibfnamefont {E.}~\bibnamefont
			{Pascual}}, \bibinfo {author} {\bibfnamefont {M.~S.}\ \bibnamefont {Ruf}}, \
		and\ \bibinfo {author} {\bibfnamefont {V.}~\bibnamefont {Sotnikov}},\
	}\bibfield  {title} {\enquote {\bibinfo {title} {{Caravel: A C++ framework
					for the computation of multi-loop amplitudes with numerical unitarity}},}\
	}\href {\doibase 10.1016/j.cpc.2021.108069} {\bibfield  {journal} {\bibinfo
			{journal} {Comput. Phys. Commun.}\ }\textbf {\bibinfo {volume} {267}},\
		\bibinfo {pages} {108069} (\bibinfo {year} {2021})},\ \Eprint
	{http://arxiv.org/abs/2009.11957} {arXiv:2009.11957 [hep-ph]} \BibitemShut
	{NoStop}%
	\bibitem [{\citenamefont {von Manteuffel}\ and\ \citenamefont
		{Schabinger}(2015)}]{vonManteuffel:2014ixa}%
	\BibitemOpen
	\bibfield  {author} {\bibinfo {author} {\bibfnamefont {Andreas}\ \bibnamefont
			{von Manteuffel}}\ and\ \bibinfo {author} {\bibfnamefont {Robert~M.}\
			\bibnamefont {Schabinger}},\ }\bibfield  {title} {\enquote {\bibinfo {title}
			{{A novel approach to integration by parts reduction}},}\ }\href {\doibase
		10.1016/j.physletb.2015.03.029} {\bibfield  {journal} {\bibinfo  {journal}
			{Phys. Lett. B}\ }\textbf {\bibinfo {volume} {744}},\ \bibinfo {pages}
		{101--104} (\bibinfo {year} {2015})},\ \Eprint
	{http://arxiv.org/abs/1406.4513} {arXiv:1406.4513 [hep-ph]} \BibitemShut
	{NoStop}%
	\bibitem [{\citenamefont {Peraro}(2016)}]{Peraro:2016wsq}%
	\BibitemOpen
	\bibfield  {author} {\bibinfo {author} {\bibfnamefont {Tiziano}\ \bibnamefont
			{Peraro}},\ }\bibfield  {title} {\enquote {\bibinfo {title} {{Scattering
					amplitudes over finite fields and multivariate functional reconstruction}},}\
	}\href {\doibase 10.1007/JHEP12(2016)030} {\bibfield  {journal} {\bibinfo
			{journal} {JHEP}\ }\textbf {\bibinfo {volume} {12}},\ \bibinfo {pages} {030}
		(\bibinfo {year} {2016})},\ \Eprint {http://arxiv.org/abs/1608.01902}
	{arXiv:1608.01902 [hep-ph]} \BibitemShut {NoStop}%
	\bibitem [{\citenamefont {Abreu}\ \emph {et~al.}(2020)\citenamefont {Abreu},
		\citenamefont {Febres~Cordero}, \citenamefont {Ita}, \citenamefont {Jaquier},
		\citenamefont {Page}, \citenamefont {Ruf},\ and\ \citenamefont
		{Sotnikov}}]{Abreu:2020lyk}%
	\BibitemOpen
	\bibfield  {author} {\bibinfo {author} {\bibfnamefont {S.}~\bibnamefont
			{Abreu}}, \bibinfo {author} {\bibfnamefont {F.}~\bibnamefont
			{Febres~Cordero}}, \bibinfo {author} {\bibfnamefont {H.}~\bibnamefont {Ita}},
		\bibinfo {author} {\bibfnamefont {M.}~\bibnamefont {Jaquier}}, \bibinfo
		{author} {\bibfnamefont {B.}~\bibnamefont {Page}}, \bibinfo {author}
		{\bibfnamefont {M.~S.}\ \bibnamefont {Ruf}}, \ and\ \bibinfo {author}
		{\bibfnamefont {V.}~\bibnamefont {Sotnikov}},\ }\bibfield  {title} {\enquote
		{\bibinfo {title} {{Two-Loop Four-Graviton Scattering Amplitudes}},}\ }\href
	{\doibase 10.1103/PhysRevLett.124.211601} {\bibfield  {journal} {\bibinfo
			{journal} {Phys. Rev. Lett.}\ }\textbf {\bibinfo {volume} {124}},\ \bibinfo
		{pages} {211601} (\bibinfo {year} {2020})},\ \Eprint
	{http://arxiv.org/abs/2002.12374} {arXiv:2002.12374 [hep-th]} \BibitemShut
	{NoStop}%
	\bibitem [{\citenamefont {Brizuela}\ \emph {et~al.}(2009)\citenamefont
		{Brizuela}, \citenamefont {Martin-Garcia},\ and\ \citenamefont
		{Mena~Marugan}}]{Brizuela:2008ra}%
	\BibitemOpen
	\bibfield  {author} {\bibinfo {author} {\bibfnamefont {David}\ \bibnamefont
			{Brizuela}}, \bibinfo {author} {\bibfnamefont {Jose~M.}\ \bibnamefont
			{Martin-Garcia}}, \ and\ \bibinfo {author} {\bibfnamefont {Guillermo~A.}\
			\bibnamefont {Mena~Marugan}},\ }\bibfield  {title} {\enquote {\bibinfo
			{title} {{xPert: Computer algebra for metric perturbation theory}},}\ }\href
	{\doibase 10.1007/s10714-009-0773-2} {\bibfield  {journal} {\bibinfo
			{journal} {Gen. Rel. Grav.}\ }\textbf {\bibinfo {volume} {41}},\ \bibinfo
		{pages} {2415--2431} (\bibinfo {year} {2009})},\ \Eprint
	{http://arxiv.org/abs/0807.0824} {arXiv:0807.0824 [gr-qc]} \BibitemShut
	{NoStop}%
	\bibitem [{\citenamefont {Nutma}(2014)}]{Nutma:2013zea}%
	\BibitemOpen
	\bibfield  {author} {\bibinfo {author} {\bibfnamefont {Teake}\ \bibnamefont
			{Nutma}},\ }\bibfield  {title} {\enquote {\bibinfo {title} {{xTras : A
					field-theory inspired xAct package for mathematica}},}\ }\href {\doibase
		10.1016/j.cpc.2014.02.006} {\bibfield  {journal} {\bibinfo  {journal}
			{Comput. Phys. Commun.}\ }\textbf {\bibinfo {volume} {185}},\ \bibinfo
		{pages} {1719--1738} (\bibinfo {year} {2014})},\ \Eprint
	{http://arxiv.org/abs/1308.3493} {arXiv:1308.3493 [cs.SC]} \BibitemShut
	{NoStop}%
	\bibitem [{\citenamefont {'t~Hooft}\ and\ \citenamefont
		{Veltman}(1972)}]{tHooft:1972tcz}%
	\BibitemOpen
	\bibfield  {author} {\bibinfo {author} {\bibfnamefont {Gerard}\ \bibnamefont
			{'t~Hooft}}\ and\ \bibinfo {author} {\bibfnamefont {M.~J.~G.}\ \bibnamefont
			{Veltman}},\ }\bibfield  {title} {\enquote {\bibinfo {title} {{Regularization
					and Renormalization of Gauge Fields}},}\ }\href {\doibase
		10.1016/0550-3213(72)90279-9} {\bibfield  {journal} {\bibinfo  {journal}
			{Nucl. Phys. B}\ }\textbf {\bibinfo {volume} {44}},\ \bibinfo {pages}
		{189--213} (\bibinfo {year} {1972})}\BibitemShut {NoStop}%
	\bibitem [{\citenamefont {Berends}\ and\ \citenamefont
		{Giele}(1988)}]{Berends:1987me}%
	\BibitemOpen
	\bibfield  {author} {\bibinfo {author} {\bibfnamefont {Frits~A.}\
			\bibnamefont {Berends}}\ and\ \bibinfo {author} {\bibfnamefont {W.~T.}\
			\bibnamefont {Giele}},\ }\bibfield  {title} {\enquote {\bibinfo {title}
			{{Recursive Calculations for Processes with n Gluons}},}\ }\href {\doibase
		10.1016/0550-3213(88)90442-7} {\bibfield  {journal} {\bibinfo  {journal}
			{Nucl. Phys. B}\ }\textbf {\bibinfo {volume} {306}},\ \bibinfo {pages}
		{759--808} (\bibinfo {year} {1988})}\BibitemShut {NoStop}%
	\bibitem [{\citenamefont {Kawai}\ \emph {et~al.}(1986)\citenamefont {Kawai},
		\citenamefont {Lewellen},\ and\ \citenamefont {Tye}}]{Kawai:1985xq}%
	\BibitemOpen
	\bibfield  {author} {\bibinfo {author} {\bibfnamefont {H.}~\bibnamefont
			{Kawai}}, \bibinfo {author} {\bibfnamefont {D.~C.}\ \bibnamefont {Lewellen}},
		\ and\ \bibinfo {author} {\bibfnamefont {S.~H.~H.}\ \bibnamefont {Tye}},\
	}\bibfield  {title} {\enquote {\bibinfo {title} {{A Relation Between Tree
					Amplitudes of Closed and Open Strings}},}\ }\href {\doibase
		10.1016/0550-3213(86)90362-7} {\bibfield  {journal} {\bibinfo  {journal}
			{Nucl. Phys. B}\ }\textbf {\bibinfo {volume} {269}},\ \bibinfo {pages}
		{1--23} (\bibinfo {year} {1986})}\BibitemShut {NoStop}%
	\bibitem [{\citenamefont {Bern}\ \emph {et~al.}(2008)\citenamefont {Bern},
		\citenamefont {Carrasco},\ and\ \citenamefont {Johansson}}]{Bern:2008qj}%
	\BibitemOpen
	\bibfield  {author} {\bibinfo {author} {\bibfnamefont {Z.}~\bibnamefont
			{Bern}}, \bibinfo {author} {\bibfnamefont {J.~J.~M.}\ \bibnamefont
			{Carrasco}}, \ and\ \bibinfo {author} {\bibfnamefont {Henrik}\ \bibnamefont
			{Johansson}},\ }\bibfield  {title} {\enquote {\bibinfo {title} {{New
					Relations for Gauge-Theory Amplitudes}},}\ }\href {\doibase
		10.1103/PhysRevD.78.085011} {\bibfield  {journal} {\bibinfo  {journal} {Phys.
				Rev. D}\ }\textbf {\bibinfo {volume} {78}},\ \bibinfo {pages} {085011}
		(\bibinfo {year} {2008})},\ \Eprint {http://arxiv.org/abs/0805.3993}
	{arXiv:0805.3993 [hep-ph]} \BibitemShut {NoStop}%
	\bibitem [{\citenamefont {Bern}\ \emph {et~al.}(2010)\citenamefont {Bern},
		\citenamefont {Carrasco},\ and\ \citenamefont {Johansson}}]{Bern:2010ue}%
	\BibitemOpen
	\bibfield  {author} {\bibinfo {author} {\bibfnamefont {Zvi}\ \bibnamefont
			{Bern}}, \bibinfo {author} {\bibfnamefont {John Joseph~M.}\ \bibnamefont
			{Carrasco}}, \ and\ \bibinfo {author} {\bibfnamefont {Henrik}\ \bibnamefont
			{Johansson}},\ }\bibfield  {title} {\enquote {\bibinfo {title} {{Perturbative
					Quantum Gravity as a Double Copy of Gauge Theory}},}\ }\href {\doibase
		10.1103/PhysRevLett.105.061602} {\bibfield  {journal} {\bibinfo  {journal}
			{Phys. Rev. Lett.}\ }\textbf {\bibinfo {volume} {105}},\ \bibinfo {pages}
		{061602} (\bibinfo {year} {2010})},\ \Eprint {http://arxiv.org/abs/1004.0476}
	{arXiv:1004.0476 [hep-th]} \BibitemShut {NoStop}%
	\bibitem [{\citenamefont {Johansson}\ and\ \citenamefont
		{Ochirov}(2019)}]{Johansson:2019dnu}%
	\BibitemOpen
	\bibfield  {author} {\bibinfo {author} {\bibfnamefont {Henrik}\ \bibnamefont
			{Johansson}}\ and\ \bibinfo {author} {\bibfnamefont {Alexander}\ \bibnamefont
			{Ochirov}},\ }\bibfield  {title} {\enquote {\bibinfo {title} {{Double copy
					for massive quantum particles with spin}},}\ }\href {\doibase
		10.1007/JHEP09(2019)040} {\bibfield  {journal} {\bibinfo  {journal} {JHEP}\
		}\textbf {\bibinfo {volume} {09}},\ \bibinfo {pages} {040} (\bibinfo {year}
		{2019})},\ \Eprint {http://arxiv.org/abs/1906.12292} {arXiv:1906.12292
		[hep-th]} \BibitemShut {NoStop}%
	\bibitem [{\citenamefont {Bautista}\ and\ \citenamefont
		{Guevara}(2021)}]{Bautista:2019evw}%
	\BibitemOpen
	\bibfield  {author} {\bibinfo {author} {\bibfnamefont {Yilber~Fabian}\
			\bibnamefont {Bautista}}\ and\ \bibinfo {author} {\bibfnamefont {Alfredo}\
			\bibnamefont {Guevara}},\ }\bibfield  {title} {\enquote {\bibinfo {title}
			{{On the double copy for spinning matter}},}\ }\href {\doibase
		10.1007/JHEP11(2021)184} {\bibfield  {journal} {\bibinfo  {journal} {JHEP}\
		}\textbf {\bibinfo {volume} {11}},\ \bibinfo {pages} {184} (\bibinfo {year}
		{2021})},\ \Eprint {http://arxiv.org/abs/1908.11349} {arXiv:1908.11349
		[hep-th]} \BibitemShut {NoStop}%
	\bibitem [{\citenamefont {Cheung}\ and\ \citenamefont
		{Remmen}(2017{\natexlab{a}})}]{Cheung:2016say}%
	\BibitemOpen
	\bibfield  {author} {\bibinfo {author} {\bibfnamefont {Clifford}\
			\bibnamefont {Cheung}}\ and\ \bibinfo {author} {\bibfnamefont {Grant~N.}\
			\bibnamefont {Remmen}},\ }\bibfield  {title} {\enquote {\bibinfo {title}
			{{Twofold Symmetries of the Pure Gravity Action}},}\ }\href {\doibase
		10.1007/JHEP01(2017)104} {\bibfield  {journal} {\bibinfo  {journal} {JHEP}\
		}\textbf {\bibinfo {volume} {01}},\ \bibinfo {pages} {104} (\bibinfo {year}
		{2017}{\natexlab{a}})},\ \Eprint {http://arxiv.org/abs/1612.03927}
	{arXiv:1612.03927 [hep-th]} \BibitemShut {NoStop}%
	\bibitem [{\citenamefont {Cheung}\ and\ \citenamefont
		{Remmen}(2017{\natexlab{b}})}]{Cheung:2017kzx}%
	\BibitemOpen
	\bibfield  {author} {\bibinfo {author} {\bibfnamefont {Clifford}\
			\bibnamefont {Cheung}}\ and\ \bibinfo {author} {\bibfnamefont {Grant~N.}\
			\bibnamefont {Remmen}},\ }\bibfield  {title} {\enquote {\bibinfo {title}
			{{Hidden Simplicity of the Gravity Action}},}\ }\href {\doibase
		10.1007/JHEP09(2017)002} {\bibfield  {journal} {\bibinfo  {journal} {JHEP}\
		}\textbf {\bibinfo {volume} {09}},\ \bibinfo {pages} {002} (\bibinfo {year}
		{2017}{\natexlab{b}})},\ \Eprint {http://arxiv.org/abs/1705.00626}
	{arXiv:1705.00626 [hep-th]} \BibitemShut {NoStop}%
	\bibitem [{\citenamefont {K\"alin}\ and\ \citenamefont
		{Porto}(2020{\natexlab{a}})}]{Kalin:2020mvi}%
	\BibitemOpen
	\bibfield  {author} {\bibinfo {author} {\bibfnamefont {Gregor}\ \bibnamefont
			{K\"alin}}\ and\ \bibinfo {author} {\bibfnamefont {Rafael~A.}\ \bibnamefont
			{Porto}},\ }\bibfield  {title} {\enquote {\bibinfo {title} {{Post-Minkowskian
					Effective Field Theory for Conservative Binary Dynamics}},}\ }\href {\doibase
		10.1007/JHEP11(2020)106} {\bibfield  {journal} {\bibinfo  {journal} {JHEP}\
		}\textbf {\bibinfo {volume} {11}},\ \bibinfo {pages} {106} (\bibinfo {year}
		{2020}{\natexlab{a}})},\ \Eprint {http://arxiv.org/abs/2006.01184}
	{arXiv:2006.01184 [hep-th]} \BibitemShut {NoStop}%
	\bibitem [{\citenamefont {Britto}\ \emph {et~al.}(2022)\citenamefont {Britto},
		\citenamefont {Gonzo},\ and\ \citenamefont {Jehu}}]{Britto:2021pud}%
	\BibitemOpen
	\bibfield  {author} {\bibinfo {author} {\bibfnamefont {Ruth}\ \bibnamefont
			{Britto}}, \bibinfo {author} {\bibfnamefont {Riccardo}\ \bibnamefont
			{Gonzo}}, \ and\ \bibinfo {author} {\bibfnamefont {Guy~R.}\ \bibnamefont
			{Jehu}},\ }\bibfield  {title} {\enquote {\bibinfo {title} {{Graviton particle
					statistics and coherent states from classical scattering amplitudes}},}\
	}\href {\doibase 10.1007/JHEP03(2022)214} {\bibfield  {journal} {\bibinfo
			{journal} {JHEP}\ }\textbf {\bibinfo {volume} {03}},\ \bibinfo {pages} {214}
		(\bibinfo {year} {2022})},\ \Eprint {http://arxiv.org/abs/2112.07036}
	{arXiv:2112.07036 [hep-th]} \BibitemShut {NoStop}%
	\bibitem [{\citenamefont {Chetyrkin}\ and\ \citenamefont
		{Tkachov}(1981)}]{Chetyrkin:1981qh}%
	\BibitemOpen
	\bibfield  {author} {\bibinfo {author} {\bibfnamefont {K.~G.}\ \bibnamefont
			{Chetyrkin}}\ and\ \bibinfo {author} {\bibfnamefont {F.~V.}\ \bibnamefont
			{Tkachov}},\ }\bibfield  {title} {\enquote {\bibinfo {title} {{Integration by
					Parts: The Algorithm to Calculate beta Functions in 4 Loops}},}\ }\href
	{\doibase 10.1016/0550-3213(81)90199-1} {\bibfield  {journal} {\bibinfo
			{journal} {Nucl. Phys. B}\ }\textbf {\bibinfo {volume} {192}},\ \bibinfo
		{pages} {159--204} (\bibinfo {year} {1981})}\BibitemShut {NoStop}%
	\bibitem [{\citenamefont {Smirnov}\ and\ \citenamefont
		{Chuharev}(2020)}]{Smirnov:2019qkx}%
	\BibitemOpen
	\bibfield  {author} {\bibinfo {author} {\bibfnamefont {A.~V.}\ \bibnamefont
			{Smirnov}}\ and\ \bibinfo {author} {\bibfnamefont {F.~S.}\ \bibnamefont
			{Chuharev}},\ }\bibfield  {title} {\enquote {\bibinfo {title} {{FIRE6:
					Feynman Integral REduction with Modular Arithmetic}},}\ }\href {\doibase
		10.1016/j.cpc.2019.106877} {\bibfield  {journal} {\bibinfo  {journal}
			{Comput. Phys. Commun.}\ }\textbf {\bibinfo {volume} {247}},\ \bibinfo
		{pages} {106877} (\bibinfo {year} {2020})},\ \Eprint
	{http://arxiv.org/abs/1901.07808} {arXiv:1901.07808 [hep-ph]} \BibitemShut
	{NoStop}%
	\bibitem [{Spi()}]{SpinBBHResults}%
	\BibitemOpen
	\href@noop {} {}\bibinfo {howpublished} {Supplemental material},\ \bibinfo
	{note} {includes three appendices on full-theory amplitudes, polarization
		tensors in non-relativistic EFT, and on the PN expansion of our coefficients.
		We also include five computer files, including amplitudes computed in the
		full theory [\texttt{FullTheory\_Amplitudes.m}], amplitudes computed from the
		EFT [\texttt{EFT\_Amplitudes.m}], translations between relativistic and
		non-relativistic tensors in Eqs.~(2) and (6) [\texttt{Tensor\_expansion.m}],
		and the matching coefficients defined in Eq.~(7) [\texttt{Coefficients.m}] as
		well as non-relativistic expansions thereof
		[\texttt{Coefficients\_Expanded.m}].}\BibitemShut {Stop}%
	\bibitem [{\citenamefont {Holstein}\ and\ \citenamefont
		{Ross}(2008)}]{Holstein:2008sx}%
	\BibitemOpen
	\bibfield  {author} {\bibinfo {author} {\bibfnamefont {Barry~R.}\
			\bibnamefont {Holstein}}\ and\ \bibinfo {author} {\bibfnamefont {Andreas}\
			\bibnamefont {Ross}},\ }\bibfield  {title} {\enquote {\bibinfo {title} {{Spin
					Effects in Long Range Gravitational Scattering}},}\ }\href@noop {} {\
		(\bibinfo {year} {2008})},\ \Eprint {http://arxiv.org/abs/0802.0716}
	{arXiv:0802.0716 [hep-ph]} \BibitemShut {NoStop}%
	\bibitem [{\citenamefont {Cristofoli}\ \emph {et~al.}(2020)\citenamefont
		{Cristofoli}, \citenamefont {Damgaard}, \citenamefont {Di~Vecchia},\ and\
		\citenamefont {Heissenberg}}]{Cristofoli:2020uzm}%
	\BibitemOpen
	\bibfield  {author} {\bibinfo {author} {\bibfnamefont {Andrea}\ \bibnamefont
			{Cristofoli}}, \bibinfo {author} {\bibfnamefont {Poul~H.}\ \bibnamefont
			{Damgaard}}, \bibinfo {author} {\bibfnamefont {Paolo}\ \bibnamefont
			{Di~Vecchia}}, \ and\ \bibinfo {author} {\bibfnamefont {Carlo}\ \bibnamefont
			{Heissenberg}},\ }\bibfield  {title} {\enquote {\bibinfo {title}
			{{Second-order Post-Minkowskian scattering in arbitrary dimensions}},}\
	}\href {\doibase 10.1007/JHEP07(2020)122} {\bibfield  {journal} {\bibinfo
			{journal} {JHEP}\ }\textbf {\bibinfo {volume} {07}},\ \bibinfo {pages} {122}
		(\bibinfo {year} {2020})},\ \Eprint {http://arxiv.org/abs/2003.10274}
	{arXiv:2003.10274 [hep-th]} \BibitemShut {NoStop}%
	\bibitem [{\citenamefont {Levi}\ and\ \citenamefont
		{Steinhoff}(2017)}]{Levi:2017kzq}%
	\BibitemOpen
	\bibfield  {author} {\bibinfo {author} {\bibfnamefont {Michele}\ \bibnamefont
			{Levi}}\ and\ \bibinfo {author} {\bibfnamefont {Jan}\ \bibnamefont
			{Steinhoff}},\ }\bibfield  {title} {\enquote {\bibinfo {title} {{EFTofPNG: A
					package for high precision computation with the Effective Field Theory of
					Post-Newtonian Gravity}},}\ }\href {\doibase 10.1088/1361-6382/aa941e}
	{\bibfield  {journal} {\bibinfo  {journal} {Class. Quant. Grav.}\ }\textbf
		{\bibinfo {volume} {34}},\ \bibinfo {pages} {244001} (\bibinfo {year}
		{2017})},\ \Eprint {http://arxiv.org/abs/1705.06309} {arXiv:1705.06309
		[gr-qc]} \BibitemShut {NoStop}%
	\bibitem [{\citenamefont {Herrmann}\ \emph
		{et~al.}(2021{\natexlab{b}})\citenamefont {Herrmann}, \citenamefont
		{Parra-Martinez}, \citenamefont {Ruf},\ and\ \citenamefont
		{Zeng}}]{Herrmann:2021lqe}%
	\BibitemOpen
	\bibfield  {author} {\bibinfo {author} {\bibfnamefont {Enrico}\ \bibnamefont
			{Herrmann}}, \bibinfo {author} {\bibfnamefont {Julio}\ \bibnamefont
			{Parra-Martinez}}, \bibinfo {author} {\bibfnamefont {Michael~S.}\
			\bibnamefont {Ruf}}, \ and\ \bibinfo {author} {\bibfnamefont {Mao}\
			\bibnamefont {Zeng}},\ }\bibfield  {title} {\enquote {\bibinfo {title}
			{{Gravitational Bremsstrahlung from Reverse Unitarity}},}\ }\href {\doibase
		10.1103/PhysRevLett.126.201602} {\bibfield  {journal} {\bibinfo  {journal}
			{Phys. Rev. Lett.}\ }\textbf {\bibinfo {volume} {126}},\ \bibinfo {pages}
		{201602} (\bibinfo {year} {2021}{\natexlab{b}})},\ \Eprint
	{http://arxiv.org/abs/2101.07255} {arXiv:2101.07255 [hep-th]} \BibitemShut
	{NoStop}%
	\bibitem [{\citenamefont {K\"alin}\ and\ \citenamefont
		{Porto}(2020{\natexlab{b}})}]{Kalin:2019rwq}%
	\BibitemOpen
	\bibfield  {author} {\bibinfo {author} {\bibfnamefont {Gregor}\ \bibnamefont
			{K\"alin}}\ and\ \bibinfo {author} {\bibfnamefont {Rafael~A.}\ \bibnamefont
			{Porto}},\ }\bibfield  {title} {\enquote {\bibinfo {title} {{From Boundary
					Data to Bound States}},}\ }\href {\doibase 10.1007/JHEP01(2020)072}
	{\bibfield  {journal} {\bibinfo  {journal} {JHEP}\ }\textbf {\bibinfo
			{volume} {01}},\ \bibinfo {pages} {072} (\bibinfo {year}
		{2020}{\natexlab{b}})},\ \Eprint {http://arxiv.org/abs/1910.03008}
	{arXiv:1910.03008 [hep-th]} \BibitemShut {NoStop}%
	\bibitem [{\citenamefont {K\"alin}\ and\ \citenamefont
		{Porto}(2020{\natexlab{c}})}]{Kalin:2019inp}%
	\BibitemOpen
	\bibfield  {author} {\bibinfo {author} {\bibfnamefont {Gregor}\ \bibnamefont
			{K\"alin}}\ and\ \bibinfo {author} {\bibfnamefont {Rafael~A.}\ \bibnamefont
			{Porto}},\ }\bibfield  {title} {\enquote {\bibinfo {title} {{From boundary
					data to bound states. Part II. Scattering angle to dynamical invariants (with
					twist)}},}\ }\href {\doibase 10.1007/JHEP02(2020)120} {\bibfield  {journal}
		{\bibinfo  {journal} {JHEP}\ }\textbf {\bibinfo {volume} {02}},\ \bibinfo
		{pages} {120} (\bibinfo {year} {2020}{\natexlab{c}})},\ \Eprint
	{http://arxiv.org/abs/1911.09130} {arXiv:1911.09130 [hep-th]} \BibitemShut
	{NoStop}%
	\bibitem [{\citenamefont {Cho}\ \emph {et~al.}(2022{\natexlab{b}})\citenamefont
		{Cho}, \citenamefont {K\"alin},\ and\ \citenamefont {Porto}}]{Cho:2021arx}%
	\BibitemOpen
	\bibfield  {author} {\bibinfo {author} {\bibfnamefont {Gihyuk}\ \bibnamefont
			{Cho}}, \bibinfo {author} {\bibfnamefont {Gregor}\ \bibnamefont {K\"alin}}, \
		and\ \bibinfo {author} {\bibfnamefont {Rafael~A.}\ \bibnamefont {Porto}},\
	}\bibfield  {title} {\enquote {\bibinfo {title} {{From boundary data to bound
					states. Part III. Radiative effects}},}\ }\href {\doibase
		10.1007/JHEP04(2022)154} {\bibfield  {journal} {\bibinfo  {journal} {JHEP}\
		}\textbf {\bibinfo {volume} {04}},\ \bibinfo {pages} {154} (\bibinfo {year}
		{2022}{\natexlab{b}})},\ \Eprint {http://arxiv.org/abs/2112.03976}
	{arXiv:2112.03976 [hep-th]} \BibitemShut {NoStop}%
	\bibitem [{\citenamefont {Chiodaroli}\ \emph {et~al.}(2022)\citenamefont
		{Chiodaroli}, \citenamefont {Johansson},\ and\ \citenamefont
		{Pichini}}]{Chiodaroli:2021eug}%
	\BibitemOpen
	\bibfield  {author} {\bibinfo {author} {\bibfnamefont {Marco}\ \bibnamefont
			{Chiodaroli}}, \bibinfo {author} {\bibfnamefont {Henrik}\ \bibnamefont
			{Johansson}}, \ and\ \bibinfo {author} {\bibfnamefont {Paolo}\ \bibnamefont
			{Pichini}},\ }\bibfield  {title} {\enquote {\bibinfo {title} {{Compton
					black-hole scattering for s \ensuremath{\leq} 5/2}},}\ }\href {\doibase
		10.1007/JHEP02(2022)156} {\bibfield  {journal} {\bibinfo  {journal} {JHEP}\
		}\textbf {\bibinfo {volume} {02}},\ \bibinfo {pages} {156} (\bibinfo {year}
		{2022})},\ \Eprint {http://arxiv.org/abs/2107.14779} {arXiv:2107.14779
		[hep-th]} \BibitemShut {NoStop}%
	\bibitem [{\citenamefont {Cheung}\ and\ \citenamefont
		{Solon}(2020)}]{Cheung:2020sdj}%
	\BibitemOpen
	\bibfield  {author} {\bibinfo {author} {\bibfnamefont {Clifford}\
			\bibnamefont {Cheung}}\ and\ \bibinfo {author} {\bibfnamefont {Mikhail~P.}\
			\bibnamefont {Solon}},\ }\bibfield  {title} {\enquote {\bibinfo {title}
			{{Tidal Effects in the Post-Minkowskian Expansion}},}\ }\href {\doibase
		10.1103/PhysRevLett.125.191601} {\bibfield  {journal} {\bibinfo  {journal}
			{Phys. Rev. Lett.}\ }\textbf {\bibinfo {volume} {125}},\ \bibinfo {pages}
		{191601} (\bibinfo {year} {2020})},\ \Eprint
	{http://arxiv.org/abs/2006.06665} {arXiv:2006.06665 [hep-th]} \BibitemShut
	{NoStop}%
	\bibitem [{\citenamefont {K\"alin}\ \emph
		{et~al.}(2020{\natexlab{b}})\citenamefont {K\"alin}, \citenamefont {Liu},\
		and\ \citenamefont {Porto}}]{Kalin:2020lmz}%
	\BibitemOpen
	\bibfield  {author} {\bibinfo {author} {\bibfnamefont {Gregor}\ \bibnamefont
			{K\"alin}}, \bibinfo {author} {\bibfnamefont {Zhengwen}\ \bibnamefont {Liu}},
		\ and\ \bibinfo {author} {\bibfnamefont {Rafael~A.}\ \bibnamefont {Porto}},\
	}\bibfield  {title} {\enquote {\bibinfo {title} {{Conservative Tidal Effects
					in Compact Binary Systems to Next-to-Leading Post-Minkowskian Order}},}\
	}\href {\doibase 10.1103/PhysRevD.102.124025} {\bibfield  {journal} {\bibinfo
			{journal} {Phys. Rev. D}\ }\textbf {\bibinfo {volume} {102}},\ \bibinfo
		{pages} {124025} (\bibinfo {year} {2020}{\natexlab{b}})},\ \Eprint
	{http://arxiv.org/abs/2008.06047} {arXiv:2008.06047 [hep-th]} \BibitemShut
	{NoStop}%
	\bibitem [{\citenamefont {Bern}\ \emph
		{et~al.}(2021{\natexlab{c}})\citenamefont {Bern}, \citenamefont
		{Parra-Martinez}, \citenamefont {Roiban}, \citenamefont {Sawyer},\ and\
		\citenamefont {Shen}}]{Bern:2020uwk}%
	\BibitemOpen
	\bibfield  {author} {\bibinfo {author} {\bibfnamefont {Zvi}\ \bibnamefont
			{Bern}}, \bibinfo {author} {\bibfnamefont {Julio}\ \bibnamefont
			{Parra-Martinez}}, \bibinfo {author} {\bibfnamefont {Radu}\ \bibnamefont
			{Roiban}}, \bibinfo {author} {\bibfnamefont {Eric}\ \bibnamefont {Sawyer}}, \
		and\ \bibinfo {author} {\bibfnamefont {Chia-Hsien}\ \bibnamefont {Shen}},\
	}\bibfield  {title} {\enquote {\bibinfo {title} {{Leading Nonlinear Tidal
					Effects and Scattering Amplitudes}},}\ }\href {\doibase
		10.1007/JHEP05(2021)188} {\bibfield  {journal} {\bibinfo  {journal} {JHEP}\
		}\textbf {\bibinfo {volume} {05}},\ \bibinfo {pages} {188} (\bibinfo {year}
		{2021}{\natexlab{c}})},\ \Eprint {http://arxiv.org/abs/2010.08559}
	{arXiv:2010.08559 [hep-th]} \BibitemShut {NoStop}%
	\bibitem [{\citenamefont {Cheung}\ \emph {et~al.}(2021)\citenamefont {Cheung},
		\citenamefont {Shah},\ and\ \citenamefont {Solon}}]{Cheung:2020gbf}%
	\BibitemOpen
	\bibfield  {author} {\bibinfo {author} {\bibfnamefont {Clifford}\
			\bibnamefont {Cheung}}, \bibinfo {author} {\bibfnamefont {Nabha}\
			\bibnamefont {Shah}}, \ and\ \bibinfo {author} {\bibfnamefont {Mikhail~P.}\
			\bibnamefont {Solon}},\ }\bibfield  {title} {\enquote {\bibinfo {title}
			{{Mining the Geodesic Equation for Scattering Data}},}\ }\href {\doibase
		10.1103/PhysRevD.103.024030} {\bibfield  {journal} {\bibinfo  {journal}
			{Phys. Rev. D}\ }\textbf {\bibinfo {volume} {103}},\ \bibinfo {pages}
		{024030} (\bibinfo {year} {2021})},\ \Eprint
	{http://arxiv.org/abs/2010.08568} {arXiv:2010.08568 [hep-th]} \BibitemShut
	{NoStop}%
	\bibitem [{\citenamefont {Bini}\ \emph
		{et~al.}(2020{\natexlab{d}})\citenamefont {Bini}, \citenamefont {Damour},\
		and\ \citenamefont {Geralico}}]{Bini:2020flp}%
	\BibitemOpen
	\bibfield  {author} {\bibinfo {author} {\bibfnamefont {Donato}\ \bibnamefont
			{Bini}}, \bibinfo {author} {\bibfnamefont {Thibault}\ \bibnamefont {Damour}},
		\ and\ \bibinfo {author} {\bibfnamefont {Andrea}\ \bibnamefont {Geralico}},\
	}\bibfield  {title} {\enquote {\bibinfo {title} {{Scattering of tidally
					interacting bodies in post-Minkowskian gravity}},}\ }\href {\doibase
		10.1103/PhysRevD.101.044039} {\bibfield  {journal} {\bibinfo  {journal}
			{Phys. Rev. D}\ }\textbf {\bibinfo {volume} {101}},\ \bibinfo {pages}
		{044039} (\bibinfo {year} {2020}{\natexlab{d}})},\ \Eprint
	{http://arxiv.org/abs/2001.00352} {arXiv:2001.00352 [gr-qc]} \BibitemShut
	{NoStop}%
	\bibitem [{\citenamefont {Haddad}\ and\ \citenamefont
		{Helset}(2020)}]{Haddad:2020que}%
	\BibitemOpen
	\bibfield  {author} {\bibinfo {author} {\bibfnamefont {Kays}\ \bibnamefont
			{Haddad}}\ and\ \bibinfo {author} {\bibfnamefont {Andreas}\ \bibnamefont
			{Helset}},\ }\bibfield  {title} {\enquote {\bibinfo {title} {{Tidal effects
					in quantum field theory}},}\ }\href {\doibase 10.1007/JHEP12(2020)024}
	{\bibfield  {journal} {\bibinfo  {journal} {JHEP}\ }\textbf {\bibinfo
			{volume} {12}},\ \bibinfo {pages} {024} (\bibinfo {year} {2020})},\ \Eprint
	{http://arxiv.org/abs/2008.04920} {arXiv:2008.04920 [hep-th]} \BibitemShut
	{NoStop}%
	\bibitem [{\citenamefont {Aoude}\ \emph {et~al.}(2021)\citenamefont {Aoude},
		\citenamefont {Haddad},\ and\ \citenamefont {Helset}}]{Aoude:2020ygw}%
	\BibitemOpen
	\bibfield  {author} {\bibinfo {author} {\bibfnamefont {Rafael}\ \bibnamefont
			{Aoude}}, \bibinfo {author} {\bibfnamefont {Kays}\ \bibnamefont {Haddad}}, \
		and\ \bibinfo {author} {\bibfnamefont {Andreas}\ \bibnamefont {Helset}},\
	}\bibfield  {title} {\enquote {\bibinfo {title} {{Tidal effects for spinning
					particles}},}\ }\href {\doibase 10.1007/JHEP03(2021)097} {\bibfield
		{journal} {\bibinfo  {journal} {JHEP}\ }\textbf {\bibinfo {volume} {03}},\
		\bibinfo {pages} {097} (\bibinfo {year} {2021})},\ \Eprint
	{http://arxiv.org/abs/2012.05256} {arXiv:2012.05256 [hep-th]} \BibitemShut
	{NoStop}%
	\bibitem [{\citenamefont {Mougiakakos}\ \emph {et~al.}(2022)\citenamefont
		{Mougiakakos}, \citenamefont {Riva},\ and\ \citenamefont
		{Vernizzi}}]{Mougiakakos:2022sic}%
	\BibitemOpen
	\bibfield  {author} {\bibinfo {author} {\bibfnamefont {Stavros}\ \bibnamefont
			{Mougiakakos}}, \bibinfo {author} {\bibfnamefont {Massimiliano~Maria}\
			\bibnamefont {Riva}}, \ and\ \bibinfo {author} {\bibfnamefont {Filippo}\
			\bibnamefont {Vernizzi}},\ }\bibfield  {title} {\enquote {\bibinfo {title}
			{{Gravitational Bremsstrahlung with tidal effects in the post-Minkowskian
					expansion}},}\ }\href@noop {} {\  (\bibinfo {year} {2022})},\ \Eprint
	{http://arxiv.org/abs/2204.06556} {arXiv:2204.06556 [hep-th]} \BibitemShut
	{NoStop}%
	\bibitem [{\citenamefont {Luna}\ \emph {et~al.}(2016)\citenamefont {Luna},
		\citenamefont {Monteiro}, \citenamefont {Nicholson}, \citenamefont
		{O'Connell},\ and\ \citenamefont {White}}]{Luna:2016due}%
	\BibitemOpen
	\bibfield  {author} {\bibinfo {author} {\bibfnamefont {Andr\'es}\
			\bibnamefont {Luna}}, \bibinfo {author} {\bibfnamefont {Ricardo}\
			\bibnamefont {Monteiro}}, \bibinfo {author} {\bibfnamefont {Isobel}\
			\bibnamefont {Nicholson}}, \bibinfo {author} {\bibfnamefont {Donal}\
			\bibnamefont {O'Connell}}, \ and\ \bibinfo {author} {\bibfnamefont
			{Chris~D.}\ \bibnamefont {White}},\ }\bibfield  {title} {\enquote {\bibinfo
			{title} {{The double copy: Bremsstrahlung and accelerating black holes}},}\
	}\href {\doibase 10.1007/JHEP06(2016)023} {\bibfield  {journal} {\bibinfo
			{journal} {JHEP}\ }\textbf {\bibinfo {volume} {06}},\ \bibinfo {pages} {023}
		(\bibinfo {year} {2016})},\ \Eprint {http://arxiv.org/abs/1603.05737}
	{arXiv:1603.05737 [hep-th]} \BibitemShut {NoStop}%
	\bibitem [{\citenamefont {Goldberger}\ and\ \citenamefont
		{Ridgway}(2017)}]{Goldberger:2016iau}%
	\BibitemOpen
	\bibfield  {author} {\bibinfo {author} {\bibfnamefont {Walter~D.}\
			\bibnamefont {Goldberger}}\ and\ \bibinfo {author} {\bibfnamefont
			{Alexander~K.}\ \bibnamefont {Ridgway}},\ }\bibfield  {title} {\enquote
		{\bibinfo {title} {{Radiation and the classical double copy for color
					charges}},}\ }\href {\doibase 10.1103/PhysRevD.95.125010} {\bibfield
		{journal} {\bibinfo  {journal} {Phys. Rev. D}\ }\textbf {\bibinfo {volume}
			{95}},\ \bibinfo {pages} {125010} (\bibinfo {year} {2017})},\ \Eprint
	{http://arxiv.org/abs/1611.03493} {arXiv:1611.03493 [hep-th]} \BibitemShut
	{NoStop}%
	\bibitem [{\citenamefont {Goldberger}\ \emph {et~al.}(2017)\citenamefont
		{Goldberger}, \citenamefont {Prabhu},\ and\ \citenamefont
		{Thompson}}]{Goldberger:2017frp}%
	\BibitemOpen
	\bibfield  {author} {\bibinfo {author} {\bibfnamefont {Walter~D.}\
			\bibnamefont {Goldberger}}, \bibinfo {author} {\bibfnamefont {Siddharth~G.}\
			\bibnamefont {Prabhu}}, \ and\ \bibinfo {author} {\bibfnamefont
			{Jedidiah~O.}\ \bibnamefont {Thompson}},\ }\bibfield  {title} {\enquote
		{\bibinfo {title} {{Classical gluon and graviton radiation from the
					bi-adjoint scalar double copy}},}\ }\href {\doibase
		10.1103/PhysRevD.96.065009} {\bibfield  {journal} {\bibinfo  {journal} {Phys.
				Rev. D}\ }\textbf {\bibinfo {volume} {96}},\ \bibinfo {pages} {065009}
		(\bibinfo {year} {2017})},\ \Eprint {http://arxiv.org/abs/1705.09263}
	{arXiv:1705.09263 [hep-th]} \BibitemShut {NoStop}%
	\bibitem [{\citenamefont {Goldberger}\ and\ \citenamefont
		{Ridgway}(2018)}]{Goldberger:2017vcg}%
	\BibitemOpen
	\bibfield  {author} {\bibinfo {author} {\bibfnamefont {Walter~D.}\
			\bibnamefont {Goldberger}}\ and\ \bibinfo {author} {\bibfnamefont
			{Alexander~K.}\ \bibnamefont {Ridgway}},\ }\bibfield  {title} {\enquote
		{\bibinfo {title} {{Bound states and the classical double copy}},}\ }\href
	{\doibase 10.1103/PhysRevD.97.085019} {\bibfield  {journal} {\bibinfo
			{journal} {Phys. Rev. D}\ }\textbf {\bibinfo {volume} {97}},\ \bibinfo
		{pages} {085019} (\bibinfo {year} {2018})},\ \Eprint
	{http://arxiv.org/abs/1711.09493} {arXiv:1711.09493 [hep-th]} \BibitemShut
	{NoStop}%
	\bibitem [{\citenamefont {Chester}(2018)}]{Chester:2017vcz}%
	\BibitemOpen
	\bibfield  {author} {\bibinfo {author} {\bibfnamefont {David}\ \bibnamefont
			{Chester}},\ }\bibfield  {title} {\enquote {\bibinfo {title} {{Radiative
					double copy for Einstein-Yang-Mills theory}},}\ }\href {\doibase
		10.1103/PhysRevD.97.084025} {\bibfield  {journal} {\bibinfo  {journal} {Phys.
				Rev. D}\ }\textbf {\bibinfo {volume} {97}},\ \bibinfo {pages} {084025}
		(\bibinfo {year} {2018})},\ \Eprint {http://arxiv.org/abs/1712.08684}
	{arXiv:1712.08684 [hep-th]} \BibitemShut {NoStop}%
	\bibitem [{\citenamefont {Goldberger}\ \emph {et~al.}(2018)\citenamefont
		{Goldberger}, \citenamefont {Li},\ and\ \citenamefont
		{Prabhu}}]{Goldberger:2017ogt}%
	\BibitemOpen
	\bibfield  {author} {\bibinfo {author} {\bibfnamefont {Walter~D.}\
			\bibnamefont {Goldberger}}, \bibinfo {author} {\bibfnamefont {Jingping}\
			\bibnamefont {Li}}, \ and\ \bibinfo {author} {\bibfnamefont {Siddharth~G.}\
			\bibnamefont {Prabhu}},\ }\bibfield  {title} {\enquote {\bibinfo {title}
			{{Spinning particles, axion radiation, and the classical double copy}},}\
	}\href {\doibase 10.1103/PhysRevD.97.105018} {\bibfield  {journal} {\bibinfo
			{journal} {Phys. Rev. D}\ }\textbf {\bibinfo {volume} {97}},\ \bibinfo
		{pages} {105018} (\bibinfo {year} {2018})},\ \Eprint
	{http://arxiv.org/abs/1712.09250} {arXiv:1712.09250 [hep-th]} \BibitemShut
	{NoStop}%
	\bibitem [{\citenamefont {Shen}(2018)}]{Shen:2018ebu}%
	\BibitemOpen
	\bibfield  {author} {\bibinfo {author} {\bibfnamefont {Chia-Hsien}\
			\bibnamefont {Shen}},\ }\bibfield  {title} {\enquote {\bibinfo {title}
			{{Gravitational Radiation from Color-Kinematics Duality}},}\ }\href {\doibase
		10.1007/JHEP11(2018)162} {\bibfield  {journal} {\bibinfo  {journal} {JHEP}\
		}\textbf {\bibinfo {volume} {11}},\ \bibinfo {pages} {162} (\bibinfo {year}
		{2018})},\ \Eprint {http://arxiv.org/abs/1806.07388} {arXiv:1806.07388
		[hep-th]} \BibitemShut {NoStop}%
	\bibitem [{\citenamefont {Goldberger}\ and\ \citenamefont
		{Li}(2020)}]{Goldberger:2019xef}%
	\BibitemOpen
	\bibfield  {author} {\bibinfo {author} {\bibfnamefont {Walter~D.}\
			\bibnamefont {Goldberger}}\ and\ \bibinfo {author} {\bibfnamefont {Jingping}\
			\bibnamefont {Li}},\ }\bibfield  {title} {\enquote {\bibinfo {title}
			{{Strings, extended objects, and the classical double copy}},}\ }\href
	{\doibase 10.1007/JHEP02(2020)092} {\bibfield  {journal} {\bibinfo  {journal}
			{JHEP}\ }\textbf {\bibinfo {volume} {02}},\ \bibinfo {pages} {092} (\bibinfo
		{year} {2020})},\ \Eprint {http://arxiv.org/abs/1912.01650} {arXiv:1912.01650
		[hep-th]} \BibitemShut {NoStop}%
	\bibitem [{\citenamefont {Almeida}\ \emph {et~al.}(2020)\citenamefont
		{Almeida}, \citenamefont {Foffa},\ and\ \citenamefont
		{Sturani}}]{Almeida:2020mrg}%
	\BibitemOpen
	\bibfield  {author} {\bibinfo {author} {\bibfnamefont {Gabriel~Luz}\
			\bibnamefont {Almeida}}, \bibinfo {author} {\bibfnamefont {Stefano}\
			\bibnamefont {Foffa}}, \ and\ \bibinfo {author} {\bibfnamefont {Riccardo}\
			\bibnamefont {Sturani}},\ }\bibfield  {title} {\enquote {\bibinfo {title}
			{{Classical Gravitational Self-Energy from Double Copy}},}\ }\href {\doibase
		10.1007/JHEP11(2020)165} {\bibfield  {journal} {\bibinfo  {journal} {JHEP}\
		}\textbf {\bibinfo {volume} {11}},\ \bibinfo {pages} {165} (\bibinfo {year}
		{2020})},\ \Eprint {http://arxiv.org/abs/2008.06195} {arXiv:2008.06195
		[gr-qc]} \BibitemShut {NoStop}%
	\bibitem [{\citenamefont {Laddha}\ and\ \citenamefont
		{Sen}(2018{\natexlab{a}})}]{Laddha:2018rle}%
	\BibitemOpen
	\bibfield  {author} {\bibinfo {author} {\bibfnamefont {Alok}\ \bibnamefont
			{Laddha}}\ and\ \bibinfo {author} {\bibfnamefont {Ashoke}\ \bibnamefont
			{Sen}},\ }\bibfield  {title} {\enquote {\bibinfo {title} {{Gravity Waves from
					Soft Theorem in General Dimensions}},}\ }\href {\doibase
		10.1007/JHEP09(2018)105} {\bibfield  {journal} {\bibinfo  {journal} {JHEP}\
		}\textbf {\bibinfo {volume} {09}},\ \bibinfo {pages} {105} (\bibinfo {year}
		{2018}{\natexlab{a}})},\ \Eprint {http://arxiv.org/abs/1801.07719}
	{arXiv:1801.07719 [hep-th]} \BibitemShut {NoStop}%
	\bibitem [{\citenamefont {Laddha}\ and\ \citenamefont
		{Sen}(2018{\natexlab{b}})}]{Laddha:2018myi}%
	\BibitemOpen
	\bibfield  {author} {\bibinfo {author} {\bibfnamefont {Alok}\ \bibnamefont
			{Laddha}}\ and\ \bibinfo {author} {\bibfnamefont {Ashoke}\ \bibnamefont
			{Sen}},\ }\bibfield  {title} {\enquote {\bibinfo {title} {{Logarithmic Terms
					in the Soft Expansion in Four Dimensions}},}\ }\href {\doibase
		10.1007/JHEP10(2018)056} {\bibfield  {journal} {\bibinfo  {journal} {JHEP}\
		}\textbf {\bibinfo {volume} {10}},\ \bibinfo {pages} {056} (\bibinfo {year}
		{2018}{\natexlab{b}})},\ \Eprint {http://arxiv.org/abs/1804.09193}
	{arXiv:1804.09193 [hep-th]} \BibitemShut {NoStop}%
	\bibitem [{\citenamefont {Laddha}\ and\ \citenamefont
		{Sen}(2019)}]{Laddha:2018vbn}%
	\BibitemOpen
	\bibfield  {author} {\bibinfo {author} {\bibfnamefont {Alok}\ \bibnamefont
			{Laddha}}\ and\ \bibinfo {author} {\bibfnamefont {Ashoke}\ \bibnamefont
			{Sen}},\ }\bibfield  {title} {\enquote {\bibinfo {title} {{Observational
					Signature of the Logarithmic Terms in the Soft Graviton Theorem}},}\ }\href
	{\doibase 10.1103/PhysRevD.100.024009} {\bibfield  {journal} {\bibinfo
			{journal} {Phys. Rev. D}\ }\textbf {\bibinfo {volume} {100}},\ \bibinfo
		{pages} {024009} (\bibinfo {year} {2019})},\ \Eprint
	{http://arxiv.org/abs/1806.01872} {arXiv:1806.01872 [hep-th]} \BibitemShut
	{NoStop}%
	\bibitem [{\citenamefont {Sahoo}\ and\ \citenamefont
		{Sen}(2019)}]{Sahoo:2018lxl}%
	\BibitemOpen
	\bibfield  {author} {\bibinfo {author} {\bibfnamefont {Biswajit}\
			\bibnamefont {Sahoo}}\ and\ \bibinfo {author} {\bibfnamefont {Ashoke}\
			\bibnamefont {Sen}},\ }\bibfield  {title} {\enquote {\bibinfo {title}
			{{Classical and Quantum Results on Logarithmic Terms in the Soft Theorem in
					Four Dimensions}},}\ }\href {\doibase 10.1007/JHEP02(2019)086} {\bibfield
		{journal} {\bibinfo  {journal} {JHEP}\ }\textbf {\bibinfo {volume} {02}},\
		\bibinfo {pages} {086} (\bibinfo {year} {2019})},\ \Eprint
	{http://arxiv.org/abs/1808.03288} {arXiv:1808.03288 [hep-th]} \BibitemShut
	{NoStop}%
	\bibitem [{\citenamefont {Ciafaloni}\ \emph {et~al.}(2019)\citenamefont
		{Ciafaloni}, \citenamefont {Colferai},\ and\ \citenamefont
		{Veneziano}}]{Ciafaloni:2018uwe}%
	\BibitemOpen
	\bibfield  {author} {\bibinfo {author} {\bibfnamefont {Marcello}\
			\bibnamefont {Ciafaloni}}, \bibinfo {author} {\bibfnamefont {Dimitri}\
			\bibnamefont {Colferai}}, \ and\ \bibinfo {author} {\bibfnamefont {Gabriele}\
			\bibnamefont {Veneziano}},\ }\bibfield  {title} {\enquote {\bibinfo {title}
			{{Infrared features of gravitational scattering and radiation in the eikonal
					approach}},}\ }\href {\doibase 10.1103/PhysRevD.99.066008} {\bibfield
		{journal} {\bibinfo  {journal} {Phys. Rev. D}\ }\textbf {\bibinfo {volume}
			{99}},\ \bibinfo {pages} {066008} (\bibinfo {year} {2019})},\ \Eprint
	{http://arxiv.org/abs/1812.08137} {arXiv:1812.08137 [hep-th]} \BibitemShut
	{NoStop}%
	\bibitem [{\citenamefont {Laddha}\ and\ \citenamefont
		{Sen}(2020)}]{Laddha:2019yaj}%
	\BibitemOpen
	\bibfield  {author} {\bibinfo {author} {\bibfnamefont {Alok}\ \bibnamefont
			{Laddha}}\ and\ \bibinfo {author} {\bibfnamefont {Ashoke}\ \bibnamefont
			{Sen}},\ }\bibfield  {title} {\enquote {\bibinfo {title} {{Classical proof of
					the classical soft graviton theorem in D\ensuremath{>}4}},}\ }\href {\doibase
		10.1103/PhysRevD.101.084011} {\bibfield  {journal} {\bibinfo  {journal}
			{Phys. Rev. D}\ }\textbf {\bibinfo {volume} {101}},\ \bibinfo {pages}
		{084011} (\bibinfo {year} {2020})},\ \Eprint
	{http://arxiv.org/abs/1906.08288} {arXiv:1906.08288 [gr-qc]} \BibitemShut
	{NoStop}%
	\bibitem [{\citenamefont {P.~V.}\ and\ \citenamefont
		{Manu}(2020)}]{PV:2019uuv}%
	\BibitemOpen
	\bibfield  {author} {\bibinfo {author} {\bibfnamefont {Athira}\ \bibnamefont
			{P.~V.}}\ and\ \bibinfo {author} {\bibfnamefont {A.}~\bibnamefont {Manu}},\
	}\bibfield  {title} {\enquote {\bibinfo {title} {{Classical double copy from
					Color Kinematics duality: A proof in the soft limit}},}\ }\href {\doibase
		10.1103/PhysRevD.101.046014} {\bibfield  {journal} {\bibinfo  {journal}
			{Phys. Rev. D}\ }\textbf {\bibinfo {volume} {101}},\ \bibinfo {pages}
		{046014} (\bibinfo {year} {2020})},\ \Eprint
	{http://arxiv.org/abs/1907.10021} {arXiv:1907.10021 [hep-th]} \BibitemShut
	{NoStop}%
	\bibitem [{\citenamefont {Saha}\ \emph {et~al.}(2020)\citenamefont {Saha},
		\citenamefont {Sahoo},\ and\ \citenamefont {Sen}}]{Saha:2019tub}%
	\BibitemOpen
	\bibfield  {author} {\bibinfo {author} {\bibfnamefont {Arnab~Priya}\
			\bibnamefont {Saha}}, \bibinfo {author} {\bibfnamefont {Biswajit}\
			\bibnamefont {Sahoo}}, \ and\ \bibinfo {author} {\bibfnamefont {Ashoke}\
			\bibnamefont {Sen}},\ }\bibfield  {title} {\enquote {\bibinfo {title} {{Proof
					of the classical soft graviton theorem in $D$ = 4}},}\ }\href {\doibase
		10.1007/JHEP06(2020)153} {\bibfield  {journal} {\bibinfo  {journal} {JHEP}\
		}\textbf {\bibinfo {volume} {06}},\ \bibinfo {pages} {153} (\bibinfo {year}
		{2020})},\ \Eprint {http://arxiv.org/abs/1912.06413} {arXiv:1912.06413
		[hep-th]} \BibitemShut {NoStop}%
	\bibitem [{\citenamefont {Manu}\ \emph {et~al.}(2021)\citenamefont {Manu},
		\citenamefont {Ghosh}, \citenamefont {Laddha},\ and\ \citenamefont
		{Athira}}]{Manu:2020zxl}%
	\BibitemOpen
	\bibfield  {author} {\bibinfo {author} {\bibfnamefont {A.}~\bibnamefont
			{Manu}}, \bibinfo {author} {\bibfnamefont {Debodirna}\ \bibnamefont {Ghosh}},
		\bibinfo {author} {\bibfnamefont {Alok}\ \bibnamefont {Laddha}}, \ and\
		\bibinfo {author} {\bibfnamefont {P.~V.}\ \bibnamefont {Athira}},\ }\bibfield
	{title} {\enquote {\bibinfo {title} {{Soft radiation from scattering
					amplitudes revisited}},}\ }\href {\doibase 10.1007/JHEP05(2021)056}
	{\bibfield  {journal} {\bibinfo  {journal} {JHEP}\ }\textbf {\bibinfo
			{volume} {05}},\ \bibinfo {pages} {056} (\bibinfo {year} {2021})},\ \Eprint
	{http://arxiv.org/abs/2007.02077} {arXiv:2007.02077 [hep-th]} \BibitemShut
	{NoStop}%
	\bibitem [{\citenamefont {Sahoo}(2020)}]{Sahoo:2020ryf}%
	\BibitemOpen
	\bibfield  {author} {\bibinfo {author} {\bibfnamefont {Biswajit}\
			\bibnamefont {Sahoo}},\ }\bibfield  {title} {\enquote {\bibinfo {title}
			{{Classical Sub-subleading Soft Photon and Soft Graviton Theorems in Four
					Spacetime Dimensions}},}\ }\href {\doibase 10.1007/JHEP12(2020)070}
	{\bibfield  {journal} {\bibinfo  {journal} {JHEP}\ }\textbf {\bibinfo
			{volume} {12}},\ \bibinfo {pages} {070} (\bibinfo {year} {2020})},\ \Eprint
	{http://arxiv.org/abs/2008.04376} {arXiv:2008.04376 [hep-th]} \BibitemShut
	{NoStop}%
\end{thebibliography}
\end{document}